\DeclareFontFamily{OT1}{msb}{}{}
\DeclareFontShape{OT1}{msb}{m}{n}
 {  <5> <6> <7> <8> <9> <10> gen * msbm
      <10.95><12><14.4><17.28><20.74><24.88>msbm10}{}
\DeclareMathAlphabet{\bubble}{OT1}{msb}{m}{n}

\newfont{\bbd}{msbm10 scaled\magstep1}

\documentclass[russian]{article}
\usepackage[T2A]{fontenc}
\usepackage[cp866]{inputenc}
\usepackage[russian,english]{babel}
\usepackage{amsfonts}

\parskip=\medskipamount
\begin{document}

\def\l#1#2{\raisebox{.0ex}{$\displaystyle
  \mathop{#1}^{{\scriptstyle #2}\rightarrow}$}}
\def\r#1#2{\raisebox{.0ex}{$\displaystyle
\mathop{#1}^{\leftarrow {\scriptstyle #2}}$}}

\newcommand{\p}[1]{(\ref{#1})}

\newcommand{\sect}[1]{\setcounter{equation}{0}\section{#1}}

\makeatletter
\def\eqnarray{\stepcounter{equation}\let\@currentlabel=\theequation
\global\@eqnswtrue
\global\@eqcnt\z@\tabskip\@centering\let\\=\@eqncr
$$\halign to \displaywidth\bgroup\@eqnsel\hskip\@centering
  $\displaystyle\tabskip\z@{##}$&\global\@eqcnt\@ne
  \hfil$\displaystyle{\hbox{}##\hbox{}}$\hfil
  &\global\@eqcnt\tw@ $\displaystyle\tabskip\z@
  {##}$\hfil\tabskip\@centering&\llap{##}\tabskip\z@\cr}
\makeatother

\renewcommand{\thefootnote}{\fnsymbol{footnote}}


\begin{center}
{\LARGE {\bf Logical foundation of theoretical physics}}\\[0.3cm]

{}~

{\large G. A. Quznetsov\\
e-mail: gunn@chelcom.ru, lak@csu.ru, gunn@mail.ru}

{}~\\
\quad \\

\end{center}


{}~

\centerline{{\bf Abstract}}

\noindent

This is the logical foundation for for Relativity Theory, Probability Theory, 
and for Quantum Theory. Contents is the following:

1 Introduction. 2 Classical logic. 3 Time and space. 3.1 Recorders. 3.2 Time. 
3.3 Space. 3.4 Relativity. 4. Probability. 4.1 В-functions. 4.2 Independent 
tests. 4.3 Function of probability. 4.4 Conditional probability. 4.5 Classical 
probability 4.6 B-functions and classical propositional logic. 4.7 Consistency 
of the probability function. 4.7.1 Nonstandard numbers. 4.7.2 Model. 5. Quantum 
theory. 5.1 Events and the moving equations. 5.2 Rotations of the x5Ox4 system 
and В-bosonn. 5.3 Masses. 5.4 The one-mass states, particles and antiparticles. 
5.5 The two-masses states. 5.5.1 Neutrinno. 5.5.2 Electroweak transformations. 
5.6 Rotations of the Cartesian coordinates system and quarrks. 5.7 Gustatory 
pentads. 5.8 Two events. 5.9 The dimension of physical space. 5.10 The events-
probability interpretation of Quantum Theory. 6. Conclusion.

\vskip 2mm

\newpage


{}~

\begin{center}
{\LARGE {\bf Логические основания теоретической физики}}\\[0.3cm]
Г. A. Кузнецов {\\
e-mail: gunn@chelcom.ru, lak@csu.ru, gunn@mail.ru}~\\
\quad \\
\end{center}

{}~

\centerline{{\bf Аннотация}}
\noindent

Это логические основания для теории относительности, теории вероятностей и 
квантовой теории. 

\newpage


\tableofcontents

\pagestyle{plain}
\renewcommand{\thefootnote}{\arabic{footnote}}
\setcounter{footnote}{0}

\section{Введение}

Как мы знаем, всякая физическая теория состоит из базисных понятий 
и постулатов и из заключений, полученных из этих базисных элементов 
классической логикой. Мы имеем последовательность теорий, из которых каждая 
следующая объясняет предыдущую по правилам этой логики. То есть базисные 
термины и постулаты каждой следующей теории более логичны, чем базисные понятия 
и аксиомы предыдущей. И когда эти базисные элементы теории станут абсолютно 
логичными, то есть когда станут терминами и правилами классической логики, тогда 
теоретическая физика закончится, потому что это уже будет не физика, а логика. 
Похоже, что такая ситуация не за горами.

Содержание этой работы следующее:

Во второй главе на основании понятия истинности \cite{TAR44} строится вариант 
пропозициональной логики с булевой функцией.

В третьей главе пространственно-временные понятия и отношения выводятся из 
логических свойств информации:

  Всякая информация, полученная от физического прибора $a$, может быть 
выражена множеством $\widehat{a}$ предложений какого-либо языка. $\widehat{a}$ 
называется {\it рекордером прибора} $a$. Некоторые системы рекордеров образуют 
структуры, подобные часам. Следующие результаты получаются из логических 
свойств множества рекордеров \cite{QQ}:

  Во-первых, все такие часы "идут в одну сторону", т.е. если событие, 
выра-жаемое предложением $A$, предшествует событию, выраженному предложением 
$B$, по каким-нибудь из этих часов, то так же для остальных.

Во-вторых, время, определенное такими часами, оказывается необратимым, т. е. 
ни один рекордер не может получить информацию о том, что какое-либо событие 
произошло, до тех пор, пока это событие действительно не произойдет. Таким 
образом, никто не может вернуться в прошлое или получить информацию из 
будущего.

В третьих, множество рекордеров естественным образом погружается в 
метри-ческое пространство; т. е. все аксиомы метрического пространства 
получаются из логических свойств множества рекордеров.

В-четвертых, если это метрическое пространство окажется эвклидовым, то 
соответствующее "пространство-время" рекордеров подчиняется преобразованиям 
полной группы Пуанкаре. Т. е. в этом случае специальная теория относительности 
следует из логических свойств информации. Если это метрическое пространство 
не эвклидово, то любая нелинейная геометрия может быть сформулирована на 
пространстве рекордеров, и любой вариант общей теории относительности может 
быть реализован на этом пространстве.

В четвертой главе вводится понятие "событие" как понятие пропозициональной 
логики. Определяются операции над событиями и понятие "физическое событие".

В параграфе 4.1 вводится понятие "B-функции" \cite{SO}, удовлетворяющей всем 
свойствам классической функции вероятности.

В параграфе 4.2 для таких функций строится схема независимых испытаний и 
выводится формула Бернулли \cite{BER13}.

В параграфе 4.3 определяется понятие "P-функции", для которой выводится Закон 
Больших Чисел в форме Бернулли. Т.е. получается функция, подчиняюща-яся всем 
свойствам классической функции вероятности и имеющая статистический смысл.

В параграфе 4.4 для такой функции определяется понятие "условной вероят-ности". 
А и параграфе 4.5 - понятие "классической вероятности".

В параграфе 4.6 устанавливается связь между вероятностью и классической 
пропозициональной логикой.

В параграфе 4.7 четвертой главы строится версия нестандартного анализа 
\cite{DVS}, подходящего для доказательства непротиворечивости введенного 
понятия функции вероятности, и доказывается эта непротиворечивость.

В пятой главе основные понятия и соотношения квантовой теории 
представ-ляются как понятия и отношения той части теории вероятности, которая 
относится к точечным событиям в пространстве-времени \cite{AFLB}:

  В параграфе 5.1 вероятности, связывающие точечные события, выражаются 
спинорными функциями и {\it операторами рождения и уничтожения вероятностей}, 
аналогичными полевым операторам квантовой теории поля. Здесь же определяет-ся понятие 
клиффордового множества матриц, в частности - понятие легкой клиффордовой 
пентады, понятие цветных и вкусовых клиффордовых пентад\footnote{В параграфах 
5.2-5.6 рассматриваются уравнения движения, содержащие только элементы легкой 
пентады. Такие уравнения здесь называются {\it лептоннными}.}, и для спинорных 
функций получаются уравнения движения в форме уравнений Дирака с 
дополнительными полями, одни из которых образуют массовые члены, а другие ведут 
себя как калибровочные поля.

  В уравнение Дирака входят только четыре элемента пентады 
Клиффорда. Три из этих элементов соответствуют 
трем пространственным координатам, а четвертая - или образует 
массовый член, или соответствует четвертой - временной - 
координате. 

  Но пентада Клиффорда содержит пять элементов. 
Повидимому есть смысл дополнить уравнение Дирака еще одним 
массовым членом с пятым элементом пентады. То есть массовая 
часть уравнения Дирака будет содержать два члена. Более того, 
если этим двум массовым членам клиффордовой пентады поставить
в соответствие две дополнительных квазипространственных координаты, 
то получится однородное уравнение Дирака, в которое все пять 
элементов клиффор-довой пентады и все пять пространственных координат 
входят одинаковым образом. В таком пятимерном пространстве все 
локальные скорости по модулю равны единице.

   В параграфе 5.2 показано, что переопределенное подобным образом 
уравнение движения инвариантно относительно поворотов в 2-пространстве 
четвертой и пятой координат. Это преобразование определяет поле, аналогичное 
$B$-бозонному полю.

В параграфе 5.3 определяется понятие, соответствующее квантово-теоретиче-скому 
понятию массы:

  Так как значения вероятностей не определяются абсолютно точно, 
то дополни-тельные поля, образующие массовые члены представимы достаточно
тонкими слоями. А величины тех двух пространственных координат, которые 
вводятся для однородности массовых членов, можно ограничить большим числом без 
ограничения общности. В этом случае спектр масс получается дискретным, 
и в каждой точке 3+1 пространства-времени либо находится только одна масса, 
либо эта точка пустая.

  Масса представлена квадратным корнем из суммы квадратов двух целых чисел:

\[
m_0=\sqrt{n_0^2+s_0^2} 
\]

из двух массовых членов уравнения движения. Но это уравнение инвариантно 
относительно поворотов в 2-пространстве четвертой и пятой координаты. 
То есть масса сама должна быть выражена натуральным
числом, и числа в массовых членах должны оставаться целыми при поворотах.

  Следовательно, масса и составляющие ее числа массовых членов должны 
составлять пифагорову тройку $\left\langle m_0;n_0,s_0\right\rangle $ 
\cite{Pf}. Здесь $m_0$ называется {\it отцом} тройки. При поворотах в 
2-пространстве четвертой и пятой координаты одна пифагорова тройка должна 
заменяться на другую с тем же самым отцом. То есть для заданной 
точности определения угла поворота должно существовать такое {\it семейство 
пифагоровых троек} с одним и тем же отцом, что при повороте на этот угол 
одна тройка семейства заменяется другой тройкой из этого же семейства.
Для любой точности определения углов поворотов в дальних частях натурального 
ряда существуют отцы, обеспечивающие такую точность. Я полагаю, что эти 
семейства пифагоровых троек соответствуют семействам элементарных частиц.

  В параграфе 5.4 операторы рождения и уничтожения частиц определяются как 
Фурье-преобразования соответствующих операторов вероятности, а античастицы -
стандартным образом.
 
  В параграфе 5.5 рассматриваются все унитарные преобразования на двух-массовых
функциях, сохраняющие 4-вектор тока вероятности. Среди таких преобразований
есть соответствующие электрослабым калибровочным полям. Эти электрослабые
унитарные преобразования тоже выражаются поворотами в 2-пространстве четвертой
и пятой пространственных координат. Частицы, аналогичные нейтрино 
({\it нейтринно}), появляются в результате таких преобразова-ний. {\it Нейтринно} 
оказываются существенно связанными со своими {\it лептоннами}.

  Уравнения движения инвариантны относительно этих преобразований, и в 
результате получаются поля, аналогичные $W$-бозонным. Безмассовое поле

\[
F_{\mu ,\nu }=\partial _\mu W_\nu -\partial _\nu W_\mu -i\frac{g_2}2\left(
W_\mu W_\nu -W_\nu W_\mu \right)
\]

вводится обычным образом, но при решении уравнений Эйлера-Лагранжа получается,
что хотя $F_{\mu ,\nu }$ безмассовое, его образующие $W_\mu $ локально ведут 
себя как массивные поля.

  Безмассовое поле $A$ и массивное поле $Z$ определяются стандартным образом по
полям $B$ и $W$.

  В параграфе 5.6 показано, что лептоннное уравнение движения инвариантно 
относительно поворотов в 3-пространстве первых обычных пространственных 
координат и лоренцевых поворо-тов в 3+1 пространстве-времени. Уравнения 
движения, соствленные элементами цветных пентад, при таких поворотах 
переме-шиваются между собой. То есть частицы, соответствующие пентадам разных 
цветов, неразделимы в пространстве и времени (конфайнмент). Таких частиц 
в семействе два сорта по три цвета - всего шесть. Я называю их {\it кваррками}.

В параграфе 5.7 показано, что для вкусовых пентад в нынешней квантовой теории 
нет применения.

В параграфе 5.8 рассматривается ситуация с парой точечных событий и etc.

В параграфе 5.9 устанавливается соответствие между линейным пространством 
спиноров и нормированной алгеброй с делением и на основании теорем Гурвица и 
Фробениуса \cite{O1}, \cite{O2} получается вывод о размерности пространства 
точечных событий. 

В параграфе 5.10 развивается идея (Bergson, Whitehead, Capek, Whipple jr. and 
J. Jeans \cite{Brg}) интерпретации квантовой теории событиями.
 
\section{Классическая логика}

Рассматриваем повествовательные предложения русского языка.

{\bf Определение 2.1.1} Предложение $\ll \Theta \gg $ {\it истинно}, если и 
только если $\Theta $ \cite{TAR44}.

Например, предложение $\ll $Идет дождь$\gg $ истинно, если и только если идет 
дождь.

{\bf Определение 2.1.2} Предлоджение $\ll \Theta \gg $ {\it ложно}, если и только если 
неверно, что $\Theta $.

{\bf Определение 2.1.3} Предложения $A$ и $B$ {\it равны} ($A=B$), если $A$ истинно, 
если и только если $B$ истинно.

Далее я использую обычные понятия классической пропозициональной логики 
\cite{MEN63}.

{\bf Определение 2.1.4} Предложение $C$ называется {\it конъюнкцией} предложений 
$A$ и $B$ ($C=\left( A\&B\right) $), если $C$ истинно, если и только если 
$A$ истинно, и $B$ истинно.

{\bf Определение 2.1.5} Предложение $C$ называется {\it отрицанием} предложения 
$A$ ($C=\left( \neg A\right) $), если $C$ истинно, если и только если $A$ ложно.


{\bf Теорема 2.1.1}

1) $(A\&A)=A$;

2) $(A\&B)=(B\& A)$;

3) $(A\& (B\& C))=((A\& B)\& C)$;

4) если $T$ истинное предложение, то для каждого предложения $A$: 
$(A\& T)=A$.

5) если $F$ ложное предложение, то $(A\& F)=F$.

{\bf Доказательство Теоремы 2.1.1: } Из Определения 2.1.1, 
2.1.2, 2.1.3, 2.1.4.  $_{\bf \Box }$

{\bf Определение 2.1.6} Каждая функция $\mathfrak{g}$, определенная в множестве  
предложе-ний, имеющая область значений на двух-элементном множестве 
$\left\{ 0;1\right\}$ называется {\it булевой функцией} если:

1) $\mathfrak{g}\left( \neg A\right) =1-$ ${%
\mathfrak{g}}\left( A\right) $ для каждого предложения $A$;

2) $\mathfrak{g}\left( A\& B\right) ={%
\mathfrak{g}}\left( A\right)\cdot\mathfrak{g}\left( B\right) $ для всех 
предложений $A$ и $B$.

{\bf Определение 2.1.7} Множество $\Im $ предложений называется {\it базисным 
множе-ством}, если для каждого элемента $A$ этого множества существуют булевы 
функции ${\mathfrak{g}}_1$ и $\mathfrak{g}_2$, которые подчиняются следующим 
условиям:

1) $\mathfrak{g}_1\left( A\right) \neq \mathfrak{g}_2\left( A\right) $;

2) $\mathfrak{g}_1\left( B\right) =\mathfrak{g}_2\left( B\right) $ для 
каждого элемента $B$ множества $\Im $, такого что $B\neq A$.

{\bf Определение 2.1.8} Множество $\left[ \Im \right]$ предложений называется 
{\it пропозициональ-ным замыканием} множества $\Im$, если выполняются следующие 
условия:

1) если $A\in \Im $, то $A\in \left[ \Im \right ] $;

2) если  $A\in \left[ \Im \right ]$, то $\left(\neg A\right)\in \left[ \Im \right ] $;

3) если $A\in \left[ \Im \right ]$ и $B\in \left[ \Im \right ] $, то $%
\left( A\& B\right) \in \left[ \Im \right ] $;

4) кроме перечисленных первыми тремя пунктами этого определения, никаких других 
элементов множества $\left[ \Im \right ]$ нет.

В следующем тексте элементы множества $\left[ \Im \right] $ называются 
$\mathit{\Im }$-{\it предложениями}.

{\bf Определение 2.1.9} $\Im $-предложение $A$ называется {\it тавтологией}, 
если для всех булевых функций $\mathfrak{g}$:

\[
\mathfrak{g}(A)=1\mbox{.} 
\]

{\bf Определение 2.1.10} {\it Дизъюнкция} и {\it импликация} определяются обычным 
об-разом:

$
\begin{array}{c}
\left( A\vee B\right)\stackrel{def}{=}\left( \neg \left( \left( \neg A\right) \&\left( \neg
B\right) \right) \right) \mbox{,} \\ 
\left( A\Rightarrow B\right)\stackrel{def}{=}\left( \neg \left( A\&\left( \neg B\right)
\right) \right) \mbox{.}
\end{array}
$

По этому определению и по Определениям 2.1.4 и 2.1.5:

$\left( A\vee B\right) $ ложно, если и только если $A$ ложно и $B$ ложно.

$\left( A\Rightarrow B\right) $ ложно, если и только если $A$ истинно, а $B$ ложно.

{\bf Определение 2.1.11} $\Im $-предложение называется {\it пропозициональной 
аксио-мой}, \cite{MEN63}, если это предложение находится в одной из следующих 
форм:

\textbf{A1}. $\left( A\Rightarrow \left( B\Rightarrow A\right) \right) $;

\textbf{A2. }$\left( \left( A\Rightarrow \left( B\Rightarrow C\right)
\right) \Rightarrow \left( \left( A\Rightarrow B\right) \Rightarrow \left(
A\Rightarrow C\right) \right) \right) $;

\textbf{A3}. $\left( \left(\left(\neg B\right)\Rightarrow\left(\neg A\right)\right)
\Rightarrow \left( \left(\left(\neg B\right)\Rightarrow A\right) \Rightarrow
B\right) \right) $.

Пусть $\Im $ - некоторое базисное множество. Ниже я рассматриваю только 
$\Im $-предложения. 

{\bf Определение 2.1.12} Предложение $B$ получается из предложений $\left( A\Rightarrow
B\right) $ и $A$ по логическому правилу {\it modus ponens}.

{\bf Определение 2.1.13} \cite{MEN63} Последовательность $A_1,A_2,\ldots ,A_n$ 
предложений на-зывается {\it пропозициональным выводом} предложения $A$ из 
списка гипотез $\Gamma $ (обозначение: $\Gamma \vdash A$), если $A_n=A$ и для всех 
чисел $l$ ($1\leq l\leq n$): $A_l$ есть либо пропозициональная аксиома, либо 
$A_l$ получается из каких-либо предложений $A_{l-k}$ и $A_{l-s}$ по modus 
ponens, либо $A_l\in \Gamma $.

{\bf Определение 2.1.14} Предложение называется {\it пропозиционально доказуемым} 
предложением, если это предложение есть пропозициональная аксиома или это 
предложение получается из пропозиционально доказуемых предложений по modus 
ponens.

Следовательно, если $A$ пропозиционально доказуемо, то существует 
пропози-циональный вывод вида:

\[
\vdash A.
\]

{\bf Теорема 2.1.2} \cite{MEN63} Если предложение $A$ пропозиционально доказуемо, 
то для всех булевых функций $\mathfrak{g}$: $\mathfrak{\
g}\left( A\right) =1$.

{\bf Доказательство Теоремы 2.1.2: }\cite{MEN63}  $_{\bf \Box }$

{\bf Теорема 2.1.3} {\bf (Теорема о полноте)} \cite{MEN63} Все тавтологии 
пропозиционально доказуемы.

{\bf Доказательство Теоремы 2.1.3: }\cite{MEN63}  $_{\bf \Box }$

\section{Время и пространство}
\subsection{Рекордеры}

Информация, получаемая от физических приборов, может быть выражена текстом, 
составленным повествовательными предложениями.

Пусть $\widehat{{\bf a}}$ - какой-либо объект, способный принимать, сохранять 
и/или переда-вать информацию \cite{Ku}. Множество ${\bf a}$ предложений, 
выражающих информацию объекта $\widehat{{\bf a}}$, называется {\it рекордером} 
этого объекта. Т.е. выражение: $\ll$ Предложение $\ll A\gg$ есть элемент 
множества ${\bf a}.\gg $ означает: $\ll \widehat{{\bf a}}$ имеет информацию о 
том, что произошло событие, выражаемое предложением $\ll A\gg. \gg $. Или 
короче: $\ll \widehat{{\bf a}}$ знает, что $A.\gg$. Или по обозначению: 
$\ll {\bf a}^{\bullet }\ll A\gg\gg $. 

Очевидно, выполняются следующие условия:

I. Для любого ${\bf a}$ и для каждого $A$: неверно, что ${\bf a}^{\bullet
}\left( A\&\left( \neg A\right) \right) $, т.е. ни один рекордер не содержит 
логического противоречия.

II. Для каждого ${\bf a}$, для любого $B$ и для всех $A$: если $B$ 
следует из $A$, и ${\bf a}^{\bullet }A$, то ${\bf a}%
^{\bullet }B$.

*III. Для всех ${\bf a}$, $b$ и для каждого $A$: если ${\bf a}^{\bullet }\ll 
{\bf b}^{\bullet }A\gg $, то ${\bf a}^{\bullet }A$.

Например, если прибор $\widehat{{\bf a}}$ имеет информацию о том, что 
прибор $\widehat{{\bf b}}$ имеет информацию о том, что масса частицы $%
\overleftarrow{\chi }$ равна $7$, то прибор $\widehat{{\bf a}}
$ имеет информацию о том, что масса частицы $\overleftarrow{\chi }$ равна $7$.

\subsection{Время}

Рассматриваем конечные (возможно - пустые) последовательности символов 
типа: ${\bf q}^{\bullet }$.

{\bf Определение 3.2.1} Последовательность $\alpha $ есть {\it подпоследовательность} 
после-довательности $\beta $ ($\alpha\prec\beta $), если $\alpha $ может быть 
получена из $\beta $ вычеркиванием некоторых (возможно - всех) элементов.

Обозначим: $\left( \alpha \right) ^1$ есть $\alpha $, и $\left(
\alpha \right) ^{k+1}$ есть $\alpha \left( \alpha \right) ^k$.

Поэтому, если $k\leq l$, то $\left( \alpha \right) ^k\prec \left( \alpha
\right) ^l$.

{\bf Определение 3.2.2} Последовательность $\alpha $ {\it эквивалентна} последовательности $\beta $ 
($\alpha\sim\beta $), если $\alpha $ может быть получена из $\beta $ заменой 
подпоследовательности типа $\left( {\bf a}^{\bullet }\right) ^k$ 
последова-тельностью такого же типа ($\left( {\bf a}^{\bullet }\right) ^s$).

В этом случае:

III. Если $\beta \prec \alpha $ или $\beta \sim \alpha $, то для любого $K$:

если ${\bf a}^{\bullet }K$, то $\mathbf{a}^{\bullet }\left( K\&\left(\alpha A\Rightarrow \beta A\right)\right) $.

Очевидно, III есть уточнение условия *III.

{\bf Определение 3.2.3} Число $q$ есть {\it момент}, в который ${\bf a}$ регистрирует $B$ по $%
\kappa-${\it часам} $\left\{ {\bf g}_0,A,{\bf b}_0\right\} $ (обозначение: 
${\bf q}$ есть $\left[ {\bf a}^{\bullet }B\uparrow {\bf a}%
,\left\{ {\bf g}_0,A,{\bf b}_0\right\} \right] $), если:

1. для любого $K$: если ${\bf a}^{\bullet }K$, то $\mathbf{a}^{\bullet }\left( K\&\left( \mathbf{a}^{\bullet }B\Rightarrow 
\mathbf{a}^{\bullet }\left( \mathbf{g}_0^{\bullet }\mathbf{b}_0^{\bullet
}\right) ^q\mathbf{g}_0^{\bullet }A\right) \right) $ и 

$\mathbf{a}^{\bullet }\left( K\&\left( \mathbf{a}^{\bullet }\left( \mathbf{g}%
_0^{\bullet }\mathbf{b}_0^{\bullet }\right) ^{q+1}\mathbf{g}_0^{\bullet
}A\Rightarrow \mathbf{a}^{\bullet }B\right) \right) $.

2.7.1. ${\bf a}^{\bullet }\left({\bf a}^{\bullet }B \&\left( \neg
 {\bf a}^{\bullet }\left( {\bf g}_0^{\bullet }{\bf b}_0^{\bullet
}\right) ^{q+1}{\bf g}_0^{\bullet }A\right) \right) $.

{\bf Лемма 3.2.1 }Если

\begin{equation}
q\mbox{ есть }\left[ {\bf a}^{\bullet }\alpha B\uparrow {\bf a},\left\{ {\bf g}%
_0,A,{\bf b}_0\right\} \right] ,  \label{b1}
\end{equation}

\begin{equation}
p\mbox{ есть }\left[ {\bf a}^{\bullet }\beta B\uparrow {\bf a},\left\{ {\bf g}%
_0,A,{\bf b}_0\right\} \right] ,  \label{bo2}
\end{equation}

\begin{equation}
\alpha \prec \beta \mbox{,} \label{bo3}
\end{equation} 

то

\[
q\leq p \mbox{.}
\]

{\bf Доказательство:} Из (\ref{bo2}):

\[
{\bf a}^{\bullet }\left( \left( {\bf a}^{\bullet }\beta B\right) \&\left(
\neg  {\bf a}^{\bullet }\left( {\bf g}_0^{\bullet }{\bf b}_0^{\bullet
}\right) ^{\left( p+1\right) }{\bf g_0^{\bullet }}A\right)\right) . 
\]

Отсюда и из (\ref{bo3}) по III:

\[
\mathbf{a}^{\bullet }\left( \left( \mathbf{a}^{\bullet }\beta B\&\left( \neg 
\mathbf{a}^{\bullet }\left( \mathbf{g}_0^{\bullet }\mathbf{b}_0^{\bullet
}\right) ^{\left( p+1\right) }\mathbf{g_0^{\bullet }}A\right) \right)
\&\left( \mathbf{a}^{\bullet }\beta B\Rightarrow \mathbf{a}^{\bullet }\alpha
B\right) \right) \mbox{.}
\]

Отсюда по II:

\[
{\bf a}^{\bullet }\left({\bf a}^{\bullet }\alpha B \&\left(
\neg {\bf a}^{\bullet }\left( {\bf g}_0^{\bullet }{\bf b}_0^{\bullet
}\right) ^{\left( p+1\right) }{\bf g_0^{\bullet }}A\right) \right) \mbox{.} 
\]

Отсюда и из (\ref{b1}):

\[
\mathbf{a}^{\bullet }\left( \left( \mathbf{a}^{\bullet }\alpha B\&\left(
\neg \mathbf{a}^{\bullet }\left( \mathbf{g}_0^{\bullet }\mathbf{b}%
_0^{\bullet }\right) ^{\left( p+1\right) }\mathbf{g_0^{\bullet }}A\right)
\right) \&\left( \mathbf{a}^{\bullet }\alpha B\Rightarrow \mathbf{a}%
^{\bullet }\left( \mathbf{g}_0^{\bullet }\mathbf{b}_0^{\bullet }\right) ^q%
\mathbf{g_0^{\bullet }}A\right) \right) \mbox{.}
\]

Отсюда по II:

\begin{equation}
{\bf a}^{\bullet }\left( \left( \neg {\bf a}^{\bullet }\left( {\bf g}%
_0^{\bullet }{\bf b}_0^{\bullet }\right) ^{\left( p+1\right) }{\bf %
g_0^{\bullet }}A \right) \& {\bf a}^{\bullet }\left( {\bf g}%
_0^{\bullet }{\bf b}_0^{\bullet }\right) ^q{\bf g_0^{\bullet }}A
\right)  \label{b4}
\end{equation}

Если $q>p$, т.е. $q\geq p+1$, то из (\ref{b4}) по III:

\[
\mathbf{a}^{\bullet }\left( 
\begin{array}{c}
\left( \left( \neg \mathbf{a}^{\bullet }\left( \mathbf{g}_0^{\bullet }%
\mathbf{b}_0^{\bullet }\right) ^{\left( p+1\right) }\mathbf{g_0^{\bullet }}%
A\right) \& \mathbf{a}^{\bullet }\left( \mathbf{g}_0^{\bullet }\mathbf{%
b}_0^{\bullet }\right) ^q\mathbf{g_0^{\bullet }}A \right) \& \\ 
\left( \mathbf{a}^{\bullet }\left( \mathbf{g}_0^{\bullet }\mathbf{b}%
_0^{\bullet }\right) ^q\mathbf{g_0^{\bullet }}A\Rightarrow \mathbf{a}%
^{\bullet }\left( \mathbf{g}_0^{\bullet }\mathbf{b}_0^{\bullet }\right)
^{\left( p+1\right) }\mathbf{g_0^{\bullet }}A\right) 
\end{array}
\right) \mbox{.}
\]

Отсюда по II:

\[
\mathbf{a}^{\bullet }\left( \left( \neg \mathbf{a}^{\bullet }\left( \mathbf{g%
}_0^{\bullet }\mathbf{b}_0^{\bullet }\right) ^{\left( p+1\right) }\mathbf{%
g_0^{\bullet }}A\right) \&\mathbf{a}^{\bullet }\left( \mathbf{g}_0^{\bullet }%
\mathbf{b}_0^{\bullet }\right) ^{\left( p+1\right) }\mathbf{g_0^{\bullet }}%
A\right) \mbox{,}
\]

Это противоречит условию I. Поэтому, $q\leq p$ $_{\bf \Box }$

Из Леммы 3.2.1 непосредственно следует, что если $q$ есть 
$\left[ {\bf a}^{\bullet }B\uparrow {\bf a},\left\{ {\bf g}_0,A,%
{\bf b}_0\right\} \right] $, и $p$ есть $\left[ {\bf a}^{\bullet }B\uparrow 
{\bf a},\left\{ {\bf g}_0,A,{\bf b}_0\right\} \right] $, то $q=p$. Поэтому,
выражение $\ll q$ есть $\left[ {\bf a}^{\bullet }B\uparrow {\bf a}%
,\left\{ {\bf g}_0,A,{\bf b}_0\right\} \right] \gg $ эквивалентно выражению 
$\ll q$ $=$ $\left[ {\bf a}^{\bullet }B\uparrow {\bf a},\left\{ 
{\bf g}_0,A,{\bf b}_0\right\} \right] \gg $.

{\bf Определение 3.2.4} $\kappa$-часы $\left\{ {\bf g}_1,B,{\bf b}_1\right\} $ и 
$\left\{{\bf g}_2,B,{\bf b}_2\right\} $ имеют {\it одинаковое направ-ление} для 
${\bf a}$, если выполняется следующее условие:

если

\begin{center}
$r${\ $=$ $\left[ {\bf a}^{\bullet }\left( {\bf g}_1^{\bullet }{\bf b}%
_1^{\bullet }\right) ^q{\bf g}_1^{\bullet }B\uparrow {\bf a},\left\{ {\bf g}%
_2,B,{\bf b}_2\right\} \right] $,}\\

$s$ $=$ $\left[ {\bf a}^{\bullet }\left( {\bf g}_1^{\bullet }{\bf b}%
_1^{\bullet }\right) ^p{\bf g}_1^{\bullet }B\uparrow {\bf a},\left\{ {\bf g}%
_2,B,{\bf b}_2\right\} \right] $,\\

$q<p$\mbox{,}
\end{center}

то

\begin{center}
$r\leq s$.\\
\end{center}

{\bf Теорема 3.2.1} Все $\kappa-$часы имеют одинаковое направление.

{\bf Доказательство:} Пусть

\begin{eqnarray*}
&&r\stackrel{def}{=}\left[ {\bf a}^{\bullet }\left( {\bf g}_1^{\bullet }{\bf b}_1^{\bullet
}\right) ^q{\bf g}_1^{\bullet }B\uparrow {\bf a},\left\{ {\bf g}_2,B,{\bf b}%
_2\right\} \right] \mbox{,}\\ 
&&s\stackrel{def}{=}\left[ {\bf a}^{\bullet }\left( {\bf g}_1^{\bullet }{\bf b}_1^{\bullet
}\right) ^p{\bf g}_1^{\bullet }B\uparrow {\bf a},\left\{ {\bf g}_2,B,{\bf b}%
_2\right\} \right] \mbox{,} 
\end{eqnarray*}

\[
q<p\mbox{.} 
\]

В этом случае:

\[
\left( {\bf g}_1^{\bullet }{\bf b}_1^{\bullet }\right) ^q\prec \left( {\bf g}%
_1^{\bullet }{\bf b}_1^{\bullet }\right) ^p. 
\]

Поэтому по Лемме 3.2.1:

\[
r\leq s_{\bf \Box }
\]

Следовательно, рекордер упорядочивает свои предложения по моментам. Причем, 
этот порядок линейный и не зависит от того, по которым $\kappa-$часам он 
установлен.

{\bf Определение 3.2.5} $\kappa-$часы $\left\{ {\bf g}_2,B,{\bf b}_2\right\} $ 
в $k$ {\it раз точнее} $\kappa-$часов $\left\{ {\bf g}_1,B,{\bf b}_1\right\}$ 
для рекордера ${\bf a}$, если для каждого $C$ выполняется следующее условие:

если

\begin{center}
$q_1$ $=$ $\left[ {\bf a}^{\bullet }C\uparrow {\bf a},\left\{ {\bf g}_1,B,%
{\bf b}_1\right\} \right] $,\\

$q_2$ $=$ $\left[ {\bf a}^{\bullet }C\uparrow {\bf a},\left\{ {\bf g}_2,B,%
{\bf b}_2\right\} \right] $,\\
\end{center}

то

\begin{center}
$q_1<\frac{q_2}k<q_1+1$.\\
\end{center}


{\bf Лемма 3.2.2} Если для каждого $n$: 

\begin{equation}
q{_{n-1}\ <\ }\frac{q{_n}}{k_n}{\ <\ }q{_{n-1}+1}\mbox{,}  \label{ser}
\end{equation}

то ряд

\[
q_0+\sum_{n=1}^\infty \frac{q_n-q_{n-1}k_n}{k_1\ldots k_n}  \label{ser1}
\]

сходится.

{\bf Доказательство:} Из (\ref{ser}):

\[
0\leq q_n-q_{n-1}k_n<k_n \mbox{.}
\]

Следовательно, ряд (\ref{ser1}) знакоположительный и мажорируется рядом 

\[
q_0+1+\sum_{n=1}^\infty \frac 1{k_1\ldots k_n} \mbox{,}
\]

сходимость которого элементарно проверяется признаком д'Аламбера $_{\bf \Box }$

{\bf Определенние 3.2.6} Последовательность $\widetilde{H}$ $\kappa-$часов:

\begin{center}
$\left\langle {\left\{ {\bf g}_0,A,{\bf b}_0\right\} ,\ \left\{ {\bf g}_1,A,%
{\bf b}_2\right\} ,...,\left\{ {\bf g}_j,A,{\bf b}_j\right\} ,\ ...\ }%
\right\rangle $\\
\end{center}

называется {\it абсолютно точными $\kappa-$часами} рекордера ${\bf a}$, если 
для каждого $j$ существует натуральное число $k_j$, такое что $\kappa-$часы  
$\left\{ {\bf g}_j,A,{\bf b}_j\right\} $ в $k_j$ раз точнее $\kappa-$часов 
$\left\{ {\bf g}_{j-1},A,{\bf b}_{j-1}\right\} $.

В этом случае если

\[
q_j=\left[ {\bf a}^{\bullet }C\uparrow {\bf a},\left\{ {\bf g}_j,A,{\bf b}%
_j\right\} \right] \mbox{,} 
\]

\[
t=q_0+\sum_{j=1}^\infty \frac{q_j-q_{j-1}\cdot k_j}{k_1\cdot k_2\cdot
...\cdot k_j} \mbox{,}
\]

то

\begin{center}
$t$ есть $\left[ {\bf a}^{\bullet }C\uparrow {\bf a},\widetilde{H}\right] $.\\
\end{center}

{\bf Лемма 3.2.3} Если
 
\begin{equation}
q\stackrel{def}{=}q_0+\sum_{j=1}^\infty \frac{q_j-q_{j-1}\cdot k_j}{k_1\cdot k_2\cdot
...\cdot k_j} \label{rd1}
\end{equation}

с

\[
q_{n-1}\leq \frac{q_n}{k_n}<q_{n-1}+1 \mbox{,}
\]

\begin{equation}
d\stackrel{def}{=}d_0+\sum_{j=1}^\infty \frac{d_j-d_{j-1}\cdot k_j}{k_1\cdot k_2\cdot
...\cdot k_j} \label{rd2}
\end{equation}

с

\[
d_{n-1}\leq \frac{d_n}{k_n}<d_{n-1}+1 \mbox{,}
\]

то если $q_n\leq d_n$, то $q\leq d$.

{\bf Доказательство} Частичная сумма ряда (\ref{rd1}):

$Q_u\stackrel{def}{=}q_0+\frac{q_1-q_0k_1}{k_1}+\frac{q_2-q_1k_2}{k_1k_2}+\cdots +\frac{%
q_u-q_{u-1}k_u}{k_1k_2\cdots k_u}$,

$Q_u=q_0+\frac{q_1}{k_1}-q_0+\frac{q_2}{k_1k_2}-\frac{q_1}{k_1}+\cdots +%
\frac{q_u}{k_1k_2\cdots k_u}-\frac{q_{u-1}}{k_1k_2\cdots k_{u-1}}$,

\begin{center}
$Q_u=\frac{q_u}{k_1k_2\cdots k_u}$.
\end{center}

Аналогично, частичная сумма ряда (\ref{rd2}):

\begin{center}
$D_u=\frac{d_u}{k_1k_2\cdots k_u}$.
\end{center}

Следовательно, из условия Леммы: $Q_n\leq D_n$ $_{\bf \Box }$

{\bf Лемма 3.2.4} Если 

\begin{center}
$q$ есть $\left[ \mathbf{a}^{\bullet }\alpha C\uparrow \mathbf{a}%
,\widetilde{H}\right] $,

$d$ есть $\left[ \mathbf{a}^{\bullet }\beta C\uparrow \mathbf{a},%
\widetilde{H}\right] $, и

$\alpha \prec \beta $, то

$q\leq d$
\end{center}

{\bf Доказательство} получается сразу из Лемм 3.2.1 и 3.2.3 $_{\bf \Box }$

Аналогично, - если $\alpha \sim \beta $, то $q=d$.

\subsection{Пространство}

{\bf Определение 3.3.1} Число $t$ называется {\it временем, измеренным 
рекордером ${\bf a}$ по часам $\widetilde{H}$, за которое сигнал $C$ прошел путь 
последовательности ${\bf a}^{\bullet }\alpha {\bf a}%
^{\bullet }$} (обозначение

\begin{center}
$t\stackrel{def}{=}\jmath \left( {\bf a}\widetilde{H}\right) \left( {\bf a}^{\bullet }\alpha 
{\bf a}^{\bullet }C\right)$),
\end{center}

если

\[
t=\left[ {\bf a}^{\bullet }\alpha {\bf a}%
^{\bullet }C\uparrow {\bf a},\widetilde{H}\right] -\left[ {\bf a}^{\bullet
}C\uparrow {\bf a},\widetilde{H}\right] \mbox{.}
\]

{\bf Теорема 3.3.1} 

\[
\jmath \left( {\bf a}\widetilde{H}\right) \left( {\bf a}^{\bullet }\alpha 
{\bf a}^{\bullet }C\right)\geq 0 \mbox{.}
\]

{\bf Доказательство} сразу следует из Леммы 3.2.4 $_{\bf \Box }$

Таким образом, любой "сигнал", "посланный" рекордером, "вернется" к нему не 
раньше, чем был "послан".

{\bf Определение 3.3.2} 

1) Для каждого рекордера ${\bf a}$: $\left( {\bf a}^{\bullet }\right) ^{\dagger }=\left( 
{\bf a}^{\bullet }\right) $;

2) для всех последовательностей $\alpha $ и $\beta $: $\left( \alpha \beta
\right) ^{\dagger }=\left( \beta \right) ^{\dagger }\left( \alpha \right)
^{\dagger }$.

{\bf Определение 3.3.3} Множество $\Re $ рекордеров есть {\it внутренне 
неподвижная система} для рекордера ${\bf a}$ с $\kappa-$часами 
$\widetilde{H}$ (обозначение: $\Re $ есть $ISS\left( {\bf a},\widetilde{H}\right) $), 
если для всех предложений $B$ и $C$, для всех элементов ${\bf a}_1$ и 
${\bf a}_2$ множества $\Re $ и для всех последовательностей $\alpha $, 
образованных элементами множества $\Re $, выполняются следующие условия:

1) {$\left[ {\bf a}^{\bullet }{\bf a}_2^{\bullet }{\bf a}_1^{\bullet
}C\uparrow {\bf a},\widetilde{H}\right] -\left[ {\bf a}^{\bullet }{\bf a}%
_1^{\bullet }C\uparrow {\bf a},\widetilde{H}\right] =$}

$=${$\left[ {\bf a}^{\bullet }{\bf a}_2^{\bullet }{\bf a}_1^{\bullet
}B\uparrow {\bf a},\widetilde{H}\right] -\left[ {\bf a}^{\bullet }{\bf a}%
_1^{\bullet }B\uparrow {\bf a},\widetilde{H}\right] $;}

2) {$\jmath \left( {\bf a}\widetilde{H}\right) \left( {\bf a}^{\bullet
}\alpha {\bf a}^{\bullet }C\right) =\jmath \left( {\bf a}\widetilde{H}%
\right) \left( {\bf a}^{\bullet }\alpha ^{\dagger }{\bf a}^{\bullet
}C\right) $.}

{\bf Теорема 3.3.2}

\begin{center}
$\left\{ {\bf a}\right\} - ISS\left( {\bf a},\widetilde{H}\right) $.
\end{center}

{\bf Доказательство}
 
1) Т.к. $\mathbf{a}^{\bullet }\sim \mathbf{a}^{\bullet }\mathbf{a}^{\bullet }$, 
то по Лемме 3.2.4: если обозначить:

\begin{eqnarray*}
&&p\stackrel{def}{=}\left[ \mathbf{a}^{\bullet }\mathbf{a}^{\bullet }B\uparrow \mathbf{a},%
\widetilde{H}\right] \mbox{,}\\
&&q\stackrel{def}{=}\left[ \mathbf{a}^{\bullet }\mathbf{a}^{\bullet }\mathbf{a}^{\bullet
}B\uparrow \mathbf{a},\widetilde{H}\right]\mbox{,}\\
&&r\stackrel{def}{=}\left[ \mathbf{a}^{\bullet }\mathbf{a}^{\bullet }C\uparrow \mathbf{a},%
\widetilde{H}\right]\mbox{,}\\
&&s\stackrel{def}{=}\left[ \mathbf{a}^{\bullet }\mathbf{a}^{\bullet }\mathbf{a}^{\bullet
}C\uparrow \mathbf{a},\widetilde{H}\right]\mbox{,}
\end{eqnarray*}

то $q=p$ и $s=r$. Поэтому $q-p=s-r$.

2) Т.к. любой ряд $\alpha$, составленный из элементов множества 
$\left\{ {\bf a}\right\}$, совпадает с $\alpha ^{\dagger }$, то 

\begin{center}
{$\jmath \left( \mathbf{a}\widetilde{H}\right) \left( \mathbf{a}^{\bullet
}\alpha \mathbf{a}^{\bullet }C\right) =\jmath \left( \mathbf{a}\widetilde{H}%
\right) \left( \mathbf{a}^{\bullet }\alpha ^{\dagger }\mathbf{a}^{\bullet
}C\right) $.} $_{\bf \Box }$
\end{center}

{\bf Лемма 3.3.1} Если $\left\{ {\bf a},{\bf a}_1,{\bf a}_2\right\} $ есть $%
ISS\left( {\bf a},\widetilde{H}\right) $, то

\begin{center}
$\left[ {\bf a}^{\bullet }{\bf a}_2^{\bullet }{\bf a}_1^{\bullet }{\bf a}%
_2^{\bullet }C\uparrow {\bf a},\widetilde{H}\right] -\left[ {\bf a}^{\bullet
}{\bf a}_2^{\bullet }C\uparrow {\bf a},\widetilde{H}\right] =$\\

$=\left[ {\bf a}^{\bullet }{\bf a}_1^{\bullet }{\bf a}_2^{\bullet }{\bf a}%
_1^{\bullet }B\uparrow {\bf a},\widetilde{H}\right] -\left[ {\bf a}^{\bullet
}{\bf a}_1^{\bullet }B\uparrow {\bf a},\widetilde{H}\right] $\\
\end{center}

{\bf Доказательство} Обозначим:

\begin{eqnarray*}
&&p\stackrel{def}{=}\left[ {\bf a}^{\bullet }{\bf a}_1^{\bullet }B\uparrow {\bf a},\widetilde{%
H}\right]\mbox{,}\\
&&q\stackrel{def}{=}\left[ {\bf a}^{\bullet }{\bf a}_1^{\bullet }{\bf a}_2^{\bullet }{\bf a}%
_1^{\bullet }B\uparrow {\bf a},\widetilde{H}\right] \mbox{,}\\
&&r\stackrel{def}{=}\left[ {\bf a}^{\bullet }{\bf a}_2^{\bullet }C\uparrow {\bf a},\widetilde{%
H}\right] \mbox{,}\\
&&s\stackrel{def}{=}\left[ {\bf a}^{\bullet }{\bf a}_2^{\bullet }{\bf a}_1^{\bullet }{\bf a}%
_2^{\bullet }C\uparrow {\bf a},\widetilde{H}\right] \mbox{,}\\
&&u\stackrel{def}{=}\left[ {\bf a}^{\bullet }{\bf a}_2^{\bullet }{\bf a}_1^{\bullet
}B\uparrow {\bf a},\widetilde{H}\right] \mbox{,}\\
&&w\stackrel{def}{=}\left[ {\bf a}^{\bullet }{\bf a}_1^{\bullet }{\bf a}_2^{\bullet
}C\uparrow {\bf a},\widetilde{H}\right] \mbox{.}
\end{eqnarray*}

Отсюда по Определению 3.3.3:

\[
u-p=s-w,w-r=q-u. 
\]

Отсюда:

\[
s-r=q-p_{{\bf \Box }} 
\]

{\bf Определение 3.3.4} {\it ${\bf a}\widetilde{H}(B)$-мерой} рекордеров 
${\bf a_1}$ и ${\bf a_2}$ (обозначение: \\${\ell }${$\left( {\bf a},\widetilde{H},B\right) \left( {\bf a}_1,{\bf a}%
_2\right)$) называется число:

\begin{center}
${\ell }${$\left( {\bf a},\widetilde{H},B\right) \left( {\bf a}_1,{\bf a}%
_2\right) =0.5\cdot \left( \left[ {\bf a}^{\bullet }{\bf a}_1^{\bullet }{\bf %
a}_2^{\bullet }{\bf a}_1^{\bullet }B\uparrow {\bf a},\widetilde{H}\right]
-\left[ {\bf a}^{\bullet }{\bf a}_1^{\bullet }B\uparrow {\bf a},\widetilde{H}%
\right] \right) $.}\\
\end{center}

{\bf Лемма 3.3.2} Если $\left\{ {\bf a},{\bf a}_1,{\bf a}%
_2\right\} $ - $ISS\left( {\bf a},\widetilde{H}\right) $, то для всех $B$
и $C$:

\begin{center}
${\ell }${$\left( {\bf a},\widetilde{H},B\right) \left( {\bf a}_1,{\bf a}%
_2\right) =$}${\ell }${$\left( {\bf a},\widetilde{H},C\right) \left( {\bf a}%
_1,{\bf a}_2\right) $.}\\
\end{center}

{\bf Доказательство} Обозначим:

\begin{eqnarray*}
&&p\stackrel{def}{=}\left[ {\bf a}^{\bullet }{\bf a}_1^{\bullet }B\uparrow {\bf a},\widetilde{%
H}\right] \mbox{,}\\
&&q\stackrel{def}{=}\left[ {\bf a}^{\bullet }{\bf a}_1^{\bullet }{\bf a}_2^{\bullet }{\bf a}%
_1^{\bullet }B\uparrow {\bf a},\widetilde{H}\right] \mbox{,}\\
&&r\stackrel{def}{=}\left[ {\bf a}^{\bullet }{\bf a}_1^{\bullet }C\uparrow {\bf a},\widetilde{%
H}\right] \mbox{,}\\
&&s\stackrel{def}{=}\left[ {\bf a}^{\bullet }{\bf a}_1^{\bullet }{\bf a}_2^{\bullet }{\bf a}%
_1^{\bullet }C\uparrow {\bf a},\widetilde{H}\right] \mbox{,}\\
&&u\stackrel{def}{=}\left[ {\bf a}^{\bullet }{\bf a}_2^{\bullet }{\bf a}_1^{\bullet
}B\uparrow {\bf a},\widetilde{H}\right] \mbox{,}\\
&&w\stackrel{def}{=}\left[ {\bf a}^{\bullet }{\bf a}_2^{\bullet }{\bf a}_1^{\bullet
}C\uparrow {\bf a},\widetilde{H}\right] \mbox{.}
\end{eqnarray*}

Отсюда по Определению 3.3.3:

\[
u-p=w-r,q-u=s-w. 
\]

Отсюда:

\[
q-p=s-r._{{\bf \Box }} 
\]

Поэтому можно писать вместо: $\ll {\ell }${$\left( {\bf a},\widetilde{H},B\right) \left( {\bf a}_1,{\bf a}%
_2\right)\gg$ - $\ll{\ell }${$\left( {\bf a},\widetilde{H}\right) \left( {\bf a}_1,%
{\bf a}_2\right)\gg $. 

{\bf Теорема 3.3.3}: Если $\left\{ {\bf a},{\bf a}_1,{\bf a}_2,{\bf a}_3\right\} $
есть $ISS\left( {\bf a},\widetilde{H}\right) $, то:

1) ${\ell }${$\left( {\bf a},\widetilde{H}\right) \left( {\bf a}_1,{\bf a}%
_2\right) \geq 0$;}

2) ${\ell }${$\left( {\bf a},\widetilde{H}\right) \left( {\bf a}_1,{\bf a}%
_1\right) =0$;}

3) ${\ell }${$\left( {\bf a},\widetilde{H}\right) \left( {\bf a}_1,{\bf a}%
_2\right) =$}${\ell }${$\left( {\bf a},\widetilde{H}\right) \left( {\bf a}_2,%
{\bf a}_1\right) $;}

4) ${\ell }${$\left( {\bf a},\widetilde{H}\right) \left( {\bf a}_1,{\bf a}%
_2\right) +$}${\ell }${$\left( {\bf a},\widetilde{H}\right) \left( {\bf a}_2,%
{\bf a}_3\right) \geq $}${\ell }${$\left( {\bf a},\widetilde{H}\right)
\left( {\bf a}_1,{\bf a}_3\right) $.}

{\bf Доказательство} 1) и 2) прямо следуют из Леммы 3.2.4, а 3) - из Леммы 3.3.2.

Обозначим:

\begin{eqnarray*}
&&p\stackrel{def}{=}\left[ {\bf a}^{\bullet }{\bf a}_1^{\bullet }C\uparrow {\bf a},\widetilde{%
H}\right] \mbox{,}\\
&&q\stackrel{def}{=}\left[ {\bf a}^{\bullet }{\bf a}_1^{\bullet }{\bf a}_2^{\bullet }{\bf a}%
_1^{\bullet }C\uparrow {\bf a},\widetilde{H}\right] \mbox{,}\\
&&r\stackrel{def}{=}\left[ {\bf a}^{\bullet }{\bf a}_1^{\bullet }{\bf a}_3^{\bullet }{\bf a}%
_1^{\bullet }C\uparrow {\bf a},\widetilde{H}\right] \mbox{,}\\
&&s\stackrel{def}{=}\left[ {\bf a}^{\bullet }{\bf a}_2^{\bullet }{\bf a}_1^{\bullet
}C\uparrow {\bf a},\widetilde{H}\right] \mbox{,}\\
&&u\stackrel{def}{=}\left[ {\bf a}^{\bullet }{\bf a}_2^{\bullet }{\bf a}_3^{\bullet }{\bf a}%
_2^{\bullet }{\bf a}_1^{\bullet }B\uparrow {\bf a},\widetilde{H}\right] \mbox{,}\\
&&w=\left[ {\bf a}^{\bullet }{\bf a}_1^{\bullet }{\bf a}_2^{\bullet }{\bf a}%
_3^{\bullet }{\bf a}_2^{\bullet }{\bf a}_1^{\bullet }C\uparrow {\bf a},%
\widetilde{H}\right] \mbox{.}
\end{eqnarray*}

Отсюда по Определению 3.3.3:

\[
w-u=q-s, 
\]

Поэтому

\[
w-p=\left( q-p\right) +\left( u-s\right) . 
\]

И по Лемме 3.2.4:

\[
w\geq r. 
\]

Следовательно:

\[
\left( q-p\right) +\left( u-s\right) \geq r-p_{{\bf \Box }} 
\]

Таким образом, {\bf все четыре аксиомы метрического пространства} \cite{MSP} 
{\bf выполняются для} ${\ell }${$\left( {\bf a},\widetilde{H}\right) $ 
{\bf во внутренне неподвижной системе рекорде-ров.} Следовательно, 
${\ell }${$\left( {\bf a},\widetilde{H}\right) $ играет роль длины расстояния 
в этом пространстве. 

{\bf Определение 3.3.5} Множество рекордеров $\Re$ {\it вырождено в луч 
${\bf a}{\bf b_1}$ и точку ${\bf a_1}$}, если найдется $C$, для которого 
выполняются условия:

1) Для любой последовательности $\alpha$, соствленной из элементов множества 
$\Re$, и для любого $K$: если ${\bf a}^{\bullet }K$, то 

\begin{center}
$\mathbf{a}^{\bullet }\left( K\&\left( \alpha \mathbf{a}_1^{\bullet
}C\Rightarrow \alpha \mathbf{b}_1^{\bullet }\mathbf{a}_1^{\bullet }C\right)
\right) $.
\end{center}

2) Существует последовательность $\beta$, составленная из элементов множества 
$\Re$, и предложение $S$ такие, что 
$\mathbf{a}^{\bullet }\left( \beta \mathbf{b}_1^{\bullet }C\&S\right) $ и 
неверно, что
$\mathbf{a}^{\bullet }\left( \beta \mathbf{a}_1^{\bullet }\mathbf{b}%
_1^{\bullet }C\&S\right) $

Далее рассматриваем только невырожденные множества рекордеров.

{\bf Определение 3.3.6} $B$ произошло в {\it одном месте} с ${\bf a}_1$ для 
${\bf a}$ (обозна-чение:$\natural \left( {\bf a}\right) \left( {\bf a}_1,B\right) $), 
если для каждой последовательности $\alpha $ и для любого предло-жения $K$ 
выполняется следующее условие:

если ${\bf a}^{\bullet }K$, то $\mathbf{a}^{\bullet }\left( K\&\left( \alpha B\Rightarrow \alpha \mathbf{a}%
_1^{\bullet }B\right) \right) $

{\bf Теорема 3.3.4} 

\begin{center}
$\natural \left( {\bf a}\right) \left( {\bf a}_1,{\bf a}%
_1^{\bullet }B\right) $.
\end{center}

{\bf Доказательство} Т.к. $\alpha {\bf a}_1^{\bullet }\sim \alpha {\bf a}%
_1^{\bullet }{\bf a}_1^{\bullet }$, то по III: если ${\bf a}_1^{\bullet }K$,
то

\[
\mathbf{a}_1^{\bullet }\left( K\&\left( \alpha \mathbf{a}_1^{\bullet
}B\Rightarrow \alpha \mathbf{a}_1^{\bullet }\mathbf{a}_1^{\bullet }B\right)
\right) _{\mathbf{\Box }}
\]

{\bf Теорема 3.3.5} Если

\begin{equation}
\natural \left( {\bf a}\right) \left( {\bf a}_1,B\right) ,  \label{bo23}
\end{equation}

\begin{equation}
\natural \left( {\bf a}\right) \left( {\bf a}_2,B\right) ,  \label{bo24}
\end{equation}

то

\[
\natural \left( {\bf a}\right) \left( {\bf a}_2,{\bf a}_1^{\bullet }B\right)
. 
\]

{\bf Доказательство} Пусть ${\bf a}^{\bullet }K$. В этом случае из (\ref{bo24}):

\[
\mathbf{a}^{\bullet }\left( K\&\left( \alpha \mathbf{a}_1^{\bullet
}B\Rightarrow \alpha \mathbf{a}_1^{\bullet }\mathbf{a}_2^{\bullet }B\right)
\right) \mbox{.}
\]

Из (\ref{bo23}):

\[
\mathbf{a}^{\bullet }\left( \left( K\&\left( \alpha \mathbf{a}_1^{\bullet
}B\Rightarrow \alpha \mathbf{a}_1^{\bullet }\mathbf{a}_2^{\bullet }B\right)
\right) \&\left( \alpha \mathbf{a}_1^{\bullet }\mathbf{a}_2^{\bullet
}B\Rightarrow \alpha \mathbf{a}_1^{\bullet }\mathbf{a}_2^{\bullet }\mathbf{a}%
_1^{\bullet }B\right) \right) \mbox{.}
\]

Отсюда по II:

\[
\mathbf{a}^{\bullet }\left( K\&\left( \alpha \mathbf{a}_1^{\bullet
}B\Rightarrow \alpha \mathbf{a}_1^{\bullet }\mathbf{a}_2^{\bullet }\mathbf{a}%
_1^{\bullet }B\right) \right) \mbox{.}
\]

Отсюда по III:

\[
\mathbf{a}^{\bullet }\left( \left( K\&\left( \alpha \mathbf{a}_1^{\bullet
}B\Rightarrow \alpha \mathbf{a}_1^{\bullet }\mathbf{a}_2^{\bullet }\mathbf{a}%
_1^{\bullet }B\right) \right) \&\left( \alpha \mathbf{a}_1^{\bullet }\mathbf{%
a}_2^{\bullet }\mathbf{a}_1^{\bullet }B\Rightarrow \alpha \mathbf{a}%
_2^{\bullet }\mathbf{a}_1^{\bullet }B\right) \right) \mbox{.}
\]

Отсюда по II:

\[
\mathbf{a}^{\bullet }\left( K\&\left( \alpha \mathbf{a}_1^{\bullet
}B\Rightarrow \alpha \mathbf{a}_2^{\bullet }\mathbf{a}_1^{\bullet }B\right)
\right) _{\mathbf{\Box }}
\]

{\bf Лемма 3.3.3} Если

\begin{equation}
\natural \left( {\bf a}\right) \left( {\bf a}_1,B\right) \mbox{,}  \label{bo25}
\end{equation}

\begin{equation}
t\ =\ \left[ {\bf a}^{\bullet }\alpha B\uparrow {\bf a},\widetilde{H}\right]
\mbox{,}  \label{bo26}
\end{equation}

то

\[
t\ =\ \left[ {\bf a}^{\bullet }\alpha {\bf a}_1^{\bullet }B\uparrow {\bf a},%
\widetilde{H}\right] \mbox{.} 
\]

{\bf Доказательство} Обозначим:

\[
t_j\stackrel{def}{=}\left[ {\bf a}^{\bullet }\alpha B\uparrow {\bf a},\left\{ {\bf g}_j,A,%
{\bf b}_j\right\} \right] \mbox{.} 
\]

Поэтому:

\[
{\bf a}^{\bullet } \left( {\bf a}^{\bullet }\alpha B \&\left(
\neg {\bf a}^{\bullet }\left( {\bf g}_j^{\bullet }{\bf b}_j^{\bullet
}\right) ^{t_j+1}{\bf g}_j^{\bullet }A \right) \right) \mbox{,} 
\]

и из (\ref{bo25}):

\[
\mathbf{a}^{\bullet }\left( \left( \mathbf{a}^{\bullet }\alpha B\&\left(
\neg \mathbf{a}^{\bullet }\left( \mathbf{g}_j^{\bullet }\mathbf{b}%
_j^{\bullet }\right) ^{t_j+1}\mathbf{g}_j^{\bullet }A\right) \right)
\&\left( \mathbf{a}^{\bullet }\alpha B\Rightarrow \mathbf{a}^{\bullet
}\alpha \mathbf{a}_1^{\bullet }B\right) \right) \mbox{.}
\]

Отсюда по II:

\begin{equation}
{\bf a}^{\bullet }\left( {\bf a}^{\bullet }\alpha {\bf a}_1^{\bullet
}B \&\left( \neg {\bf a}^{\bullet }\left( {\bf g}_j^{\bullet }%
{\bf b}_j^{\bullet }\right) ^{t_j+1}{\bf g}_j^{\bullet }A \right)
\right) \mbox{,}  \label{bo27}
\end{equation}

Пусть ${\bf a}^{\bullet }K$. В этом случае из (\ref{bo26}):

\[
\mathbf{a}^{\bullet }\left( K\&\left( \mathbf{a}^{\bullet }\alpha
B\Rightarrow \mathbf{a}^{\bullet }\left( \mathbf{g}_j^{\bullet }\mathbf{b}%
_j^{\bullet }\right) ^{t_j}\mathbf{g}_j^{\bullet }A\right) \right) \mbox{.}
\]

Поэтому по III:

\[
\mathbf{a}^{\bullet }\left( \left( K\&\left( \mathbf{a}^{\bullet }\alpha
B\Rightarrow \mathbf{a}^{\bullet }\left( \mathbf{g}_j^{\bullet }\mathbf{b}%
_j^{\bullet }\right) ^{t_j}\mathbf{g}_j^{\bullet }A\right) \right) \&\left( 
\mathbf{a}^{\bullet }\alpha \mathbf{a}_1^{\bullet }B\Rightarrow \mathbf{a}%
^{\bullet }\alpha B\right) \right) \mbox{.}
\]

Отсюда по II:

\begin{equation}
\mathbf{a}^{\bullet }\left( K\&\left( \mathbf{a}^{\bullet }\alpha \mathbf{a}%
_1^{\bullet }B\Rightarrow \mathbf{a}^{\bullet }\left( \mathbf{g}_j^{\bullet }%
\mathbf{b}_j^{\bullet }\right) ^{t_j}\mathbf{g}_j^{\bullet }A\right) \right) %
\mbox{.}  \label{bo28}
\end{equation}

Отсюда и из (\ref{bo25}):

\[
\mathbf{a}^{\bullet }\left( \left( K\&\left( \mathbf{a}^{\bullet }\left( 
\mathbf{g}_j^{\bullet }\mathbf{b}_j^{\bullet }\right) ^{t_j+1}\mathbf{g}%
_j^{\bullet }A\Rightarrow \mathbf{a}^{\bullet }\alpha B\right) \right)
\&\left( \mathbf{a}^{\bullet }\alpha B\Rightarrow \mathbf{a}^{\bullet
}\alpha \mathbf{a}_1^{\bullet }B\right) \right) \mbox{.}
\]

Отсюда по II:

\[
\mathbf{a}^{\bullet }\left( K\&\left( \mathbf{a}^{\bullet }\left( \mathbf{g}%
_j^{\bullet }\mathbf{b}_j^{\bullet }\right) ^{t_j+1}\mathbf{g}_j^{\bullet
}A\Rightarrow \mathbf{a}^{\bullet }\alpha \mathbf{a}_1^{\bullet }B\right)
\right) \mbox{.}
\]

Отсюда и из (\ref{bo27}), (\ref{bo28}) для всех $j$:

\[
t_j=\left[ {\bf a}^{\bullet }\alpha {\bf a}_1^{\bullet }B\uparrow {\bf a}%
,\left\{ {\bf g}_j,A,{\bf b}_j\right\} \right] . 
\]

Следовательно,

\[
t\ =\ \left[ {\bf a}^{\bullet }\alpha {\bf a}_1^{\bullet }B\uparrow {\bf a},%
\widetilde{H}\right] _{\bf \Box }
\]

{\bf Теорема 3.3.6} Если $\left\{ {\bf a},{\bf a}_1,{\bf a}_2\right\} $ - $%
ISS\left( {\bf a},\widetilde{H}\right) $,

\begin{equation}
\natural \left( {\bf a}\right) \left( {\bf a}_1,B\right) \mbox{,}  \label{bo29}
\end{equation}

\begin{equation}
\natural \left( {\bf a}\right) \left( {\bf a}_2,B\right) \mbox{,}  \label{bo30}
\end{equation}

то

\[
\ell \left( {\bf a},\widetilde{H}\right) \left( {\bf a}_1,{\bf a}_2\right) =0%
\mbox{.} 
\]

{\bf Доказательство} Обозначим:

\[
t\stackrel{def}{=}\left[ {\bf a}^{\bullet }B\uparrow {\bf a},\widetilde{H}\right] . 
\]

Отсюда по Лемме 3.3.3:

из (\ref{bo29}):

\[
t\ =\ \left[ {\bf a}^{\bullet }{\bf a}_1^{\bullet }B\uparrow {\bf a},%
\widetilde{H}\right] \mbox{,} 
\]

из (\ref{bo30}):

\[
t=\left[ {\bf a}^{\bullet }{\bf a}_1^{\bullet }{\bf a}_2^{\bullet }B\uparrow 
{\bf a},\widetilde{H}\right] \mbox{,} 
\]

снова из (\ref{bo29}):

\[
t=\left[ {\bf a}^{\bullet }{\bf a}_1^{\bullet }{\bf a}_2^{\bullet }{\bf a}%
_1^{\bullet }B\uparrow {\bf a},\widetilde{H}\right] . 
\]

Следовательно:

\[
\ell \left( {\bf a},\widetilde{H}\right) \left( {\bf a}_1,{\bf a}_2\right)
=0.5\cdot \left( t-t\right) =0 _{{\bf \Box }} 
\]

{\bf Теорема 3.3.7} Если $\left\{ {\bf a}_1,{\bf a}_2,{\bf a}_3\right\} $ - $%
ISS\left( {\bf a},\widetilde{H}\right) $ и существует предложение $B$ такое, 
что:

\begin{equation}
\natural \left( {\bf a}\right) \left( {\bf a}_1,B\right) \mbox{,}  \label{bo31}
\end{equation}

\begin{equation}
\natural \left( {\bf a}\right) \left( {\bf a}_2,B\right) \mbox{,}  \label{bo32}
\end{equation}

то

\[
{\ell \left( {\bf a},\widetilde{H}\right) \left( {\bf a}_3,{\bf a}_2\right)
=\ell \left( {\bf a},\widetilde{H}\right) \left( {\bf a}_3,{\bf a}_1\right) \mbox{.}%
} 
\]

{\bf Доказательство} По Теореме 3.3.6 из (\ref{bo31}) и (\ref{bo32}):

\begin{equation}
\ell \left( {\bf a},\widetilde{H}\right) \left( {\bf a}_1,{\bf a}_2\right)
=0\mbox{;}  \label{bo33}
\end{equation}

По Теореме 3.3.3:

\[
{\ell \left( {\bf a},\widetilde{H}\right) \left( {\bf a}_1,{\bf a}_2\right)
+\ell \left( {\bf a},\widetilde{H}\right) \left( {\bf a}_2,{\bf a}_3\right)
\geq \ell \left( {\bf a},\widetilde{H}\right) \left( {\bf a}_1,{\bf a}%
_3\right) }\mbox{,} 
\]

поэтому из (\ref{bo33}):

\[
{\ell \left( {\bf a},\widetilde{H}\right) \left( {\bf a}_2,{\bf a}_3\right)
\geq \ell \left( {\bf a},\widetilde{H}\right) \left( {\bf a}_1,{\bf a}%
_3\right) }\mbox{,} 
\]

т.е., по Теореме 3.3.3:

\begin{equation}
{\ell \left( {\bf a},\widetilde{H}\right) \left( {\bf a}_3,{\bf a}_2\right)
\geq \ell \left( {\bf a},\widetilde{H}\right) \left( {\bf a}_1,{\bf a}%
_3\right) }\mbox{.}  \label{bo34}
\end{equation}

Из

\[
{\ell \left( {\bf a},\widetilde{H}\right) \left( {\bf a}_3,{\bf a}_1\right)
+\ell \left( {\bf a},\widetilde{H}\right) \left( {\bf a}_1,{\bf a}_2\right)
\geq \ell \left( {\bf a},\widetilde{H}\right) \left( {\bf a}_3,{\bf a}%
_2\right) }\mbox{:} 
\]

\[
{\ell \left( {\bf a},\widetilde{H}\right) \left( {\bf a}_3,{\bf a}_1\right)
\geq \ell \left( {\bf a},\widetilde{H}\right) \left( {\bf a}_3,{\bf a}%
_2\right) }\mbox{.} 
\]

Отсюда и из (\ref{bo34}):

\[
{\ell \left( {\bf a},\widetilde{H}\right) \left( {\bf a}_3,{\bf a}_1\right)
=\ell \left( {\bf a},\widetilde{H}\right) \left( {\bf a}_3,{\bf a}_2\right) }%
 _{{\bf \Box }} 
\]

{\bf Определение 3.3.7} Вещественное число $t$ есть момент предложения $B$ в {\it 
системе отсчета} $\left( \Re {\bf a}\widetilde{H}\right)$ (обозначение: 
$t=\left[ B\mid \Re {\bf a}\widetilde{H}\right] $), если

1) $\Re $ - $ISS\left( {\bf a},\widetilde{H}\right) $,

2) найдется рекордер ${\bf b}$ такой, что: ${\bf b}\in \Re $ и $\natural
\left( {\bf a}\right) \left( {\bf b},B\right) $),

3) $t=${$\left[ {\bf a}^{\bullet }B\uparrow {\bf a},\widetilde{H}\right] -$}$%
{\ell }${$\left( {\bf a},\widetilde{H}\right) \left( {\bf a},{\bf b}\right) $%
.}

{\bf Лемма 3.3.4}

\[
{\left[ {\bf a}^{\bullet }B\uparrow {\bf a},\widetilde{H}\right] =}\left[ 
{\bf a}^{\bullet }B\mid \Re {\bf a}\widetilde{H}\right] \mbox{.} 
\]

{\bf Доказательство} Пусть $\Re $ - $ISS\left( {\bf a},\widetilde{H}\right) $, 
${\bf a}_1\in \Re $ и

\begin{equation}
\natural \left( {\bf a}\right) \left( {\bf a}_1,{\bf a}^{\bullet }B\right) %
\mbox{.}  \label{bo35}
\end{equation}

По Теореме 3.3.4:

\[
\natural \left( {\bf a}\right) \left( {\bf a},{\bf a}^{\bullet }B\right) %
\mbox{.} 
\]

Отсюда и из (\ref{bo35}) по Теореме 3.3.6:

\[
\ell \left( {\bf a},\widetilde{H}\right) \left( {\bf a},{\bf a}_1\right) =0%
\mbox{,} 
\]

поэтому

\[
\left[ {\bf a}^{\bullet }B\mid \Re {\bf a}\widetilde{H}\right] ={\left[ {\bf %
a}^{\bullet }B\uparrow {\bf a},\widetilde{H}\right] -}\ell \left( {\bf a},%
\widetilde{H}\right) \left( {\bf a},{\bf a}_1\right) ={\left[ {\bf a}%
^{\bullet }B\uparrow {\bf a},\widetilde{H}\right] } _{{\bf \Box }} 
\]

{\bf Определение 3.3.8} Вещественное число $z$ есть {\it длина расстояния} 
между $B$ и $C$ в системе отсчета $\left( \Re {\bf a}\widetilde{H}\right) $
(обозначение: $z={\ell }\left( \Re {\bf a}\widetilde{H}\right) \left( B,C\right)$%
, если

1) $\Re $ - $ISS\left( {\bf a},\widetilde{H}\right) $,

2) существуют рекордеры ${\bf a}_1$ и ${\bf a}_2$ такие, что: ${\bf a}_1\in
\Re $, ${\bf a}_2\in \Re $, $\natural \left( {\bf a}\right) \left( {\bf a}%
_1,B\right) $) и $\natural \left( {\bf a}\right) \left( {\bf a}_2,C\right) 
$),

3) $z={\ell }${$\left( {\bf a},\widetilde{H}\right) \left( {\bf a}_2,{\bf a}%
_1\right) $.}

Из Теоремы 3.3.3: {\bf такая длина расстояния подчиняется всем аксиомам 
метрического пространства.}

\subsection{Относительность}

{\bf Определение 3.4.1} Рекордеры ${\bf a}_1$ и ${\bf a}_2$ {\it одинаково 
принимают сигнал о} $B$, для рекордера ${\bf a}$, если 

\begin{center}
$\ll\natural \left( {\bf a}\right) \left( {\bf a}_2,{\bf a}_1^{\bullet }B\right)
\gg $ = $\ll \natural \left( {\bf a}\right) \left( {\bf a}_1,%
{\bf a}_2^{\bullet }B\right) \gg $.
\end{center}

{\bf Определение 3.4.2} Множество рекордеров называем {\it однородным 
прост-ранством рекордеров}, если все его элементы одинаково принимают все 
сигналы.

{\bf Определение 3.4.3} Вещественное число $c$ есть {\it скорость 
распространения информации} о $B$ к рекордеру ${\bf a}_1$ в системе 
отсчета $\left( \Re {\bf a}\widetilde{H}\right) $ если

\begin{center}
$c=\frac{{\ell }\left( \Re \mathbf{a}\widetilde{H}\right) \left( B,\mathbf{a}%
_1^{\bullet }B\right) }{\left[ \mathbf{a}_1^{\bullet }B\mid \Re \mathbf{a}%
\widetilde{H}\right] -\left[ B\mid \Re \mathbf{a}\widetilde{H}\right] }$. 
\end{center}

{\bf Теорема 3.4.1} Во всех однородных пространствах:

\begin{center}
$c=1$.\\
\end{center}

{\bf Доказательство} Пусть $c$ - скорость распространения информации о $B$ к 
рекордеру ${\bf a}_1$ в системе отсчета $\left( \Re {\bf a}\widetilde{H}\right)$. 
Т.е.: если

\[
\Re \mbox{ есть }ISS\left( {\bf a},\widetilde{H}\right) , 
\]

\begin{eqnarray}
&&z\stackrel{def}{=}{\ell }\left( \Re {\bf a}\widetilde{H}\right) \left( B,{\bf a}_1^{\bullet
}B\right) \mbox{,}  \label{bo36}\\
&&t_1\stackrel{def}{=}\left[ B\mid \Re {\bf a}\widetilde{H}\right] \mbox{,}  \label{bo37}\\
&&t_2\stackrel{def}{=}\left[ {\bf a}_1^{\bullet }B\mid \Re {\bf a}\widetilde{H}\right] ,
\label{bo38}
\end{eqnarray}

то

\begin{equation}
c=\frac{z}{t_2-t_1}\mbox{.}
\end{equation}

Из (\ref{bo36}): существуют элементы ${\bf b}_1$ и ${\bf b}_2$ множества $\Re$ 
такие, что:

\begin{equation}
\natural \left( {\bf a}\right) \left( {\bf b}_1,B\right) \mbox{,}
\label{b40}
\end{equation}

\begin{equation}
\natural \left( {\bf a}\right) \left( {\bf b}_2,{\bf a}_2^{\bullet }B\right) %
\mbox{,}  \label{b41}
\end{equation}

\begin{equation}
z={\ell \left( {\bf a},\widetilde{H}\right) \left( {\bf b}_1,{\bf b}%
_2\right) }\mbox{.}  \label{b42}
\end{equation}

Из (\ref{bo37}) и (\ref{bo38}): существуют элементы ${\bf b}_1^{\prime }$ и 
${\bf b}_2^{\prime }$ множества $\Re $ такие, что:

\begin{equation}
\natural \left( {\bf a}\right) \left( {\bf b}_1^{\prime },B\right) \mbox{,}
\label{b43}
\end{equation}

\begin{equation}
\natural \left( {\bf a}\right) \left( {\bf b}_2^{\prime },{\bf a}_2^{\bullet
}B\right) \mbox{,}  \label{b44}
\end{equation}

\begin{equation}
t_1={\left[ {\bf a}^{\bullet }B\uparrow {\bf a},\widetilde{H}\right] -\ell
\left( {\bf a},\widetilde{H}\right) \left( {\bf a},{\bf b}_1^{\prime
}\right) }\mbox{,}  \label{b45}
\end{equation}

\begin{equation}
t_2={\left[ {\bf a}^{\bullet }{\bf a}_2^{\bullet }B\uparrow {\bf a},%
\widetilde{H}\right] -\ell \left( {\bf a},\widetilde{H}\right) \left( {\bf a}%
,{\bf b}_2^{\prime }\right) }\mbox{.}  \label{b46}
\end{equation}

Из (\ref{bo36}), (\ref{b40}), (\ref{b43}) по Теореме 3.3.7:

\begin{equation}
{\ell \left( {\bf a},\widetilde{H}\right) \left( {\bf a},{\bf b}_1\right)
=\ell \left( {\bf a},\widetilde{H}\right) \left( {\bf a},{\bf b}_1^{\prime
}\right) }\mbox{.}  \label{b47}
\end{equation}

Аналогично из (\ref{bo36}), (\ref{b41}), (\ref{b44}):

\begin{equation}
{\ell \left( {\bf a},\widetilde{H}\right) \left( {\bf a},{\bf b}_2\right)
=\ell \left( {\bf a},\widetilde{H}\right) \left( {\bf a},{\bf b}_2^{\prime
}\right) }\mbox{.}  \label{b48}
\end{equation}

Из (\ref{b45}), (\ref{b40}), (\ref{b47}) по Лемме 3.3.3:

\begin{equation}
t_1={\left[ {\bf a}^{\bullet }{\bf b}_1^{\bullet }B\uparrow {\bf a},%
\widetilde{H}\right] -\ell \left( {\bf a},\widetilde{H}\right) \left( {\bf a}%
,{\bf b}_1\right) }\mbox{.}  \label{b49}
\end{equation}

Из (\ref{b41}) по Лемме 3.3.3:

\begin{equation}
{\left[ {\bf a}^{\bullet }{\bf a}_2^{\bullet }B\uparrow {\bf a},\widetilde{H}%
\right] =\left[ {\bf a}^{\bullet }{\bf b}_2^{\bullet }{\bf a}_2^{\bullet
}B\uparrow {\bf a},\widetilde{H}\right] }\mbox{.}  \label{b50}
\end{equation}

По Лемме 3.2.4:

\begin{equation}
{\left[ {\bf a}^{\bullet }{\bf b}_2^{\bullet }{\bf a}_2^{\bullet }B\uparrow 
{\bf a},\widetilde{H}\right] \geq \left[ {\bf a}^{\bullet }{\bf b}%
_2^{\bullet }B\uparrow {\bf a},\widetilde{H}\right] }\mbox{.}  \label{b51}
\end{equation}

Из (\ref{b41}):

\[
\natural \left( {\bf a}\right) \left( {\bf a}_2,{\bf b}_2^{\bullet }B\right) %
\mbox{.} 
\]

Отсюда по Лемме 3.3.3:

\begin{equation}
{\left[ {\bf a}^{\bullet }{\bf a}_2^{\bullet }{\bf b}_2^{\bullet }B\uparrow 
{\bf a},\widetilde{H}\right] =\left[ {\bf a}^{\bullet }{\bf b}_2^{\bullet
}B\uparrow {\bf a},\widetilde{H}\right] }\mbox{.}  \label{b52}
\end{equation}

Снова по Лемме 3.2.4:

\[
{\left[ {\bf a}^{\bullet }{\bf a}_2^{\bullet }{\bf b}_2^{\bullet }B\uparrow 
{\bf a},\widetilde{H}\right] \geq \left[ {\bf a}^{\bullet }{\bf a}%
_2^{\bullet }B\uparrow {\bf a},\widetilde{H}\right] }\mbox{.} 
\]

Отсюда и из (\ref{b52}), (\ref{b50}), (\ref{b51}):

\[
{\left[ {\bf a}^{\bullet }{\bf a}_2^{\bullet }B\uparrow {\bf a},\widetilde{H}%
\right] \geq \left[ {\bf a}^{\bullet }{\bf b}_2^{\bullet }B\uparrow {\bf a},%
\widetilde{H}\right] \geq \left[ {\bf a}^{\bullet }{\bf a}_2^{\bullet
}B\uparrow {\bf a},\widetilde{H}\right] }\mbox{,} 
\]

поэтому

\[
{\left[ {\bf a}^{\bullet }{\bf a}_2^{\bullet }B\uparrow {\bf a},\widetilde{H}%
\right] =\left[ {\bf a}^{\bullet }{\bf b}_2^{\bullet }B\uparrow {\bf a},%
\widetilde{H}\right] }\mbox{.} 
\]

Отсюда и из (\ref{b46}), (\ref{b48}):

\[
t_2={\left[ {\bf a}^{\bullet }{\bf b}_2^{\bullet }B\uparrow {\bf a},%
\widetilde{H}\right] -\ell \left( {\bf a},\widetilde{H}\right) \left( {\bf a}%
,{\bf b}_2\right) }\mbox{.} 
\]

Отсюда и из (\ref{b40}) по Лемме 3.3.3:

\begin{equation}
t_2={\left[ {\bf a}^{\bullet }{\bf b}_2^{\bullet }{\bf b}_1^{\bullet
}B\uparrow {\bf a},\widetilde{H}\right] -\ell \left( {\bf a},\widetilde{H}%
\right) \left( {\bf a},{\bf b}_2\right) }\mbox{.}  \label{b53}
\end{equation}

Обозначим:

\begin{eqnarray}
&&u\stackrel{def}{=}\left[ {\bf a}^{\bullet }C\uparrow {\bf a},\widetilde{H}\right]\mbox{,}
\label{b54}\\
&&d\stackrel{def}{=}\left[ {\bf a}^{\bullet }{\bf b}_1^{\bullet }{\bf a}^{\bullet }C\uparrow 
{\bf a},\widetilde{H}\right] \mbox{,}  \label{b55}\\
&&w\stackrel{def}{=}\left[ {\bf a}^{\bullet }{\bf b}_2^{\bullet }{\bf a}^{\bullet }C\uparrow 
{\bf a},\widetilde{H}\right] \mbox{,}  \label{b56}\\
&&j\stackrel{def}{=}\left[ {\bf a}^{\bullet }{\bf b}_2^{\bullet }{\bf b}_1^{\bullet }{\bf a}%
^{\bullet }C\uparrow {\bf a},\widetilde{H}\right] \mbox{,}  \label{b57}\\
&&q\stackrel{def}{=}\left[ {\bf a}^{\bullet }{\bf b}_1^{\bullet }{\bf b}_2^{\bullet }{\bf a}%
^{\bullet }C\uparrow {\bf a},\widetilde{H}\right] \mbox{,}\nonumber\\ 
&&p\stackrel{def}{=}\left[ \mathbf{a}^{\bullet }\mathbf{b}_1^{\bullet }%
\mathbf{b}_2^{\bullet }\mathbf{b}_1^{\bullet }\mathbf{a}^{\bullet }C\uparrow 
\mathbf{a},\widetilde{H}\right]\mbox{,}
\label{b58}\\ 
&&r\stackrel{def}{=}\left[ {\bf a}^{\bullet }{\bf b}_2^{\bullet }{\bf b}_1^{\bullet }{\bf b}%
_2^{\bullet }{\bf a}^{\bullet }C\uparrow {\bf a},\widetilde{H}\right] %
\mbox{.}\nonumber 
\end{eqnarray}

Т.к. $\Re$ - $ISS\left( {\bf a},\widetilde{H}\right) $, то

\begin{equation}
q-w=p-j\mbox{,}  \label{b59}
\end{equation}

\begin{equation}
j=q\mbox{.}  \label{b60}
\end{equation}

И из (\ref{b53}), (\ref{b49}), (\ref{b55}), (\ref{b57}):

\[
\left( t_2+{\ell \left( {\bf a},\widetilde{H}\right) \left( {\bf a},{\bf b}%
_2\right) }\right) -\left( t_1+{\ell \left( {\bf a},\widetilde{H}\right)
\left( {\bf a},{\bf b}_1\right) }\right) =j-d\mbox{,} 
\]

поэтому

\begin{equation}
t_2-t_1=j-d-{\ell \left( {\bf a},\widetilde{H}\right) \left( {\bf a},{\bf b}%
_2\right) +\ell \left( {\bf a},\widetilde{H}\right) \left( {\bf a},{\bf b}%
_1\right) }\mbox{.}  \label{b61}
\end{equation}

Из (\ref{b54}), (\ref{b55}), (\ref{b56}) по Лемме 3.2.4:

\[
{\ell \left( {\bf a},\widetilde{H}\right) \left( {\bf a},{\bf b}_2\right)
=0.5\cdot }\left( w-u\right) \mbox{, }{\ell \left( {\bf a},\widetilde{H}%
\right) \left( {\bf a},{\bf b}_1\right) =0.5\cdot }\left( d-u\right) \mbox{.}
\]

Отсюда и из (\ref{b59}), (\ref{b60}), (\ref{b61}):

\[
t_2-t_1=0.5\cdot \left( \left( j-d\right) +\left( j-w\right) \right)
=0.5\cdot \left( j-d+p-j\right) =0.5\cdot \left( p-d\right) \mbox{.} 
\]

Из (\ref{b58}), (\ref{b55}), (\ref{b42}):

\[
z=0.5\cdot \left( p-d\right) \mbox{.} 
\]

Следовательно,

\[
z=t_2-t_1\mbox{ } _{\bf\Box}
\]

{\bf Теорема 3.4.2} Если $\Re$ - однородное пространство, то  

\begin{center}
$\left[ {\bf a}_1^{\bullet }B\mid \Re {\bf a}\widetilde{H}\right] \geq
\left[ B\mid \Re {\bf a}\widetilde{H}\right] $.
\end{center}

{\bf Доказательство} получается сразу из Теоремы 3.4.1 $_{\bf\Box}$

Следовательно, {\bf в любом однородном пространстве любой рекордер узнает о том, 
что произошло $B$ не раньше, чем $B$ произойдет. {\it Время необратимо.}}                 

{\bf Теорема 3.4.3} Если ${\bf a}_1$ и ${\bf a}_2$ - элементы $\Re$,

\begin{eqnarray}
&&\Re -ISS\left( \mathbf{a},\widetilde{H}\right) \mbox{,}  \label{i1}\\
&&p\stackrel{def}{=}\left[ \mathbf{a}_1^{\bullet }B\mid \Re \mathbf{a}\widetilde{H}\right] \mbox{,}
\label{i2}\\
&&q\stackrel{def}{=}\left[ \mathbf{a}_2^{\bullet }\mathbf{a}_1^{\bullet }B\mid \Re \mathbf{a}%
\widetilde{H}\right] \mbox{,}  \label{i3}\\
&&z\stackrel{def}{=}\ell \left( \Re \mathbf{a}\widetilde{H}\right) \left( \mathbf{a}_1,%
\mathbf{a}_2\right)\mbox{,}\nonumber 
\end{eqnarray}

то

\[
z=q-p \mbox{.}
\]

{\bf Доказательство} По Теореме 3.4.1 из (\ref{i1}), (\ref{i2}), (\ref{i3}): 

\[
q-p={\ell }\left( \Re \mathbf{a}\widetilde{H}\right) \left( \mathbf{a}%
_1^{\bullet }B,\mathbf{a}_2^{\bullet }\mathbf{a}_1^{\bullet }B\right)\mbox{,} 
\]

т.е. по Определению 3.3.8 в $\Re$ найдутся ${\bf b_1}$ и ${\bf b_2}$ такие, 
что

\begin{equation}
\natural \left( \mathbf{a}\right) \left( \mathbf{b}_1,\mathbf{a}_1^{\bullet
}B\right) \mbox{,}  \label{i4}
\end{equation}

\begin{equation}
\natural \left( \mathbf{a}\right) \left( \mathbf{b}_2,\mathbf{a}_2^{\bullet }%
\mathbf{a}_1^{\bullet }B\right) \mbox{,}  \label{i5}
\end{equation}

\[
q-p={\ell }\left( \Re \mathbf{a}\widetilde{H}\right) \left( \mathbf{b}_1,%
\mathbf{b}_2\right) \mbox{.}
\]

Кроме того, по Теореме 3.3.4

\begin{eqnarray}
&&\natural \left( \mathbf{a}\right) \left( \mathbf{a}_1^{\bullet },\mathbf{a}%
_1^{\bullet }B\right)\mbox{,}   \label{i6} \\
&&\natural \left( \mathbf{a}\right) \left( \mathbf{a}_2^{\bullet },\mathbf{a}%
_2^{\bullet }\mathbf{a}_1^{\bullet }B\right) \mbox{.}  \nonumber
\end{eqnarray}

Отсюда и из (\ref{i5}) по Теореме 3.3.7:

\begin{equation}
{\ell }\left( \Re \mathbf{a}\widetilde{H}\right) \left( \mathbf{b}_1,\mathbf{%
b}_2\right) ={\ell }\left( \Re \mathbf{a}\widetilde{H}\right) \left( \mathbf{%
b}_1,\mathbf{a}_2\right) \mbox{.}  \label{i7}
\end{equation}

По Теореме 3.3.3:

\begin{equation}
{\ell }\left( \Re \mathbf{a}\widetilde{H}\right) \left( \mathbf{b}_1,\mathbf{%
a}_2\right) ={\ell }\left( \Re \mathbf{a}\widetilde{H}\right) \left( \mathbf{%
a}_2,\mathbf{b}_1\right) \mbox{.}  \label{i8}
\end{equation}

Снова по Теореме 3.3.7 из (\ref{i6}), (\ref{i4}):

\begin{equation}
{\ell }\left( \Re \mathbf{a}\widetilde{H}\right) \left( \mathbf{a}_2,\mathbf{%
b}_1\right) ={\ell }\left( \Re \mathbf{a}\widetilde{H}\right) \left( \mathbf{%
a}_2,\mathbf{a}_1\right) \mbox{.}  \label{i9}
\end{equation}

Снова по Теореме 3.3.3:

\[
{\ell }\left( \Re \mathbf{a}\widetilde{H}\right) \left( \mathbf{a}_2,\mathbf{%
a}_1\right) ={\ell }\left( \Re \mathbf{a}\widetilde{H}\right) \left( \mathbf{%
a}_1\mathbf{a}_2\right) \mbox{.} 
\]

Отсюда и из (\ref{i9}), (\ref{i8}), (\ref{i7}):

\[
{\ell }\left( \Re \mathbf{a}\widetilde{H}\right) \left( \mathbf{b}_1,\mathbf{%
b}_2\right) ={\ell }\left( \Re \mathbf{a}\widetilde{H}\right) \left( \mathbf{%
a}_1\mathbf{a}_2\right) _{\bf\Box}
\]

По Теореме Урысона \cite{MSP1}: любое однородное пространство гомеоморфно 
некторому множеству точек вещественного гильбертого пространства. Если этот 
гомеоморфизм не является тождественным преобразованием, то $\Re $ будет 
представлять неэвклидово пространство. В этом случае в таком "пространстве-
времени" может быть сконструирован соответствующий вариант общей теории 
относительности. В противном случае $\Re$ есть эвклидово пространство. 
В этом случае существует {\it система координат} $R^\mu $, для которой 
выполняется следующее условие:

для всех элементов ${\bf a}_1$ и ${\bf a}_2$ множества $\Re $ существуют точки 
$\bf x_1$ и $\bf x_2$ системы $R^\mu $ такие, что:

\begin{center}
${\ell }${$\left( {\bf a},\widetilde{H}\right) \left( {\bf a}_k,{\bf a}%
_s\right) =\left( \sum_{j=1}^\mu \left( x_{s,j}-x_{k,j}\right) ^2\right)
^{0.5}$.}\\
\end{center}

В этом случае $R^\mu$ называется {\it системой координат системы отсчета} 
$\left( \Re {\bf a}\widetilde{H}\right) $, а числа $\left\langle x_{k,1},x_{k,2},
\ldots ,x_{k,\mu }\right\rangle $ - {\it координатами рекордера} ${\bf a}_k$ в 
$R^\mu $.

{\bf Система координат системы отсчета определена с точностью до преобразований 
сдвига, поворота и инверсии.}

{\bf Определение 3.4.4} Числа $\left\langle x_1,x_2,\ldots ,x_\mu\right\rangle $ 
называются {\it координатами} $B$ {\it в системе координат} $R^\mu $
{\it системы отсчета} $\left( \Re {\bf a}\widetilde{H}\right)$, если существует 
рекордер ${\bf b}$, для которого: ${\bf b}\in \Re $, $\natural \left( {\bf a}\right) \left( {\bf b}%
,B\right) $ и координатами этого рекордера в $R^\mu $ являются эти числа. 

{\bf Теорема 3.4.4} В координатной системе $R^\mu $ системы отсчета $\left( \Re {\bf a}\widetilde{H}%
\right)$: если $z$ - длина расстояния между $B$ и $C$, координаты $B$ есть 
$\left( b_1, b_2, ..., b_n\right)$, координаты $C$ есть 
$\left( c_1, c_2, ..., c_3\right)$, то 

\[
z=\left( \sum_{j=1}^\mu \left( c_j-b_j\right) ^2\right) ^{0.5}\mbox{.}
\]

{\bf Доказательство} получается сразу из Определения 3.4.4 $_{\bf \Box}$

{\bf Определение 3.4.5} Числа  $\left\langle x_1,x_2,\ldots ,x_\mu \right\rangle $ 
называются {\it координатами рекор-дера} ${\bf b}$ {\it в системе координам} $R^\mu $ 
{\it в момент} $t$ {\it системы отсчета} $\left( \Re {\bf a}\widetilde{H}%
\right) $, если для каждого $B$ выполняется условие:

если $t=\left[ {\bf b}^{\bullet }B\mid \Re {\bf a}\widetilde{H}\right] $, то  
$\ll {\bf b}^{\bullet }B\gg $ имеет координаты $\left\langle
x_1,x_2,\ldots ,x_\mu \right\rangle $ в системе координат $R^\mu $ 
системы отсчета $\left( \Re {\bf a}\widetilde{H}\right) $.

{\bf Лемма 3.4.1} Если

\begin{eqnarray}
&&\tau\stackrel{def}{=}\left[{\bf b^\bullet}C\uparrow {\bf b},\left\{ {\bf g_0}, B, {\bf b_0}\right\}%
\right] \mbox{,} \label{o71}\\
&&p\stackrel{def}{=}\left[ {\bf a^{\bullet} b^{\bullet}}\left({\bf g_0^{\bullet} b_0^{\bullet}}%
\right)^{\tau}{\bf g_0^{\bullet}} B \uparrow {\bf a}, \left \{ {\bf g_1}, A,%
{\bf b_1}\right \}\right] \mbox{,} \label{o72}\\
&&q\stackrel{def}{=}\left[ {\bf a^{\bullet} b^{\bullet}}\left({\bf g_0^{\bullet} b_0^{\bullet}}%
\right)^{\tau +1}{\bf g_0^{\bullet}} B \uparrow {\bf a}, \left \{ {\bf g_1}, A,%
{\bf b_1}\right \}\right] \mbox{,} \label{o73}\\
&&t\stackrel{def}{=}\left[{\bf a^{\bullet} b^{\bullet}} C\uparrow {\bf a}, \left \{ {\bf g_1}, A%
, {\bf b_1}\right \}\right] \mbox{,} \label{o74}
\end{eqnarray}

то 

\begin{center}
$p \leq t \leq q$.
\end{center}

{\bf Доказательство} 

1) Из (\ref{o73}):

\begin{equation}
\mathbf{a}^{\bullet }\left( \mathbf{a^{\bullet }b^{\bullet }}\left( \mathbf{%
g_0^{\bullet }b_0^{\bullet }}\right) ^{\tau +1}\mathbf{g_0^{\bullet }}%
B\&\left( \neg \mathbf{a}^{\bullet }\left( \mathbf{g}_1^{\bullet }\mathbf{b}%
_1^{\bullet }\right) ^{q+1}\mathbf{g}_1^{\bullet }A\right) \right)\mbox{.}
\label{o1}
\end{equation}

Т.к. из (\ref{o71}):

\begin{center}
$\left( \mathbf{b}^{\bullet }\left( \mathbf{g}%
_0^{\bullet }\mathbf{b}_0^{\bullet }\right) ^{\tau +1}\mathbf{g}_0^{\bullet
}B\Rightarrow \mathbf{b^{\bullet }}C\right) $,
\end{center}

то из (\ref{o1}) по II:

\begin{center}
$\mathbf{a}^{\bullet }\left( \mathbf{a^{\bullet }b^{\bullet }}C\&\left( \neg 
\mathbf{a}^{\bullet }\left( \mathbf{g}_1^{\bullet }\mathbf{b}_1^{\bullet
}\right) ^{q+1}\mathbf{g}_1^{\bullet }A\right) \right)$ .
\end{center}

Отсюда по II т.к. из (\ref{o74}):

\begin{center}
$\left( \mathbf{a^{\bullet }b^{\bullet }}%
C\Rightarrow \mathbf{a}^{\bullet }\left( \mathbf{g}_1^{\bullet }\mathbf{b}%
_1^{\bullet }\right) ^t\mathbf{g}_1^{\bullet }A\right) $,
\end{center}

то

\begin{equation}
\mathbf{a}^{\bullet }\left( \mathbf{a}^{\bullet }\left( \mathbf{g}%
_1^{\bullet }\mathbf{b}_1^{\bullet }\right) ^t\mathbf{g}_1^{\bullet
}A\&\left( \neg \mathbf{a}^{\bullet }\left( \mathbf{g}_1^{\bullet }\mathbf{b}%
_1^{\bullet }\right) ^{q+1}\mathbf{g}_1^{\bullet }A\right) \right) \mbox{.}
\label{o2}
\end{equation}

Если $t>q$, то $t\geq q+1$. Следовательно по III из (\ref{o2}):

\begin{center}
$\mathbf{a}^{\bullet }\left( \mathbf{a}^{\bullet }\left( \mathbf{g}%
_1^{\bullet }\mathbf{b}_1^{\bullet }\right) ^{q+1}\mathbf{g}_1^{\bullet
}A\&\left( \neg \mathbf{a}^{\bullet }\left( \mathbf{g}_1^{\bullet }\mathbf{b}%
_1^{\bullet }\right) ^{q+1}\mathbf{g}_1^{\bullet }A\right) \right) $,
\end{center}

что противоречит I. Следовательно, $t\leq q$. 

2) Из (\ref{o74}):

\begin{equation}
\mathbf{a}^{\bullet }\left( \mathbf{a^{\bullet }b^{\bullet }}C\&\left( \neg 
\mathbf{a}^{\bullet }\left( \mathbf{g}_1^{\bullet }\mathbf{b}_1^{\bullet
}\right) ^{t+1}\mathbf{g}_1^{\bullet }A\right) \right) \mbox{.}\label{o3}
\end{equation}

Т.к. из (\ref{o71}):

\begin{center}
$\left( \mathbf{b^{\bullet }}C\Rightarrow 
\mathbf{b}^{\bullet }\left( \mathbf{g}_0^{\bullet }\mathbf{b}_0^{\bullet
}\right) ^\tau \mathbf{g}_0^{\bullet }B\right)$,
\end{center}

то из (\ref{o3}) по II:

\begin{equation} 
\mathbf{a}^{\bullet }\left( \mathbf{a^{\bullet }b}^{\bullet }\left( \mathbf{%
g}_0^{\bullet }\mathbf{b}_0^{\bullet }\right) ^\tau \mathbf{g}_0^{\bullet
}B\&\left( \neg \mathbf{a}^{\bullet }\left( \mathbf{g}_1^{\bullet }\mathbf{b}%
_1^{\bullet }\right) ^{t+1}\mathbf{g}_1^{\bullet }A\right) \right) \mbox{.}
\label{o4}
\end{equation}

Т.к. из (\ref{o72}):

\begin{center}
$\left( \mathbf{a^{\bullet }b^{\bullet }}%
\left( \mathbf{g_0^{\bullet }b_0^{\bullet }}\right) ^\tau \mathbf{%
g_0^{\bullet }}B\Rightarrow \mathbf{a}^{\bullet }\left( \mathbf{g}%
_1^{\bullet }\mathbf{b}_1^{\bullet }\right) ^p\mathbf{g}_1^{\bullet
}A\right)  $,
\end{center}

то по II из (\ref{o4}):

\begin{equation}
\mathbf{a}^{\bullet }\left( \mathbf{a}^{\bullet }\left( \mathbf{g}%
_1^{\bullet }\mathbf{b}_1^{\bullet }\right) ^p\mathbf{g}_1^{\bullet
}A\&\left( \neg \mathbf{a}^{\bullet }\left( \mathbf{g}_1^{\bullet }\mathbf{b}%
_1^{\bullet }\right) ^{t+1}\mathbf{g}_1^{\bullet }A\right) \right)\mbox{.}
\label{o5}
\end{equation}

Если $p>t$, то $p\geq t+1$. В этом случае из (\ref{o5}) по III:

\begin{center}
$\mathbf{a}^{\bullet }\left( \mathbf{a}^{\bullet }\left( \mathbf{g}%
_1^{\bullet }\mathbf{b}_1^{\bullet }\right) ^{t+1}\mathbf{g}_1^{\bullet
}A\&\left( \neg \mathbf{a}^{\bullet }\left( \mathbf{g}_1^{\bullet }\mathbf{b}%
_1^{\bullet }\right) ^{t+1}\mathbf{g}_1^{\bullet }A\right) \right)$,
\end{center}

что противоречит I. Следовательно, $p\leq t$ $_{\bf \Box}$

{\bf Теорема 3.4.5} В координатной системе $R^\mu $ системы отсчета 
$\left( \Re {\bf a}\widetilde{H}\right) $: если в каждый момент $t$:
координаты ($v$ - какое-нибудь вещественное положительное число, для которого $|v|<1$):  

${\bf b}$: $\left\langle x_{{\bf b},1}+v\cdot t,x_{{\bf b},2},x_{{\bf b}%
,3},\ldots ,x_{{\bf b},\mu }\right\rangle $;

${\bf g}_0$: $\left\langle x_{0,1}+v\cdot t,x_{0,2},x_{0,3},\ldots
,x_{0,\mu }\right\rangle $;

${\bf b}_0$: $\left\langle x_{0,1}+v\cdot t,x_{0,2}+l,x_{0,3},\ldots
,x_{0,\mu }\right\rangle $; и

$t_C=\left[ {\bf b}^{\bullet }C\mid \Re {\bf a}\widetilde{H}\right] $;

$t_D=\left[ {\bf b}^{\bullet }D\mid \Re {\bf a}\widetilde{H}\right] $;

$q_C$ $=$ $\left[ {\bf b}^{\bullet }C\uparrow {\bf b},\left\{ {\bf g}_0,A,%
{\bf b}_0\right\} \right] $;

$q_D$ $=$ $\left[ {\bf b}^{\bullet }D\uparrow {\bf b},\left\{ {\bf g}_0,A,%
{\bf b}_0\right\} \right] $,

то

\[
\lim_{l\rightarrow 0}2\cdot \frac l{\sqrt{\left( 1-v^2\right) }}\cdot \frac{%
q_D-q_C}{t_D-t_C}=1 \mbox{.}
\]

{\bf Доказательство} Обозначим:

\begin{eqnarray}
&&t_1\stackrel{def}{=}\left[{\bf b^{\bullet}}\left({\bf {g_0}^{\bullet}{b_0}^{\bullet}}\right)%
^{q_C}{\bf {g_0}^{\bullet}}B\mid \Re {\bf a}\widetilde{H}\right]\mbox{,}
\label{o78}\\
&&t_2\stackrel{def}{=}\left[{\bf b^{\bullet}}\left({\bf {g_0}^{\bullet}{b_0}^{\bullet}}\right)%
^{q_C+1}{\bf {g_0}^{\bullet}}B\mid \Re {\bf a}\widetilde{H}\right]\mbox{,}
\label{o79}\\
&&t_3\stackrel{def}{=}\left[\left({\bf {g_0}^{\bullet}{b_0}^{\bullet}}\right)%
^{q_C}{\bf {g_0}^{\bullet}}B\mid \Re {\bf a}\widetilde{H}\right]\mbox{,}
\label{o80}\\
&&t_4\stackrel{def}{=}\left[\left({\bf {g_0}^{\bullet}{b_0}^{\bullet}}\right)%
^{q_C+1}{\bf {g_0}^{\bullet}}B\mid \Re {\bf a}\widetilde{H}\right]\mbox{.}
\label{o81}
\end{eqnarray}

В этом случае координаты:

\begin{eqnarray}
&&\ll {\bf b^{\bullet}}\left({\bf {g_0}^{\bullet}{b_0}^{\bullet}}\right)%
^{q_C}{\bf {g_0}^{\bullet}}B \gg: \left\langle x_{{\bf b},1}+v\cdot t_1,x_{{\bf b},2},x_{{\bf b}%
,3},\ldots ,x_{{\bf b},\mu }\right\rangle \mbox{,}\label{o82}\\
&&\ll {\bf b^{\bullet}}\left({\bf {g_0}^{\bullet}{b_0}^{\bullet}}\right)%
^{q_C+1}{\bf {g_0}^{\bullet}}B \gg: \left\langle x_{{\bf b},1}+v\cdot t_2,x_{{\bf b},2},x_{{\bf b}%
,3},\ldots ,x_{{\bf b},\mu }\right\rangle \mbox{,}\label{o83}\\ 
&&\ll \left({\bf {g_0}^{\bullet}{b_0}^{\bullet}}\right)^{q_C}{\bf {g_0}^{\bullet}%
}B\gg: \left\langle x_{0,1}+v\cdot t_3,x_{0,2},x_{0,3},\ldots
,x_{0,\mu }\right\rangle \mbox{,}\label{o84}\\
&&\ll \left({\bf {g_0}^{\bullet}{b_0}^{\bullet}}\right)^{q_C+1}{\bf {g_0}^{\bullet}%
}B\gg: \left\langle x_{0,1}+v\cdot t_4,x_{0,2},x_{0,3},\ldots
,x_{0,\mu }\right\rangle \mbox{,}\label{o85}\\
&&\ll {\bf b^{\bullet}}C \gg: \left\langle x_{{\bf b},1}+v\cdot t_C,x_{{\bf b},2},x_{{\bf b}%
,3},\ldots ,x_{{\bf b},\mu }\right\rangle \mbox{.}\label{o86}
\end{eqnarray}

По Теореме 3.4.1 и Лемме 3.3.4 из (\ref{o78}), (\ref{o82}), (\ref{o79}), (\ref{o83}),
(\ref{o86}):

\[
\begin{array}{c}
\left[ \mathbf{a^{\bullet }b^{\bullet }}\left( \mathbf{{g_0}^{\bullet }{b_0}%
^{\bullet }}\right) ^{q_C}{\bf {g_0}^{\bullet}}B\mid \Re \mathbf{a}\widetilde{H}\right] = \\ 
\left[ \mathbf{a^{\bullet }b^{\bullet }}\left( \mathbf{{g_0}^{\bullet }{b_0}%
^{\bullet }}\right) ^{q_C}{\bf {g_0}^{\bullet}}B\uparrow \mathbf{a},\widetilde{H}\right] = \\ 
t_1+\left( \left( x_{b,1}+vt_1\right) ^2+\sum_{j+2}^\mu x_{b,j}^2\right)
^{0.5}\mbox{,}
\end{array}
\]

\[
\begin{array}{c}
\left[ \mathbf{a^{\bullet }b^{\bullet }}\left( \mathbf{{g_0}^{\bullet }{b_0}%
^{\bullet }}\right) ^{q_C+1}B\mid \Re \mathbf{a}\widetilde{H}\right] = \\ 
\left[ \mathbf{a^{\bullet }b^{\bullet }}\left( \mathbf{{g_0}^{\bullet }{b_0}%
^{\bullet }}\right) ^{q_C+1}B\uparrow \mathbf{a},\widetilde{H}\right] = \\ 
t_2+\left( \left( x_{b,1}+vt_2\right) ^2+\sum_{j=2}^\mu x_{b,j}^2\right)
^{0.5}\mbox{.}
\end{array}
\]

Отсюда по Лемме 3.4.1:

\begin{eqnarray}
&&t_1+\left( \left( x_{b,1}+vt_1\right) ^2+\sum_{j=2}^\mu x_{b,j}^2\right)
^{0.5} \nonumber\\
&\leq &t_C+\left( \left( x_{b,1}+vt_C\right) ^2+\sum_{j=2}^\mu
x_{b,j}^2\right) ^{0.5} \label{o87}\\
&\leq &t_2+\left( \left( x_{b,1}+vt_2\right) ^2+\sum_{j=2}^\mu
x_{b,j}^2\right) ^{0.5}\mbox{.}\nonumber
\end{eqnarray}

По Теореме 3.4.1 из (\ref{o78}), (\ref{o80}), (\ref{o82}), (\ref{o84}):

\begin{center}
$t_1=t_3+\left( \left( x_{0,1}+vt_3-x_{b,1}-vt_1\right) ^2+\sum_{j=2}^\mu
\left( x_{0,j}-x_{b,j}\right) ^2\right) ^{0.5}$.
\end{center}

Из (\ref{o79}), (\ref{o81}), (\ref{o83}), (\ref{o85}):

\begin{center}
$t_2=t_4+\left( \left( x_{0,1}+vt_4-x_{b,1}-vt_2\right) ^2+\sum_{j=2}^\mu
\left( x_{0,j}-x_{b,j}\right) ^2\right) ^{0.5}$.
\end{center}

Отсюда:

\begin{center}
$\left( t_1-t_3\right) ^2=v^2\left( t_1-t_3\right) ^2-2v\left(
t_1-t_3\right) \left( x_{0,1}-x_{b,1}\right) +\sum_{j=2}^\mu \left(
x_{0,j}-x_{b,j}\right) ^2$,

$\left( t_2-t_4\right) ^2=v^2\left( t_2-t_4\right) ^2-2v\left(
t_2-t_4\right) \left( x_{0,1}-x_{b,1}\right) +\sum_{j=2}^\mu \left(
x_{0,j}-x_{b,j}\right) ^2$.
\end{center}

Следовательно, 

\begin{equation}
t_2-t_4=t_1-t_3 \mbox{.}\label{o88}
\end{equation}

Обозначим:

\begin{equation}
t_5\stackrel{def}{=}\left[{\bf {b_0}^{\bullet}}\left({\bf {g_0}^{\bullet}{b_0}^{\bullet}}\right)%
^{q_C}{\bf {g_0}^{\bullet}}B\mid \Re {\bf a}\widetilde{H}\right]\mbox{.}
\label{o89}
\end{equation}

В этом случае координаты:

\begin{center}
$\ll {\bf {b_0}^{\bullet}}\left({\bf {g_0}^{\bullet}{b_0}^{\bullet}}\right)^{q_C}{\bf {g_0}^{\bullet}%
}B\gg: \left\langle x_{0,1}+v\cdot t_5,x_{0,2}+l,x_{0,3},\ldots
,x_{0,\mu }\right\rangle$ .
\end{center}

Отсюда и из (\ref{o80}), (\ref{o84}) по Теореме 3.4.1:

\begin{center}
$t_5-t_3=\left( \left( x_{0,1}+vt_5-x_{0,1}-vt_3\right)^2 +\left(
x_{0,2}+l-x_{0,2}\right)^2 +\sum_{j=3}^\mu \left( x_{0,j}-x_{0,j}\right)^2
\right) ^{0.5}$,
\end{center}

т.е.:

\begin{equation}
t_5-t_3=\frac l{\sqrt{1-v^2}} \mbox{.} \label{o90}
\end{equation}

Аналогично, из (\ref{o89}), (\ref{o81}), (\ref{o85}):

\[
t_4-t_5=\frac l{\sqrt{1-v^2}} \mbox{.}
\]

Отсюда и из (\ref{o90}):

\[
t_4-t_3=\frac {2l}{\sqrt{1-v^2}} \mbox{.}
\]

Тогда из (\ref{o88}):

\[
t_2-t_1=\frac {2l}{\sqrt{1-v^2}} \mbox{.}
\]

Следовательно, из (\ref{o87}):

\begin{eqnarray*}
&&\ t_1+\left( \left( x_{b,1}+vt_1\right) ^2+\sum_{j=2}^\mu x_{b,j}^2\right)
^{0.5} \\
\  &\leq &t_C+\left( \left( x_{b,1}+vt_C\right) ^2+\sum_{j=2}^\mu
x_{b,j}^2\right) ^{0.5} \\
\  &\leq &t_1+\frac{2l}{\sqrt{1-v^2}}+\left( \left( x_{b,1}+v\left( t_1+%
\frac{2l}{\sqrt{1-v^2}}\right) \right) ^2+\sum_{j=2}^\mu x_{b,j}^2\right)
^{0.5} \mbox{.}
\end{eqnarray*}

Или, если $l\rightarrow 0$, то $t_2\rightarrow t_1$, и 

\begin{eqnarray*}
&&\lim_{l\rightarrow 0}\left( \ t_1+\left( \left( x_{b,1}+vt_1\right)
^2+\sum_{j=2}^\mu x_{b,j}^2\right) ^{0.5}\right)  \\
&=&t_C+\left( \left( x_{b,1}+vt_C\right) ^2+\sum_{j=2}^\mu x_{b,j}^2\right)
^{0.5}\mbox{.}
\end{eqnarray*}

Т.к. при $v^2<1$ функция 

\[
f\left( t\right) =\ t+\left( \left( x_{b,1}+vt\right) ^2+\sum_{j=2}^\mu
x_{b,j}^2\right) ^{0.5}
\]

монотонна, то 

\[
\lim_{l\rightarrow 0}t_1=t_C \mbox{,}
\]

т.е.

\begin{equation}
\lim_{l\rightarrow 0}\left[ \mathbf{b^{\bullet }}\left( \mathbf{{g_0}%
^{\bullet }{b_0}^{\bullet }}\right) ^{q_C}\mathbf{{g_0}^{\bullet }}B\mid \Re 
\mathbf{a}\widetilde{H}\right] =t_C \mbox{.} \label{o92}
\end{equation}

Аналогично, 

\begin{equation}
\lim_{l\rightarrow 0}\left[ \mathbf{b^{\bullet }}\left( \mathbf{{g_0}%
^{\bullet }{b_0}^{\bullet }}\right) ^{q_D}\mathbf{{g_0}^{\bullet }}B\mid \Re 
\mathbf{a}\widetilde{H}\right] =t_D \mbox{.}\label{o93}
\end{equation}

По Теореме 3.4.1 и по (\ref{o78}) и (\ref{o79}):

\begin{eqnarray*}
&&\left[ \mathbf{b^{\bullet }}\left( \mathbf{{g_0}^{\bullet }{b_0}^{\bullet }%
}\right) ^{q_D}\mathbf{{g_0}^{\bullet }}B\mid \Re \mathbf{a}\widetilde{H}%
\right] -\left[ \mathbf{b^{\bullet }}\left( \mathbf{{g_0}^{\bullet }{b_0}%
^{\bullet }}\right) ^{q_C}\mathbf{{g_0}^{\bullet }}B\mid \Re \mathbf{a}%
\widetilde{H}\right]  \\
&=&\left( t_1+\frac{2l}{\sqrt{1-v^2}}\left( q_D-q_C\right) \right) -t_1 \\
&=&\frac{2l\left( q_D-q_C\right) }{\sqrt{1-v^2}}\mbox{.}
\end{eqnarray*}

Отсюда и из (\ref{o92}) и (\ref{o93}): 

\[
\lim_{l\rightarrow 0}\frac{2l\left( q_D-q_C\right) }{t_D-t_C}=\sqrt{1-v^2}
{\bf _\Box}
\]

{\bf Следствие из Теоремы 3.4.5:} Если обозначить: $q_D^{st}\stackrel{def}{=}q_D$ и 
$q_C^{st}\stackrel{def}{=}q_C$ для $v=0$, то

\[
\lim_{l\rightarrow 0}2l\frac{q_D^{st}-q_C^{st}}{t_D-t_C}=1 \mbox{,}
\]

т.е.:

\[
\lim_{l\rightarrow 0}\frac{q_D-q_C}{q_D^{st}-q_C^{st}}=\sqrt{1-v^2} \mbox{.}
\]

Или для абсолютно-точных часов:

\[
q_D^{st}-q_C^{st}=\frac{q_D-q_C}{\sqrt{1-v^2}} {\bf _\Box}
\]
             
Следовательно, движущиеся со скоростью $v$ $\kappa -$часы "идут" в 
$\left( 1-v^2\right) ^{-0.5}$ раз медленнее, чем покоящиеся.

{\bf Теорема 3.4.6} Пусть: $v$ ($\left| v\right| <1$) и $l$ - вещественные 
числа, а $k_i$ - натуральные.

Пусть в координатной системе $R^\mu $ системы отсчета $\left( \Re 
{\bf a}\widetilde{H}\right) $: в каждый момент $t$ координаты:

${\bf b}$: $\left\langle x_{ b,1}+v\cdot t,x_{b,2},x_{b%
,3},\ldots ,x_{b,\mu }\right\rangle $,

${\bf g}_j$: $\left\langle y_{j,1}+v\cdot t,y_{j,2},y_{j,3},\ldots
,y_{j,\mu }\right\rangle $,

${\bf u}_j$: $\left\langle y_{j,1}+v\cdot t,y_{j,2}+l/\left( k_1\cdot
\ldots \cdot k_j\right) ,y_{j,3},\ldots ,y_{j,\mu }\right\rangle $,

для всех ${\bf b_i}$: если ${\bf b_i}\in \Im $, то координаты

${\bf b_i}$: $\left\langle x_{i,1}+v\cdot t,x_{i,2},x_{i,3},\ldots ,x_{i,\mu
}\right\rangle $,

$\widetilde{T}$ есть $\left\langle {\left\{ {\bf g}_1,A,{\bf u}_1\right\} ,\
\left\{ {\bf g}_2,A,{\bf u}_2\right\} ,...,\left\{ {\bf g}_j,A,{\bf u}%
_j\right\} ,\ ...\ }\right\rangle $.

В этом случае: $\Im $ - $ISS\left( {\bf b},\widetilde{T}\right) $.

{\bf Доказательство} 

1) Обозначим:

\begin{eqnarray*}
&&p\stackrel{def}{=}\left[ \mathbf{b^{\bullet }b_1^{\bullet }}B\uparrow \mathbf{b},%
\widetilde{T}\right] \mbox{,} \\
&&q\stackrel{def}{=}\left[ \mathbf{b^{\bullet }b_2^{\bullet }b_1^{\bullet }}B\uparrow 
\mathbf{b},\widetilde{T}\right] \mbox{,} \\
&&r\stackrel{def}{=}\left[ \mathbf{b^{\bullet }b_1^{\bullet }}C\uparrow \mathbf{b},%
\widetilde{T}\right] \mbox{,} \\
&&s\stackrel{def}{=}\left[ \mathbf{b^{\bullet }b_2^{\bullet }b_1^{\bullet }}C\uparrow 
\mathbf{b},\widetilde{T}\right] \mbox{,}
\end{eqnarray*}

\begin{eqnarray}
&&t_p\stackrel{def}{=}\left[ \mathbf{b^{\bullet }b_1^{\bullet }}B\mid \Re \mathbf{a}%
\widetilde{H}\right]\mbox{,} \label{o94}  \\
&&t_q\stackrel{def}{=}\left[ \mathbf{b^{\bullet }b_2^{\bullet }b_1^{\bullet }}B\mid \Re 
\mathbf{a}\widetilde{H}\right]\mbox{,}\label{o95}  \\
&&t_r\stackrel{def}{=}\left[ \mathbf{b^{\bullet }b_1^{\bullet }}C\mid \Re \mathbf{a}%
\widetilde{H}\right]\mbox{,} \label{o96} \\
&&t_s\stackrel{def}{=}\left[ \mathbf{b^{\bullet }b_2^{\bullet }b_1^{\bullet }}B\mid \Re 
\mathbf{a}\widetilde{H}\right]\mbox{.}\label{o97} 
\end{eqnarray}

По Следствию из Теоремы 3.4.5:

\begin{eqnarray}
t_q-t_p &=&\frac{q-p}{\sqrt{1-v^2}}\mbox{,}\label{o98} \\
t_s-t_r &=&\frac{s-r}{\sqrt{1-v^2}}\mbox{.}\label{o99}
\end{eqnarray}

Из (\ref{o94}-\ref{o97}) координаты:

\begin{eqnarray}
&&\ll\mathbf{b^{\bullet }b_1^{\bullet }}B\gg:\left\langle
x_{b,1}+vt_p,x_{b,2},x_{b,3},\ldots ,x_{b,\mu }\right\rangle\mbox{,}\label{o100}  \\
&&\ll\mathbf{b^{\bullet }b_2^{\bullet }b_1^{\bullet }}B\gg:\left\langle
x_{b,1}+vt_q,x_{b,2},x_{b,3},\ldots ,x_{b,\mu }\right\rangle \mbox{,}\nonumber
\end{eqnarray}

\begin{eqnarray}
&&\ll\mathbf{b^{\bullet }b_1^{\bullet }}C\gg:\left\langle
x_{b,1}+vt_r,x_{b,2},x_{b,3},\ldots ,x_{b,\mu }\right\rangle\mbox{,}\label{o101}  \\
&&\ll\mathbf{b^{\bullet }b_2^{\bullet }b_1^{\bullet }}C\gg:\left\langle
x_{b,1}+vt_s,x_{b,2},x_{b,3},\ldots ,x_{b,\mu }\right\rangle \mbox{.}\nonumber
\end{eqnarray}

Обозначим:

\begin{eqnarray}
&&t_1\stackrel{def}{=}\left[ \mathbf{b_1^{\bullet }}B\mid \Re \mathbf{a}\widetilde{H}%
\right]\mbox{,}\label{o102}  \\
&&t_2\stackrel{def}{=}\left[ \mathbf{b_1^{\bullet }}C\mid \Re \mathbf{a}\widetilde{H}%
\right] \mbox{.}\label{o103}
\end{eqnarray}

Тогда координаты:

\begin{eqnarray*}
&&\ll\mathbf{b_1^{\bullet }}B\gg:\left\langle x_{1,1}+vt_1,x_{1,2},x_{1,3},\ldots
,x_{1,\mu }\right\rangle\mbox{,}  \\
&&\ll\mathbf{b_1^{\bullet }}C\gg:\left\langle x_{1,1}+vt_2,x_{1,2},x_{1,3},\ldots
,x_{1,\mu }\right\rangle \mbox{.}
\end{eqnarray*}

По Теореме 3.4.1 из (\ref{o101}), (\ref{o103}), (\ref{o96}):

\[
t_r-t_2=\left( \left( x_{b,1}+vt_r-x_{1,1}-vt_2\right) ^2+\sum_{j=2}^\mu
\left( x_{b,j}-x_{1,j}\right) ^2\right) ^{0.5}\mbox{.}
\]

Аналогично, из (\ref{o100}), (\ref{o102}), (\ref{o94}): 

\[
t_p-t_1=\left( \left( x_{b,1}+vt_p-x_{1,1}-vt_1\right) ^2+\sum_{j=2}^\mu
\left( x_{b,j}-x_{1,j}\right) ^2\right) ^{0.5}\mbox{.}
\]

Следовательно,

\begin{equation}
t_r-t_2=t_p-t_1\mbox{.}\label{o104}
\end{equation}
Обозначим:

\begin{eqnarray*}
&&t_3\stackrel{def}{=}\left[ \mathbf{{b_2}^{\bullet}b_1^{\bullet }}B\mid \Re \mathbf{a}\widetilde{H}%
\right]\mbox{,}  \\
&&t_4\stackrel{def}{=}\left[ \mathbf{{b_2}^{\bullet}b_1^{\bullet }}C\mid \Re \mathbf{a}\widetilde{H}%
\right] \mbox{.}
\end{eqnarray*}

Отсюда координаты:

\begin{eqnarray*}
&&\ll\mathbf{{b_2}^{\bullet}b_1^{\bullet }}B\gg:\left\langle x_{2,1}+vt_3,x_{2,2},x_{2,3},\ldots
,x_{2,\mu }\right\rangle\mbox{,}  \\
&&\ll\mathbf{{b_2}^{\bullet}b_1^{\bullet }}C\gg:\left\langle x_{2,1}+vt_4,x_{2,2},x_{2,3},\ldots
,x_{2,\mu }\right\rangle\mbox{.} 
\end{eqnarray*}

По Теореме 3.4.1:

\begin{eqnarray*}
t_3-t_1=\left( \left( x_{2,1}+vt_3-x_{1,1}-vt_1\right) ^2+\sum_{j=2}^\mu
\left( x_{2,j}-x_{1,j}\right) ^2\right) ^{0.5}\mbox{.}\\
t_4-t_2=\left( \left( x_{2,1}+vt_4-x_{1,1}-vt_2\right) ^2+\sum_{j=2}^\mu
\left( x_{2,j}-x_{1,j}\right) ^2\right) ^{0.5}\mbox{.}
\end{eqnarray*}

Отсюда:

\begin{equation}
t_3-t_4=t_1-t_2\mbox{.}\label{o105}
\end{equation}

И аналогично:

\begin{equation}
t_q-t_3=t_s-t_4\mbox{.}\label{o106}
\end{equation}

Из (\ref{o105}), (\ref{o106}), (\ref{o104}):

\begin{center}
$t_q-t_p=t_s-t_r$.
\end{center}

Отсюда и из (\ref{o99}), (\ref{o98}): 

\begin{equation}
q-p=s-r\mbox{.}\label{o107}
\end{equation}

2) Обозначим:

\begin{eqnarray*}
&&p^{\prime }\stackrel{def}{=}\left[ \mathbf{b^{\bullet }}C\uparrow \mathbf{b},\widetilde{T}%
\right]\mbox{,}  \\
&&q^{\prime }\stackrel{def}{=}\left[ \mathbf{b^{\bullet }}\alpha \mathbf{b^{\bullet }}%
C\uparrow \mathbf{b},\widetilde{T}\right]\mbox{,}  \\
&&r^{\prime }\stackrel{def}{=}\left[ \mathbf{b^{\bullet }}\alpha ^{\dagger }\mathbf{%
b^{\bullet }}C\uparrow \mathbf{b},\widetilde{T}\right]\mbox{;} 
\end{eqnarray*}
  
здесь: $\alpha$ есть $\mathbf{b_1^{\bullet }b_2^{\bullet }\ldots b_k^{\bullet }b_{k+1}^{\bullet
}\ldots b_N^{\bullet }}$.

Тогда по Определению 3.3.1:

\begin{eqnarray}
&&\jmath \left( \mathbf{b}\widetilde{T}\right) \left( \mathbf{b}^{\bullet
}\alpha \mathbf{b}^{\bullet }C\right)  =q^{\prime }-p^{\prime }\mbox{,}\label{o108} \\
&&\jmath \left( \mathbf{b}\widetilde{T}\right) \left( \mathbf{b}^{\bullet
}\alpha ^{\dagger }\mathbf{b}^{\bullet }C\right)  =r^{\prime }-p^{\prime }
\mbox{.}\label{o109}
\end{eqnarray}

Обозначим:

\begin{eqnarray}
&&t_0\stackrel{def}{=}\left[ \mathbf{b^{\bullet }}C\mid \Re \mathbf{a}\widetilde{H}\right]\mbox{,}\nonumber \\ 
&&t_1=\left[ \mathbf{b_1^{\bullet }b^{\bullet }}C\mid \Re \mathbf{a}\widetilde{%
H}\right]\mbox{,}\nonumber  \\ 
&&t_2\stackrel{def}{=}\left[ \mathbf{b_2^{\bullet }b_1^{\bullet }b^{\bullet }}C\mid \Re 
\mathbf{a}\widetilde{H}\right] \mbox{,}\nonumber \\ 
&&\cdots\mbox{,}\nonumber  \\ 
&&t_k\stackrel{def}{=}\left[ \mathbf{b_k^{\bullet }}\ldots \mathbf{b_2^{\bullet }b_1^{\bullet
}b^{\bullet }}C\mid \Re \mathbf{a}\widetilde{H}\right] \mbox{,}\label{o110} \\ 
&&t_{k+1}\stackrel{def}{=}\left[ \mathbf{b_{k+1}^{\bullet }b_k^{\bullet }}\ldots \mathbf{%
b_2^{\bullet }b_1^{\bullet }b^{\bullet }}C\mid \Re \mathbf{a}\widetilde{H}%
\right] \mbox{,}\nonumber \\ 
&&\cdots \mbox{,}\nonumber \\ 
&&t_N\stackrel{def}{=}\left[ \mathbf{b_N^{\bullet }}\ldots \mathbf{b_{k+1}^{\bullet
}b_k^{\bullet }}\ldots \mathbf{b_2^{\bullet }b_1^{\bullet }b^{\bullet }}%
C\mid \Re \mathbf{a}\widetilde{H}\right] \mbox{,}\nonumber \\ 
&&t_{N+1}\stackrel{def}{=}\left[ \mathbf{b^{\bullet }}\alpha ^{\dagger }\mathbf{b^{\bullet }}%
C\mid \Re \mathbf{a}\widetilde{H}\right] \mbox{.}\nonumber
\end{eqnarray}

Тогда из условия Теоремы координаты:

\begin{eqnarray*}
&\ll &\mathbf{b^{\bullet }}C\gg :\left\langle
x_{b,1}+vt_0,x_{b,2},x_{b,3},\ldots ,x_{b,\mu }\right\rangle \mbox{,} \\
&\ll &\mathbf{b_1^{\bullet }b^{\bullet }}C\gg :\left\langle
x_{1,1}+vt_1,x_{1,2},x_{1,3},\ldots ,x_{1,\mu }\right\rangle \mbox{,} \\
&\ll &\mathbf{b_2^{\bullet }b_1^{\bullet }b^{\bullet }}C\gg :\left\langle
x_{2,1}+vt_2,x_{2,2},x_{2,3},\ldots ,x_{2,\mu }\right\rangle \mbox{,} \\
&&\cdots \mbox{,} \\
&\ll &\mathbf{b_k^{\bullet }}\cdots \mathbf{b_2^{\bullet }b_1^{\bullet
}b^{\bullet }}C\gg :\left\langle x_{k,1}+vt_k,x_{k,2},x_{k,3},\ldots
,x_{k,\mu }\right\rangle \mbox{,} \\
&\ll &\mathbf{b_{k+1}^{\bullet }b_k^{\bullet }}\cdots \mathbf{b_2^{\bullet
}b_1^{\bullet }b^{\bullet }}C\gg :\left\langle
x_{k+1,1}+vt_{k+1},x_{k+1,2},x_{k+1,3},\ldots ,x_{k+1,\mu }\right\rangle %
\mbox{,} \\
&&\cdots \mbox{,} \\
&\ll &\mathbf{b_N^{\bullet }}\cdots \mathbf{b_{k+1}^{\bullet }b_k^{\bullet }}%
\cdots \mathbf{b_2^{\bullet }b_1^{\bullet }b^{\bullet }}C\gg :\left\langle
x_{N,1}+vt_N,x_{N,2},x_{N,3},\ldots ,x_{N,\mu }\right\rangle \mbox{,} \\
&\ll &\mathbf{b^{\bullet }}\alpha ^{\dagger }\mathbf{b^{\bullet }}C\gg
:\left\langle x_{N+1,1}+vt_{N+1},x_{N+1,2},x_{N+1,3},\ldots ,x_{N+1,\mu
}\right\rangle \mbox{.}
\end{eqnarray*}

Отсюда и из (\ref{o110}) по Теореме 3.4.1:

\begin{eqnarray*}
t_1-t_0 &=&\left( \left( x_{1,1}+vt_1-x_{b,1}-vt_0\right) ^2+\sum_{j=2}^\mu
\left( x_{1,j}-x_{b,j}\right) ^2\right) ^{0.5} \mbox{,}\\
t_2-t_1 &=&\left( \left( x_{2,1}+vt_2-x_{1,1}-vt_1\right) ^2+\sum_{j=2}^\mu
\left( x_{2,j}-x_{1,j}\right) ^2\right) ^{0.5} \mbox{,}\\
&&\ldots \mbox{,} \\
t_{k+1}-t_k &=&\left( \left( x_{k+1,1}+vt_{k+1}-x_{k,1}-vt_k\right)
^2+\sum_{j=2}^\mu \left( x_{k+1,j}-x_{k,j}\right) ^2\right) ^{0.5}\mbox{,} \\
&&\ldots \mbox{,} \\
t_{N+1}-t_N &=&\left( \left( x_{b,1}+vt_{N+1}-x_{N,1}-vt_N\right)
^2+\sum_{j=2}^\mu \left( x_{b,j}-x_{N,j}\right) ^2\right) ^{0.5}\mbox{.}
\end{eqnarray*}

Если обозначить:

\[
\rho _{a,b}^2\stackrel{def}{=}\sum_{j=1}^\mu \left( x_{b,1}-x_{a,1}\right) ^2 \mbox{,}
\]

то для каждого $k$:

\[
t_{k+1}-t_k=\frac v{1-v^2}\left( x_{k+1,1}-x_{k,1}\right) +\frac
1{1-v^2}\left( \rho _{k,k+1}^2-v^2\sum_{j=2}^\mu \left(
x_{k+1,j}-x_{k,j}\right) ^2\right) ^{0.5}\mbox{.}
\]

Отсюда:

\begin{eqnarray*}
&&t_{N+1}-t_0= \\
&=&\frac 1{1-v^2}\left( 
\begin{array}{c}
\left( \rho _{b,1}^2-v^2\sum_{j=2}^\mu \left( x_{1,j}-x_{b,j}\right)
^2\right) ^{0.5} \\ 
+\left( \rho _{N,b}^2-v^2\sum_{j=2}^\mu \left( x_{b,j}-x_{N,j}\right)
^2\right) ^{0.5} \\ 
+\sum_{k=1}^{N-1}\left( \rho _{k,k+1}^2-v^2\sum_{j=2}^\mu \left(
x_{k+1,j}-x_{k,j}\right) ^2\right) ^{0.5}
\end{array}
\right)\mbox{.} 
\end{eqnarray*}

Аналогично, если обозначить:

\begin{center}
$\tau _{N+1}\stackrel{def}{=}\left[ \mathbf{b^{\bullet }}\alpha \mathbf{b^{\bullet }}C\mid
\Re \mathbf{a}\widetilde{H}\right] $,
\end{center}

то

\begin{eqnarray*}
&&\ \tau _{N+1}-t_0= \\
\  &=&\frac 1{1-v^2}\left( 
\begin{array}{c}
\left( \rho _{1,b}^2-v^2\sum_{j=2}^\mu \left( x_{b,j}-x_{1,j}\right)
^2\right) ^{0.5} \\ 
+\left( \rho _{b,N}^2-v^2\sum_{j=2}^\mu \left( x_{N,j}-x_{b,j}\right)
^2\right) ^{0.5} \\ 
+\sum_{k=1}^{N-1}\left( \rho _{k+1,k}^2-v^2\sum_{j=2}^\mu \left(
x_{k,j}-x_{k+1,j}\right) ^2\right) ^{0.5}
\end{array}
\right) \mbox{,}
\end{eqnarray*}

т.е.

\begin{equation}
t_{N+1}-t_0=\tau_{N+1}-t_0 \mbox{.}\label{o111}
\end{equation}

По Теореме 3.4.5:

\begin{center}
$\tau _{N+1}-t_0=\frac{q^{\prime }-p^{\prime }}{\sqrt{1-v^2}}$ и $t_{N+1}-t_0=%
\frac{r^{\prime }-p^{\prime }}{\sqrt{1-v^2}}$.
\end{center}

Отсюда и из (\ref{o111}), (\ref{o108}), (\ref{o109}):

\begin{center}
$\jmath \left( \mathbf{b}\widetilde{T}\right) \left( \mathbf{b}^{\bullet
}\alpha \mathbf{b}^{\bullet }C\right) =\jmath \left( \mathbf{b}\widetilde{T}%
\right) \left( \mathbf{b}^{\bullet }\alpha ^{\dagger }\mathbf{b}^{\bullet
}C\right) $.
\end{center}

Отсюда и из (\ref{o107}) по Определению 3.3.3:  
$\Im $ - $ISS\left( {\bf b},\widetilde{T}\right) {\bf _\Box}$

Следовательно, внутренняя неподвижность сохраняется в прямолинейном равномерном 
движении.

{\bf Теорема 3.4.7}

Пусть:

1) в координатной системе $R^\mu $ системы отсчета $\left( \Re {\bf a}%
\widetilde{H}\right) $ в каждый момент $t$:

${\bf b}$ : $\left\langle x_{{\bf b},1}+v\cdot t,x_{{\bf b},2},x_{{\bf b}%
,3},\ldots ,x_{{\bf b},\mu }\right\rangle $,

${\bf g}_j$: $\left\langle y_{j,1}+v\cdot t,y_{j,2},y_{j,3},\ldots
,y_{j,\mu }\right\rangle $,

${\bf u}_j$: $\left\langle y_{j,1}+v\cdot t,y_{j,2}+l/\left( k_1\cdot
\ldots \cdot k_j\right) ,y_{j,3},\ldots ,y_{j,\mu }\right\rangle $,

для каждого рекордера ${\bf q}_i$: если ${\bf q}_i\in \Im $, то координаты

${\bf q}_i$ : $\left\langle x_{i,1}+v\cdot t,x_{i,2},x_{i,3},\ldots ,x_{i,\mu
}\right\rangle $,

$\widetilde{T}$ есть $\left\langle {\left\{ {\bf g}_1,A,{\bf u}_1\right\} ,\
\left\{ {\bf g}_2,A,{\bf u}_2\right\} ,...,\left\{ {\bf g}_j,A,{\bf u}%
_j\right\} ,\ ...\ }\right\rangle $.

$C$ : $\left\langle C_1,C_2,C_3,\ldots ,C_\mu \right\rangle $,

$D$ : $\left\langle D_1,D_2,D_3,\ldots ,D_\mu \right\rangle $,

$t_C=\left[C\mid \Re {\bf a}\widetilde{H}\right] $,

$t_D=\left[D\mid \Re {\bf a}\widetilde{H}\right] $;

2) в координатной системе $R^{\mu \prime }$ системы отсчета 
$\left( \Im {\bf b}\widetilde{T}\right) $:

$C$ : $\left\langle C_1^{\prime },C_2^{\prime },C_3^{\prime },\ldots ,C_\mu
^{\prime }\right\rangle $,

$D$ : $\left\langle D_1^{\prime },D_2^{\prime },D_3^{\prime },\ldots ,D_\mu
^{\prime }\right\rangle $,

$t_C^{\prime }=\left[C\mid \Im {\bf b}\widetilde{T}%
\right] $,

$t_D^{\prime }=\left[D\mid \Im {\bf b}\widetilde{T}%
\right] $.

В этом случае:

\begin{center}
$\left( t_D^{\prime }-t_C^{\prime }\right) ^2-\left( D_1^{\prime
}-C_1^{\prime }\right) ^2-\left( D_2^{\prime }-C_2^{\prime
}\right)^2-\ldots-\left( D_\rho ^{\prime }-C_\mu ^{\prime }\right) ^2=$\\

=$\left( t_D-t_C\right) ^2-\left( D_1-C_1\right) ^2-\left(
D_2-C_2\right)^2-\ldots -\left( D_\rho -C_\mu \right) ^2$ \mbox{.}\\
\end{center}

{\bf Доказательство} 

Обозначим:

\[
\rho _{\mathbf{a},\mathbf{b}}\stackrel{def}{=}\left( \sum_{j=1}^\mu \left( b_j-a_j\right)
^2\right) ^{0.5}\mbox{.}
\]

По Определению 3.3.8 существуют элементы ${\bf q}_C$ и ${\bf q}_D$ множества $\Im$ 
такие, что 

\begin{center}
$\natural \left( {\bf b}\right) \left( {\bf q}_C,C\right) $), 
$\natural \left( {\bf b}\right) \left( {\bf q}_D,D\right) $
\end{center}

и

\begin{center}
$\ell\left( \Im {\bf b}\widetilde{T}\right) \left( C,D\right)=%
\ell\left( {\bf b},\widetilde{T}\right) \left( {\bf q}_C,{\bf q}_D\right)$.
\end{center}

В этом случае:

\begin{center}
$t_C^{\prime }=\left[ C\mid \Im \mathbf{b}\widetilde{T}\right] =\left[ 
\mathbf{q}_C^{\bullet }C\mid \Im \mathbf{b}\widetilde{T}\right] $,

$t_D^{\prime }=\left[ D\mid \Im \mathbf{b}\widetilde{T}\right] =\left[ 
\mathbf{q}_D^{\bullet }D\mid \Im \mathbf{b}\widetilde{T}\right] $.
\end{center}

По Следствию из Теоремы 3.4.5:

\begin{center}
$\left[ \mathbf{q}_C^{\bullet }C\mid \Re \mathbf{a}\widetilde{H}\right]
=\left[ C\mid \Re \mathbf{a}\widetilde{H}\right] =t_C$,

$\left[ \mathbf{q}_D^{\bullet }D\mid \Re \mathbf{a}\widetilde{H}\right]
=\left[ D\mid \Re \mathbf{a}\widetilde{H}\right] =t_D$.
\end{center}

Обозначим:

\begin{eqnarray*}
&&\tau_1\stackrel{def}{=}\left[ \mathbf{b}^{\bullet }C\uparrow \mathbf{b},\widetilde{T}\right] 
\mbox{,}\\
&&\tau_2\stackrel{def}{=}\left[ \mathbf{b}^{\bullet }D\uparrow \mathbf{b},\widetilde{T}\right] \mbox{,}\\
&&t_1\stackrel{def}{=}\left[ \mathbf{b}^{\bullet }C\mid \Re \mathbf{a}\widetilde{H}\right] \mbox{,}\\
&&t_2\stackrel{def}{=}\left[ \mathbf{b}^{\bullet }D\mid \Re \mathbf{a}\widetilde{H}\right] \mbox{,}\\
&&t_3\stackrel{def}{=}\left[ \mathbf{b}^{\bullet }B\mid \Re \mathbf{a}\widetilde{H}\right] \mbox{,}\\
&&t_4\stackrel{def}{=}\left[ \mathbf{q}_C^{\bullet }\mathbf{b}^{\bullet }B\mid \Re \mathbf{a}%
\widetilde{H}\right] \mbox{,}\\
&&t_5\stackrel{def}{=}\left[ \mathbf{b}^{\bullet }\mathbf{q}_C^{\bullet }\mathbf{b}^{\bullet
}B\mid \Re \mathbf{a}\widetilde{H}\right] \mbox{,}\\
&&t_6\stackrel{def}{=}\left[ \mathbf{q}_D^{\bullet }\mathbf{q}_C^{\bullet }\mathbf{b}^{\bullet
}B\mid \Re \mathbf{a}\widetilde{H}\right] \mbox{,}\\
&&t_7\stackrel{def}{=}\left[ \mathbf{q}_C^{\bullet }\mathbf{q}_D^{\bullet }\mathbf{q}_C^{\bullet }%
\mathbf{b}^{\bullet }B\mid \Re \mathbf{a}\widetilde{H}\right] \mbox{,}\\
&&t_8\stackrel{def}{=}\left[ \mathbf{b}^{\bullet }\mathbf{q}_C^{\bullet }\mathbf{q}_D^{\bullet }%
\mathbf{q}_C^{\bullet }\mathbf{b}^{\bullet }B\mid \Re \mathbf{a}\widetilde{H}%
\right] \mbox{.}\\
\end{eqnarray*}

При этих обозначениях:

\begin{center}
$t_8-t_7=t_5-t_4$, т.е.: $t_8-t_5=t_7-t_4$, и
$\ell \left( \Im \mathbf{b}\widetilde{T}\right) \left( C,D\right) =0.5\left(
\left[ \mathbf{b}^{\bullet }\mathbf{q}_C^{\bullet }\mathbf{q}_D^{\bullet }%
\mathbf{q}_C^{\bullet }\mathbf{b}^{\bullet }B\uparrow \mathbf{b},\widetilde{T%
}\right] -\left[ \mathbf{b}^{\bullet }\mathbf{q}_C^{\bullet }\mathbf{b}%
^{\bullet }B\uparrow \mathbf{b},\widetilde{T}\right] \right) $,
\end{center}

т.е.:

\begin{center}
$\ell \left( \Im \mathbf{b}\widetilde{T}\right) \left( C,D\right) =0.5\left(
t_8-t_5\right) \sqrt{1-v^2}=0.5\left( t_7-t_4\right) \sqrt{1-v^2}$,\\
$\left( t_7-t_6\right) ^2=\left( x_{C,1}+vt_7-x_{D,1}-vt_6\right)
^2+\sum_{j=2}^\mu \left( x_{C,j}-x_{D,j}\right) ^2$, \\ 
$\left( t_6-t_4\right) ^2=\left( x_{D,1}+vt_6-x_{C,1}-vt_4\right)
^2+\sum_{j=2}^\mu \left( x_{C,j}-x_{D,j}\right) ^2$,
\end{center}

т.е.:

\begin{center}
$\left( t_7-t_6\right) ^2=v^2\left( t_7-t_6\right) ^2+2v\left(
x_{C,1}-x_{D,1}\right) \left( t_7-t_6\right) +\rho _{\mathbf{q}_C,\mathbf{q}%
_D}^2$,\\
$\left( t_6-t_4\right) ^2=v^2\left( t_6-t_4\right) ^2+2v\left(
x_{D,1}-x_{C,1}\right) \left( t_6-t_4\right) +\rho _{\mathbf{q}_D,\mathbf{q}%
_C}^2$.
\end{center}

Отсюда:

\begin{center}
$t_7-t_4=\frac 2{\sqrt{1-v^2}}\left( v^2\left( x_{D,1}-x_{C,1}\right)
^2+\left( 1-v^2\right) \rho _{\mathbf{q}_C,\mathbf{q}_D}^2\right) ^{0.5}$.
\end{center}

Обозначим:

\[
R_{\mathbf{a},\mathbf{b}}\stackrel{def}{=}\left( \rho _{\mathbf{a},\mathbf{b%
}}^2-v^2\sum_{j=2}^\mu \left( a_j-b_j\right) ^2\right) ^{0.5}\mbox{.}
\]

При таком обозначении:

\[
\ell \left( \Im \mathbf{b}\widetilde{T}\right) \left( C,D\right) =\frac{R_{%
\mathbf{q}_C,\mathbf{q}_D}}{\sqrt{1-v^2}}\mbox{.}
\]

Т.к.

\begin{center}
$C_1=x_{C,1}+vt_C$, $D_1=x_{D,1}+vt_D$,\\
$C_{j+1}=x_{C,j+1}$, $D_{j+1}=x_{D,j+1}$,
\end{center}

то

\[
R_{\mathbf{q}_C,\mathbf{q}_D}=\left( 
\begin{array}{c}
v^2\left( D_1-vt_D-C_1+vt_C\right) ^2 \\ 
+\left( 1-v^2\right) \left( \left( D_1-vt_D-C_1+vt_C\right)
^2+\sum_{j=2}^\mu \left( D_j-C_j\right) ^2\right) 
\end{array}
\right) ^{0.5}\mbox{,}
\]

т.е.:

\begin{equation}
R_{\mathbf{q}_C,\mathbf{q}_D}=\left( v^2\left( t_D-t_C\right) ^2-2v\left(
t_D-t_C\right) \left( D_1-C_1\right) +\rho _{C,D}^2-v^2\sum_{j=2}^\mu \left(
D_j-C_j\right) ^2\right) ^{0.5}\mbox{.}\label{r}
\end{equation}

Кроме того, по Определению 3.3.7:

\begin{equation}
t_D^{\prime }-t_C^{\prime }=\left( \tau _2-\tau _1\right) -\left( \ell \left( \mathbf{b},%
\widetilde{T}\right) \left( \mathbf{b},\mathbf{q}_D\right) -\ell \left( 
\mathbf{b},\widetilde{T}\right) \left( \mathbf{b},\mathbf{q}_C\right)
\right)\mbox{.}   \label{o112}
\end{equation}

По Следствию из Теоремы 3.4.5:

\begin{equation}
\tau _2-\tau _1=\left( t_2-t_1\right) \sqrt{1-v^2}\mbox{.}  \label{o113}
\end{equation}

По Теореме 3.4.3:

\begin{center}
$\left( t_1-t_C\right) ^2=\left( x_{\mathbf{b},1}+vt_1-C_1\right)
^2+\sum_{j=2}^\mu \left( x_{\mathbf{b},j}-C_j\right) ^2$,\\
$\left( t_2-t_D\right) ^2=\left( x_{\mathbf{b},1}+vt_2-D_1\right)
^2+\sum_{j=2}^\mu \left( x_{\mathbf{b},j}-D_j\right) ^2$.
\end{center}

Следовательно,

\begin{center}
$\left( t_1-t_C\right) ^2=v^2\left( t_1-t_C\right) ^2+2v\left( x_{\mathbf{b}%
,1}-x_{C,1}\right) \left( t_1-t_C\right) +\rho _{\mathbf{b},\mathbf{q}_C}^2$,\\
$\left( t_2-t_D\right) ^2=v^2\left( t_2-t_D\right) ^2+2v\left( x_{\mathbf{b}%
,1}-x_{D,1}\right) \left( t_2-t_D\right) +\rho _{\mathbf{b},\mathbf{q}_D}^2$.
\end{center}

Отсюда:

\begin{equation}
t_2-t_1=\left( t_D-t_C\right) +\frac v{1-v^2}\left( x_{C,1}-x_{D,1}\right) +\frac
1{1-v^2}\left( R_{\mathbf{b},\mathbf{q}_D}-R_{\mathbf{b},\mathbf{q}%
_C}\right) \mbox{.}  \label{o114}
\end{equation}

Учитывая что:

\[
\ell \left( \mathbf{b},\widetilde{T}\right) \left( \mathbf{b},\mathbf{q}%
_D\right) =\frac{R_{\mathbf{b},\mathbf{q}_D}}{\sqrt{1-v^2}}\mbox{, }
\ell \left( \mathbf{b},\widetilde{T}\right) \left( \mathbf{b},\mathbf{q}%
_C\right) =\frac{R_{\mathbf{b},\mathbf{q}_C}}{\sqrt{1-v^2}}\mbox{,}
\]

из (\ref{o112}), (\ref{o113}), (\ref{o114}):

\[
t_D^{\prime }-t_C^{\prime }=\left( t_D-t_C\right) \sqrt{1-v^2}-\frac v{%
\sqrt{1-v^2}}\left( x_{D,1}-x_{C,1}\right)\mbox{,}
\]

т.е.:

\[
t_D^{\prime }-t_C^{\prime }=\left( t_D-t_C\right) \sqrt{1-v^2}-\frac v{%
\sqrt{1-v^2}}\left( \left( D_1-C_1\right) -v\left( t_D-t_C\right) \right)\mbox{,}
\]

т.е.:

\begin{equation}
t_D^{\prime }-t_C^{\prime }=\frac{\left( t_D-t_C\right) -v\left(
D_1-C_1\right) }{\sqrt{1-v^2}}\mbox{.}  \label{oo}
\end{equation}

Отсюда:

$
\left( t_D^{\prime }-t_C^{\prime }\right) ^2-\left( \ell \left( \Im \mathbf{b%
}\widetilde{T}\right) \left( C,D\right) \right) ^2= \\ 
=\frac 1{1-v^2}\left( \left( t_D-t_C\right) ^2-2v\left( t_D-t_C\right) \left(%
D_1-C_1\right)  +v^2\left( D_1-C_1\right) ^2-R_{\mathbf{q}_C,%
\mathbf{q}_D}^2
\right) 
$,

т.е. по (\ref{r}):

\begin{center}
$\left( t_D^{\prime }-t_C^{\prime }\right) ^2-\left( D_1^{\prime
}-C_1^{\prime }\right) ^2-\left( D_2^{\prime }-C_2^{\prime }\right)
^2-\cdots -\left( D_\mu ^{\prime }-C_\mu ^{\prime }\right) ^2=\left(
t_D-t_C\right) ^2-\rho _{C,D}^2$ ${\bf _\Box}$
\end{center}

Отсюда сразу получаются пространственно-временные преобразования Лорен-ца.
\section{Вероятность}

Рассматриваем выражения каких-либо языков.

{\bf О.1:} Множество выражений называем {\it $\kappa$-объектом}, 
если любая пара элементов этого множества представляет пару синонимов. Элементы 
этого множества называем {\it именами} этого $\kappa$-объекта.

Например, множество: $\{ \ll$ сумма чисел 3 и 1 $\gg$, $\ll$ four $\gg$, 
$\ll 2+2 \gg$, $\ll IV \gg$, $\ll 2^2 \gg$, $\ll 4 \gg$, $\ll$ четверка $\gg$ 
$\}$ есть $\kappa$-объект с именем $\ll 4 \gg$.

{\bf О.2:} $\kappa$-объект, именами которого являются повествовательные 
предложения, называется {\it событием}.

Событие с именем $A$ обозначаем следующим образом: $^{\circ }A$.

{\bf О.9:} Cобытие {\it происходит}, если и только если его имена истинны.

{\bf О.10:} Cобытие ${\mathcal C}$ есть {\it произведение} событий ${\mathcal A}$ 
и ${\mathcal B}$ (обозначение: 
$\mathcal{C}=\left(\mathcal{A}\cdot \mathcal{B}\right)$), если $\mathcal{C}$ 
имеет имя вида: $\left(A \& B \right)$, где $A$ - имя события $\mathcal{A}$, а 
$B$ - имя события $\mathcal{B}$.

{\bf О.11:} События $\mathcal{A}$ и $\mathcal{B}$ {\it несовместны}, если 
событие $\left( \mathcal{A}\cdot \mathcal{B}\right)$ имеет ложное имя.  

{\bf О.12:} Cобытие $\mathcal{C}$ {\it противоположно} к событию 
$\mathcal{A}$ (обозначение: $\mathcal{C} = \overline{\mathcal{A}}$), если 
$\mathcal{C}$ имеет имя вида: 
$\left(\neg A \right)$, где $A$ - имя события $\mathcal{A}$. 

{\bf О.13:} Cобытие $\mathcal{C}$ есть {\it сумма} событий $\mathcal{A}$ и 
$\mathcal{B}$ (обозначение: $\mathcal{C}=\left(\mathcal{A}+\mathcal{B}\right)$), 
если $\mathcal{C}$ имеет имя вида: $\left(A \vee B \right)$, где $A$ - имя 
события $\mathcal{A}$, а $B$ - имя события $\mathcal{B}$.

  Далее все (если специально не оговаривается) происходит в координатной 
системе $R^\mu $ системы отсчета $\left( \Re {\bf a}\widetilde{H}\right) $.

  Предложения вида $\ll$В момент $t$ $A$ имеет координаты ${\bf x}\gg$ я 
обозначаю как $A\left( t,\mathbf{x}\right)$. 

{\bf О.14:} Событие, именем которого является предложение вида
$A\left( t,\mathbf{x}\right)$ я называю {\it точечным событием}.

{\bf О.15:} Событие $\mathcal{C}$ я называю {\it физическим} событием, если и 
только если $\mathcal{C}$ есть точечное событие, или $\mathcal{C}$ есть 
произведение или сумма каких-либо физических событий, или $\mathcal{C}$ 
противоположно к какому-либо физическому событию. Кроме перечисленных, никаких 
физических событий нет.

\subsection{B-функции}

{\bf О.16:} Каждая функция $\mathfrak{b}\left( \mathcal{X}\right)$, 
определенная на множестве физических событий, имеющая область значений на отрезке 
$\left[ 0;1\right] $ числовой оси, называется {\it B-функцией}, если существует 
событие $\mathcal{C}$, для которого:

\[
\mathfrak{b}\left( \mathcal{C}\right) =1,
\]

и для любых событий $\mathcal{A}$ и $\mathcal{B}$:

\[
\mathfrak{b}\left( \mathcal{A}\cdot \mathcal{B}\right) +\mathfrak{b}\left( %
\mathcal{A}\cdot \overline{\mathcal{B}}\right) =\mathfrak{b}\left( \mathcal{A}\right).  
\]

{\bf Т.2:} Для любой B-функции $\mathfrak{b}$:

1) для всех событий $\mathcal{A}$ и $\mathcal{B}$: $\mathfrak{b}\left( \mathcal{A}\cdot \mathcal{B}\right)
\leq \mathfrak{b}\left( \mathcal{A}\right) $;

2) для каждого события $\mathcal{A}$: если $\mathcal{T}$ происходит, то $\mathfrak{b}%
\left( \mathcal{A}\right) +\mathfrak{b}\left(\overline{\mathcal{A}}\right)=\mathfrak{b}%
\left( \mathcal{T}\right) $;

3) для каждого события $\mathcal{A}$: если $\mathcal{T}$ происходит, то $\mathfrak{b}%
\left( \mathcal{A}\right) \leq \mathfrak{b}\left( \mathcal{T}\right) $;

4) если $\mathcal{T}$ происходит, то

\begin{center}
$\mathfrak{b}\left( \mathcal{T}\right) =1$;  
\end{center}

5) для каждого события $\mathcal{A}$:

\begin{center}
$\mathfrak{b}\left( \mathcal{A}\right) +\mathfrak{b}\left(\overline{\mathcal{A}}\right) =1$;
\end{center}

6) если событие $\mathcal{F}$ имеет ложное имя, то 

\begin{center}
$\mathfrak{b}\left( \mathcal{F}\right) = 0$.
\end{center}

{\bf Доказательство Т.2:}

1) Из О.16.

2) По пунктам 4 и 2 Т.1 и О.9:

\[
\mathfrak{b}\left( \mathcal{T}\cdot \mathcal{A}\right) +\mathfrak{b}\left( \mathcal{T}\cdot 
\overline{\mathcal{A}}\right) =\mathfrak{b}\left( \mathcal{A}\right) +\mathfrak{b}\left( 
\overline{\mathcal{A}}\right) . 
\]

3) Из предыдущего пункта этой Теоремы.

4) Из предыдущего пункта этой Теоремы и из О.16.

5) Из пункта 2 этой Теоремы.

6) Из пункта 5 Т.1, из О.10 и О.16:

\[
\mathfrak{b}\left( \mathcal{F}\cdot \mathcal{A}\right) +\mathfrak{b}\left( \mathcal{F}\cdot 
\overline{\mathcal{A}}\right) =\mathfrak{b}\left( \mathcal{F}\right) +\mathfrak{b}\left( 
\mathcal{F}\right) =\mathfrak{b}\left( \mathcal{F}\right)  _{\bf \Box } 
\]

{\bf Т.3:}

\[
\mathfrak{b}\left( \mathcal{A}+ \mathcal{B}\right) =\mathfrak{b}\left( \mathcal{A}\right) +%
\mathfrak{b}\left( \mathcal{B}\right) -\mathfrak{b}\left( \mathcal{A}\cdot \mathcal{B}\right) \mbox{.}
\]

{\bf Доказательство Т.3: }по О.13, O.8 и по пункту 5 Т.2:

\[
\mathfrak{b}\left( \mathcal{A}+ \mathcal{B}\right) =1-\mathfrak{b}\left(\overline{\mathcal{A}}%
\cdot\overline{\mathcal{B}}\right) . 
\]

По О.16:

\[
\mathfrak{b}\left( \mathcal{A}+ \mathcal{B}\right) =1-\mathfrak{b}\left(\overline{\mathcal{A}}%
\right) +\mathfrak{b}\left(\overline{\mathcal{A}}\cdot \mathcal{B}\right) =\mathfrak{b}%
\left( \mathcal{A}\right) +\mathfrak{b}\left( \mathcal{B}\right) -\mathfrak{b}\left( \mathcal{A}\cdot
\mathcal{B}\right)  _{\bf \Box } 
\]

{\bf Т.4} Если предложения $\mathcal{A}$ и $\mathcal{B}$ несовместны, то

\[
\mathfrak{b}\left( \mathcal{A} + \mathcal{B}\right) =\mathfrak{b}\left( \mathcal{A}\right) +%
\mathfrak{b}\left( \mathcal{B}\right) \mbox{.} 
\]

{\bf Доказательство Т.4:} Из пункта 6 T.2, из T.3 по О.11 $_{\bf \Box }$

{\bf Т.5:} Если $\mathfrak{b}\left( \mathcal{A}\cdot \mathcal{B}\right) =\mathfrak{\
b}\left( \mathcal{A}\right) \cdot \mathfrak{b}\left( \mathcal{B}\right) $, то $\mathfrak{b}%
\left( \mathcal{A}\cdot \overline{\mathcal{B}}\right) =\mathfrak{b}\left( \mathcal{A}\right) \cdot {%
\mathfrak{b}}\left( \overline{\mathcal{B}}\right) $.

{\bf Доказательство Т.5:} По Определению O.16:

\[
\mathfrak{b}\left( \mathcal{A}\cdot \overline{\mathcal{B}} \right) =\mathfrak{b}\left(
\mathcal{A}\right) -\mathfrak{b}\left( \mathcal{A}\cdot \mathcal{B}\right) \mbox{.} 
\]

Поэтому,

\[
\mathfrak{b}\left( \mathcal{A}\cdot \overline{\mathcal{B}}\right) =\mathfrak{b}\left(
\mathcal{A}\right) -\mathfrak{b}\left( \mathcal{A}\right) \cdot \mathfrak{b}\left( \mathcal{B}\right) =%
{\mathfrak{b}}\left( \mathcal{A}\right) \cdot \left( 1-\mathfrak{b}\left( \mathcal{B}\right)
\right) \mbox{.} 
\]

Поэтому, по пункту 5 Т.2:

\[
\mathfrak{b}\left( \mathcal{A}\cdot \overline{\mathcal{B}} \right) =\mathfrak{b}\left(
\mathcal{A}\right) \cdot \mathfrak{b}\left(\overline{\mathcal{B}} \right)  _{\bf \Box } 
\]

{\bf O.17:} События $\mathcal{A}$ и $\mathcal{B}$ {\it независимы} для $B$-функции $\mathfrak{b}$, 
если 

\[
\mathfrak{b}\left( \mathcal{A}\cdot \mathcal{B}\right) =\mathfrak{\
b}\left( \mathcal{A}\right) \cdot \mathfrak{b}\left( \mathcal{B}\right) \mbox{.}
\]  

{\bf Т.6:} $\mathfrak{b}\left( \mathcal{A}\cdot \overline{\mathcal{A}}\cdot \mathcal{B}\right) =0$.

{\bf Доказательство Т.6: }По О.16 и по пунктам 2 и 3 Т.1:

\[
\mathfrak{b}\left( \mathcal{A}\cdot \overline{\mathcal{A}} \cdot \mathcal{B}\right) =\mathfrak{b}%
\left( \mathcal{A}\cdot \mathcal{B}\right) -\mathfrak{b}\left( \mathcal{A}\cdot \mathcal{A}\cdot \mathcal{B}\right) , 
\]

поэтому, из пункта 1 Т.1:

\[
\mathfrak{b}\left( \mathcal{A}\cdot \overline{\mathcal{A}} \cdot \mathcal{B}\right) =\mathfrak{b}%
\left( \mathcal{A}\cdot \mathcal{B}\right) -\mathfrak{b}\left( \mathcal{A}\cdot \mathcal{B}\right)  _{\bf \Box } 
\]

{\bf Т.7:}

\[
\mathrm{P}\left( \mathcal{A}\cdot \left( \mathcal{B} + \mathcal{C}\right) \right) = %
\mathrm{P}\left( \mathcal{A}\cdot \mathcal{B}\right) +\mathrm{P}\left( \mathcal{A}\cdot \mathcal{C}\right) -%
\mathrm{P}\left( \mathcal{A}\cdot \mathcal{B}\cdot \mathcal{C}\right) \mbox{.}
\]

{\bf Доказательство Т.7:}

По О.8, O.13 и О.16:

$\mathrm{P}\left( \mathcal{A}\cdot \left( \mathcal{B} + \mathcal{C}\right) \right) =\mathrm{P}\left(
\mathcal{A}\cdot \overline{\overline{\mathcal{B}}\cdot \overline{\mathcal{C}}}\right) =%
\mathrm{P}\left( \mathcal{A}\right) -\mathrm{P}\left( \mathcal{A}\cdot \overline{\mathcal{B}}\cdot 
\overline{\mathcal{C}}\right) =\mathrm{P}\left( \mathcal{A}\right) -\mathrm{P}\left( \mathcal{A}\cdot 
\overline{\mathcal{B}}\right) +\mathrm{P}\left( \mathcal{A}\cdot \overline{\mathcal{B}}\cdot \mathcal{C}\right) =
\mathrm{P}\left( \mathcal{A}\cdot \mathcal{B}\right) +\mathrm{P}\left( \mathcal{A}\cdot \mathcal{C}\right) -%
\mathrm{P}\left( \mathcal{A}\cdot \mathcal{B}\cdot \mathcal{C}\right)  _{\bf \Box }$

\subsection{Независимые испытания}

Пусть ${\bf N}$ - множество натуральных чисел.

{\bf О.18} Пусть $s(n)$ - функция, определенная на ${\bf N}$, имеющая область 
значений в множестве событий.

В этом случае событие $\mathcal{C}$ называется {\it [s]-серией 
ранга $r$ с V-числом $k$}, если $\mathcal{C}$, $r$ и $k$ подчиняются одному из 
следующих условий:

1) $r=1$ и $k=1$, $\mathcal{C}\stackrel{def}{=}s\left( 1\right) $, или $k=0$, 
$\mathcal{C}\stackrel{def}{=}\overline{s\left(1\right)} $;

2) $\mathcal{B}$ есть [s]-серия ранга $r-1$ с V-числом $k-1$ и

\[
\mathcal{C}\stackrel{def}{=}\left( \mathcal{B}\cdot s\left( r\right) \right) \mbox{,} 
\]

или $\mathcal{B}$ есть [s]-серия ранга $r-1$ с V-числом $k$ и

\[
\mathcal{C}\stackrel{def}{=}\left( \mathcal{B}\cdot \overline{s\left( r\right)} \right) \mbox{.} 
\]

Обозначим $[s](r,k)$ множество $[s]$-серий ранга $r$ с 
V-числом $k$.

Например, если $s\left( n\right) $ есть последовательность $\mathcal{B}_n$, то 
события:

$\left( \mathcal{B}_1\cdot \mathcal{B}_2\cdot \overline{\mathcal{B}_3}\right) $, $\left( \mathcal{B}_1\cdot%
\overline{\mathcal{B}_2}\cdot \mathcal{B}_3\right) $, $\left(\overline{\mathcal{B}_1}\cdot \mathcal{B}_2\cdot
\mathcal{B}_3\right) $

являются элементами множества $[s](3,2)$, а $\left( \mathcal{B}_1\cdot \mathcal{B}_2\cdot\overline{%
\mathcal{B}_3}\cdot \mathcal{B}_4\cdot\overline{\mathcal{B}_5}\right) \in [s](5,3)$.

{\bf О.19:} Функция $s(n)$ {\it независима} для B-функции 
$\mathfrak{b}$, если:

\[
\mathfrak{b}\left( \mathcal{A}\right) =\prod\limits_{n=1}^r\mathfrak{b}\left(
s\left( n\right) \right) 
\]

для любого $\mathcal{A}$, такого что  $\mathcal{A}\in $ $[s](r,r)$. 

{\bf О.20:} Пусть $s(n)$ - функция, определенная на ${\bf N}$, имеющая область 
значений в множестве событий.

В этом случае событие $\mathcal{C}$ называется {\it [s]-суммой ранга $r$ с 
V-числом $k$} (обозначим: $\mathfrak{t}[s](r,k)$), если $\mathcal{C}$ есть сумма 
всех элементов множества $[s](r,k)$.

Например, если $s\left( n\right) $ есть событие $\mathcal{C}_n$, то:

$\left( \overline{\mathcal{C}_1}\cdot \overline{\mathcal{C}_2}\cdot \overline{\mathcal{C}_3}\right) ={%
\mathfrak{t}}[s]\left( 3,0\right) $,

$\mathfrak{t}[s]\left( 3,1\right) =\\=\left( \left( \mathcal{C}_1\cdot \overline{\mathcal{C}_2}%
\cdot \overline{\mathcal{C}_3}\right) + \left(\overline{\mathcal{C}_1}\cdot \mathcal{C}_2\cdot 
\overline{\mathcal{C}_3}\right) + \left(\overline{\mathcal{C}_1}\cdot \overline{\mathcal{C}_2}\cdot
\mathcal{C}_3\right) \right) $,

$\mathfrak{t}[s]\left( 3,2\right) =\left( \left( \mathcal{C}_1\cdot \mathcal{C}_2\cdot 
\overline{\mathcal{C}_3}\right) + \left(\overline{\mathcal{C}_1}\cdot \mathcal{C}_2\cdot \mathcal{C}_3\right)
+ \left( \mathcal{C}_1\cdot\overline{\mathcal{C}_2}\cdot \mathcal{C}_3\right) \right) $,

$\left( \mathcal{C}_1\cdot \mathcal{C}_2\cdot \mathcal{C}_3\right) =\mathfrak{t}[s]\left( 3,3\right)$.

{\bf О.21:} Пусть функция $s_{\mathcal{A}}(n)$, определенная на ${\bf N}$, 
имеющая значения в множестве событий, независимая для B-функции $\mathfrak{b}$, 
подчиняется условию: 

\begin{center}
$\mathfrak{b}\left(s_{\mathcal{A}}(n)\right)=\mathfrak{b}\left(\mathcal{A}\right)$.
\end{center} 

для любого $n$.

В этом случае  $[s_\mathcal{A}]$-серию ранга $r$ с V-числом $k$ называем {\it 
последовательностью $r$ независимых для B-функции $\mathfrak{b}$ 
$[s_\mathcal{A}]$-испытаний события $\mathcal{A}$ с результатом $k$}.

А число $k/r$ называем {\it частотой события} $\mathcal{A}$ в этой последовательности
(обозначение: $\nu _r\left[ s_\mathcal{A}\right]\stackrel{def}{=}k/r$). 

{\bf Т.8:} {\bf (формула Бернулли} \cite{BER13}{\bf )} Если $s(n)$ 
независима для B-функции $\mathfrak{b}$ и существует вещественное число $p$, 
такое что для всех $n$: $\mathfrak{b}\left( s\left( n\right)\right) =p$, то

\[
\mathfrak{b}\left( \mathfrak{t}\left[ s\right] \left( r,k\right)
\right) =\frac{r!}{k!\cdot \left( r-k\right) !}\cdot p^k\cdot \left(
1-p\right) ^{r-k}\mbox{.} 
\]

{\bf Доказательство Т.8: }По О.19 и по Т.5: 
если $\mathcal{B}\in \left[ s\right] \left( r,k\right) $, то:

\[
\mathfrak{b}\left( \mathcal{B}\right) =p^k\cdot \left( 1-p\right) ^{r-k}\mbox{.} 
\]

Т.к. $\left[ s\right] \left( r,k\right) $ содержит ${r!}/\left({k!\cdot
\left( r-k\right) !}\right)$ элементов, то по T.4, T.5, T.6 эта 
Теорема выполняется  $_{\bf \Box }$

{\bf О.22:} Пусть функция $s(n)$ определена на ${\bf N}$ и имеет область 
значений в множестве событий.

В этом случае событие $\mathfrak{T}[s](r,k,l)$, здесь $r,k,l$ - натуральные 
числа, определяет-ся следующим образом: 

1) $\mathfrak{T}[s](r,k,k)\stackrel{def}{=}\mathfrak{t}[s](r,k)$,

2) $\mathfrak{T}[s](r,k,l+1)\stackrel{def}{=}(\mathfrak{T}[s](r,k,l) + \mathfrak{t}[s](r,l+1)) $.

{\bf О.23:} Если $a$ и $b$ - вещественные числа, и $k-1<a\leq k$
и $l\leq b<l+1$, то $\mathfrak{T}[s](r,a,b)=\mathfrak{T}[s](r,k,l)$.

{\bf Т.9:} 

\[
\mathfrak{T}[s_\mathcal{A}](r,a,b)=^{\circ }\ll \frac ar \leq \nu _r\left[ s_\mathcal{A}\right]%
 \leq \frac br\gg \mbox{.} 
\]

{\bf Доказательство Т.9: }По О.23: существует натуральные числа $r$ и $k$, 
для которых: $k-1<a\leq k$, и $l\leq b<l+1$, и 
$\mathfrak{T}[s_\mathcal{A}](r,a,b)=\mathfrak{T}[s_\mathcal{A}](r,k,l)$. 

Индукция по $l$:

1. Пусть $l=k$.

В этом случае по О.22 и О.21:

\[
\mathfrak{T}[s_\mathcal{A}](r,k,k)=\mathfrak{t}[s_\mathcal{A}](r,k)=^{\circ}\ll \nu _r\left[ s_\mathcal{A}\right]
 =\frac kr\gg \mbox{.} 
\]

2. Пусть $n$ - некоторое натуральное число.

Индуктивное допущение: Допустим

\[
\mathfrak{T}[s_\mathcal{A}](r,k,k+n)=^{\circ}\ll \frac kr\leq \nu _r\left[ s_\mathcal{A}\right] \leq \frac {k+n}r\gg \mbox{.} 
\]

по О.22:

\[
\mathfrak{T}[s_\mathcal{A}](r,k,k+n+1)=(\mathfrak{T}[s_\mathcal{A}](r,k,k+n) + \mathfrak{t}%
[s_\mathcal{A}](r,k+n+1))\mbox{.} 
\]

По индуктивному допущению, по О.21 и по О.5:

\[
\mathfrak{T}[s_\mathcal{A}](r,k,k+n+1)= 
\]

\[
=(^{\circ}\ll \frac kr\leq \nu _r\left[ s_\mathcal{A}\right]  \leq \frac
{k+n}r\gg + ^{\circ}\ll \nu _r\left[ s_\mathcal{A}\right]  =\frac {k+n+1}r\gg)%
\mbox{.} 
\]

Поэтому, по О.13 и О.8:

\[
\mathfrak{T}[s_\mathcal{A}](r,k,k+n+1)=^{\circ}\ll \frac kr\leq \nu _r\left[ s_\mathcal{A}\right]%
 \leq \frac {k+n+1}r\gg  _{\bf \Box } 
\]

{\bf Т.10:} Если $s(n)$ независима для B-функции ${\mathfrak{b}}$ и 
существует вещественное число $p$, такое что ${\mathfrak{b}}\left( s\left( n\right) \right) =p$ 
для всех $n$, то

\[
\mathfrak{b}\left( \mathfrak{T}[s](r,a,b)\right) =\sum_{a\leq k\leq
b}\frac {r!}{k!\cdot \left( r-k\right) !}\cdot p^k\cdot \left( 1-p\right)
^{r-k}\mbox{.} 
\]

{\bf Доказательство Т.10: }Это сразу следует из Т.8 по Т.4  $_{\bf \Box }$

{\bf Т.11:} Если $s(n)$ независима для B-функции ${\mathfrak{b}}$ и 
существует вещественное число $p$, такое что ${\mathfrak{b}}\left( s\left( n\right) \right) =p$ 
для всех $n$, то

\[
\mathfrak{b}\left( \mathfrak{T}[s](r,r\cdot \left( p-\varepsilon
\right) ,r\cdot \left( p+\varepsilon \right) )\right) \geq 1-\frac{p\cdot
\left( 1-p\right) }{r\cdot \varepsilon ^2} 
\]

для любого положительного вещественного числа $\varepsilon $.

{\bf Доказательство Т.11:} Так как

\[
\sum_{k=0}^r\left( k-r\cdot p\right) ^2\cdot \frac{r!}{k!\cdot \left(
r-k\right) !}\cdot p^k\cdot \left( 1-p\right) ^{r-k}=r\cdot p\cdot \left(
1-p\right) ,
\]

то если

\[
J=\left\{ k\in \mathbf{N}|0\leq k\leq r\cdot \left( p-\varepsilon \right)
\right\} \cup \left\{ k\in \mathbf{N}|r\cdot \left( p+\varepsilon \right)
\leq k\leq r\right\} ,
\]

то

\[
\sum_{k\in J}\frac {r!}{k!\cdot \left( r-k\right) !}\cdot p^k\cdot \left(
1-p\right) ^{r-k}\leq \frac {p\cdot \left( 1-p\right) }{r\cdot \varepsilon
^2}\mbox{.} 
\]

Поэтому, по пункту 5 Т.2 эта Теорема выполняется  $_{\bf \Box }$

Поэтому

\begin{equation}
\lim\limits_{r\rightarrow \infty }\mathfrak{b}\left( \mathfrak{T}[s](r,r%
\cdot \left( p-\varepsilon \right) ,r\cdot \left( p+\varepsilon \right) )%
\right) =1 \label{epsl} 
\end{equation}

для всех как угодно малых положительных чисел $\varepsilon $.

\subsection{Функция вероятности}

{\bf О.24:} $n${\it -вход} ${\bf S_n}$ для ${\bf N}$ 
определяется рекурсивно: 

1) ${\bf S}_1\stackrel{def}{=}\left\{ 1\right\} $; 

2) ${\bf S}_{\left( n+1\right) }\stackrel{def}{=}{\bf S}_n\cup \left\{ n+1\right\} $. 

{\bf О.25: }Если ${\bf S}_n$ есть $n$-вход для ${\bf N}$, и  $%
{\bf {A}}\subseteq {\bf N}$, то $\left\| {\bf {A}}\cap {\bf S}_n\right\| $ есть 
количество элементов множества ${\bf {A}}\cap {\bf S}_n$, и если

\[ 
\varpi _n\left( {\bf {A}}\right) =\frac{\left\| {\bf {A}}\cap {\bf S}_n\right\| }%
n\mbox{,} 
\] 

то $\varpi _n\left( {\bf {A}}\right) $ есть {\it частота} множества $%
{\bf {A}}$ на $n$-входе ${\bf S}_n$. 

{\bf Т.12:} 

1) $\varpi _n({\bf N})=1$; 

2) $\varpi _n(\emptyset )=0$; 

3) $\varpi _n({\bf {A}})+\varpi _n({\bf N}-{\bf {A}})=1$; 

4) $\varpi _n({\bf {A}}\cap {\bf B})+\varpi _n({\bf {A}}\cap ({\bf N}-{\bf B}%
))=\varpi _n({\bf {A}})$. 

{\bf Доказательство Т.12:} Из О.24 и О.25  $_{\bf \Box }$ 

{\bf О.26: }Если "$\lim $" есть Коши-Вейерштрасса "limit", то:

\[ 
{\bf \Phi ix}\stackrel{def}{=}\left\{ {\bf {A}}\subseteq {\bf N}|\lim_{n\rightarrow \infty 
}\varpi _n({\bf {A}})=1\right\} \mbox{.} 
\] 

{\bf Т.13: }${\bf \Phi ix}$ есть фильтр \cite{DVS}, т.е.: 

1) ${\bf N}\in {\bf \Phi ix}$, 

2) $\emptyset \notin {\bf \Phi ix}$, 

3) если ${\bf {A}}\in {\bf \Phi ix}$, и ${\bf B}\in {\bf \Phi ix}$, то $({\bf {A}%
}\cap {\bf B})\in {\bf \Phi ix}$ ; 

4) если ${\bf {A}}\in {\bf \Phi ix}$, и ${\bf {A}}\subseteq {\bf B}$, то ${\bf B}%
\in {\bf \Phi ix}$. 

{\bf Доказательство Т.13:}

1), 2) и 4) получается сразу из свойств Коши-Вейерштрасса предела по О.24, О.25 
и О.26.

3): Из пункта 3 Т.12: 

\[ 
\lim_{n\rightarrow \infty }\varpi _n({\bf N}-{\bf B})=0\mbox{.} 
\] 

Из пункта 4 Т.12: 

\[ 
\varpi _n({\bf {A}}\cap ({\bf N}-{\bf B}))\leq \varpi _n({\bf N}-{\bf B})%
\mbox{.} 
\] 

Следовательно, 

\[ 
\lim_{n\rightarrow \infty }\varpi _n\left( {\bf {A}}\cap ({\bf N}-{\bf B}%
)\right) =0\mbox{.} 
\] 

Следовательно, 

\[ 
\lim_{n\rightarrow \infty }\varpi _n\left( {\bf {A}}\cap {\bf B}\right) 
=\lim_{n\rightarrow \infty }\varpi _n({\bf {A}})  _{\bf \Box }
\] 

{\bf О.27: }Функцию $\mathcal{C}(n)$, определенную на ${\bf N}$ и имеющую значения в 
множестве событий, называем {\it суперсобытием} $\mathcal{C}(n)$. 

{\bf О.28:} Суперсобытие $\mathcal{C}(n)$ {\it происходит}, если  
$\left\{n \in {\bf N} |\mathcal{C}\left( n\right)\mbox{ происходит} \right\} \in %
\mathbf{\Phi ix}$.

{\bf О.29:} $B$-функция $\rm P$ называется {\it $P$-функцией}, если 
$\rm P$ подчиняется следующему условию: 

Для любого суперсобытия $\mathcal{C}(n)$: если 

\[
\lim_{n\rightarrow \infty }\rm P\left( \mathcal{C}\left( n\right) \right)
=1 \mbox{,}
\]

то $\mathcal{C}(n)$ происходит. 

{\bf T.14:} Если ${\rm P}$ - P-функция, то происходит событие 

\[
^{\circ}\ll {\rm P}\left(\mathcal{A}\right)-\varepsilon\leq \nu _n\left[ s_%
\mathcal{A}\right] \leq {\rm P}\left(\mathcal{A}\right)+\varepsilon\gg \
\]

для любого как угодно малого положительного $\varepsilon$ и для любого 
физического события $\mathcal{A}$.

{\bf Доказательство T.14:} Из О.21, O.29 и \ref{epsl}:

\[
\lim\limits_{n\rightarrow \infty }{\rm P}\left( \mathfrak{T}[s_\mathcal{A}](n,n%
\cdot \left({\rm P}\left(\mathcal{A}\right)-\varepsilon \right) ,
n\cdot \left({\rm P}\left(\mathcal{A}\right)+\varepsilon \right) )\right) =1
\]

Следовательно, по О.29 и по Т.9 происходит суперсобытие 

\[
^{\circ}\ll\frac {n\cdot \left({\rm P}\left(\mathcal{A}\right)-\varepsilon%
\right)}n\leq \nu _n\left[ s_\mathcal{A}\right] \leq \frac {n\cdot \left%
({\rm P}\left(\mathcal{A}\right)+\varepsilon\right)}n\gg \
\]

для любого как угодно малого положительного $\varepsilon$ $_\Box$

Т.е. почти для всех больших $n$:

\[
\nu _n\left[ s_\mathcal{A}\right] ={\rm P}\left(\mathcal{A}\right)\pm %
\varepsilon \mbox{.}
\]

Таким образом, P-функциии, подчиняющиеся всем свойствам функции 
вероят-ности, имеют статистический смысл. Следовательно, их значения могут быть 
измерены однозначно статистическим экспериментом. Т.е. только одна такая 
функция существует. Она определяет объективную вероятность физических событий, 
не зависящую от наблюдателя.

\subsection{Условная вероятность}

{\bf Определение 2.5.1:} {\it Условной вероятностью} $\mathcal{B}$ по $\mathcal{C}$ называется: 

\begin{equation}
{\rm P}\left( \mathcal{B}/\mathcal{C}\right) \stackrel{def}{=}\frac{{\rm P}\left( \mathcal{C}\cdot\mathcal{B}\right) 
}{{\rm P}\left( \mathcal{C}\right) }\mbox{.}\label{CP}
\end{equation}

{\bf Теорема 2.5.1} Функция условной вероятности есть B-функция.

{\bf Доказательство Теоремы 2.5.1} Из Определения 2.5.1:

\begin{center}
${\rm P}\left( \mathcal{C}/\mathcal{C}\right) =\frac{{\rm P}\left( \mathcal{C}\cdot\mathcal{C}\right) }{{\rm P}%
\left( \mathcal{C}\right) }$.
\end{center}

Отсюда по пункту 1 Теоремы 2.1.1:

\begin{center}
${\rm P}\left( \mathcal{C}/\mathcal{C}\right) =\frac{{\rm P}\left( \mathcal{C}\right) }{{\rm P}\left(
\mathcal{C}\right) }=1$.
\end{center}

Также из Определения 2.5.1:

\begin{center}
${\rm P}\left( \left( \mathcal{A}\cdot\mathcal{B}\right) /\mathcal{C}\right) +{\rm P}\left( \left(
\mathcal{A}\cdot \overline{ \mathcal{B}} \right) /\mathcal{C}\right) =\frac{{\rm P}\left( \mathcal{C}\cdot\left(
\mathcal{A}\cdot\mathcal{B}\right) \right) }{{\rm P}\left( \mathcal{C}\right) }+\frac{{\rm P}\left(
\mathcal{C}\cdot\left( \mathcal{A}\cdot \overline {\mathcal{B}} \right) \right) }{{\rm P}\left( \mathcal{C}\right) }
$.
\end{center}

Отсюда:

\begin{center}
${\rm P}\left( \left( \mathcal{A}\cdot\mathcal{B}\right) /\mathcal{C}\right) +{\rm P}\left( \left(
\mathcal{A}\cdot \overline{\mathcal{B}} \right) /\mathcal{C}\right) =\frac{{\rm P}\left( \mathcal{C}\cdot\left(
\mathcal{A}\cdot\mathcal{B}\right) \right) +{\rm P}\left( \mathcal{C}\cdot\left( \mathcal{A}\cdot \overline{\mathcal{B}}
\right) \right) }{{\rm P}\left( \mathcal{C}\right) }$.
\end{center}

По пункту 3 Теоремы 2.1.1:

\begin{center}
${\rm P}\left( \left( \mathcal{A}\cdot\mathcal{B}\right) /\mathcal{C}\right) +{\rm P}\left( \left(
\mathcal{A}\cdot \overline{\mathcal{B}} \right) /\mathcal{C}\right) =\frac{{\rm P}\left( \left(
\mathcal{C}\cdot\mathcal{A}\right) \cdot\mathcal{B}\right) +{\rm P}\left( \left( \mathcal{C}\cdot\mathcal{A}\right) \cdot \overline%
{\mathcal{B}} \right) }{{\rm P}\left( \mathcal{C}\right) }$.
\end{center}

Отсюда по О.16:

\begin{center}
${\rm P}\left( \left( \mathcal{A}\cdot\mathcal{B}\right) /\mathcal{C}\right) +{\rm P}\left( \left(
\mathcal{A}\cdot \overline{\mathcal{B}} \right) /\mathcal{C}\right) =\frac{{\rm P}\left( \mathcal{C}\cdot\mathcal{A}\right) 
}{{\rm P}\left( \mathcal{C}\right) }$.
\end{center}

Отсюда по Определению 2.5.1:

\begin{center}
${\rm P}\left( \left( \mathcal{A}\cdot\mathcal{B}\right) /\mathcal{C}\right) +{\rm P}\left( \left(
\mathcal{A}\cdot \overline{\mathcal{B}} \right) /\mathcal{C}\right) ={\rm P}\left( \mathcal{A}/\mathcal{C}\right) $ 
$_{\bf \Box }$
\end{center}

\subsection{Классическая вероятность}

{\bf Определение 2.6.1} $\left\{ \mathcal{B}_1,\mathcal{B}_2,\ldots ,\mathcal{B}_n\right\} $ есть {\it 
полная система}, если выполняются следующие условия:

1. если $k\neq s$, то $\left( \mathcal{B}_k\cdot  \mathcal{B}_s\right)$ 
несовместны;

2. происходит событие $\left( \mathcal{B}_1 + \mathcal{B}_2 + \ldots  + \mathcal{B}_n\right)$.

{\bf Определение 2.6.2} $\mathcal{B}$ {\it благоприятствует событию} $\mathcal{A}$, если 
$\left(\mathcal{B}\cdot\overline{\mathcal{A}}\right)$ несовмес-тны, и $\mathcal{B}$ не благоприятствует 
предложению $\mathcal{A}$, если $\left(\mathcal{B}\cdot \mathcal{A}\right)$ несовместны.

Пусть 

1. $\left\{ \mathcal{B}_1,\mathcal{B}_2,\ldots ,\mathcal{B}_n\right\}$ - полная система;

2. для $k\in \left\{ 1,2,\ldots ,n\right\} $ и $s\in \left\{ 1,2,\ldots ,n%
\right\} $: $\mathrm{P}\left( \mathcal{B}_k\right) =\mathrm{P}\left( \mathcal{B}_s\right) $;

3. если $1\leq k\leq m$, то $\mathcal{B}_k$ благоприятствует предложению $\mathcal{A}$, а если 
$m+1\leq s\leq n$, то $\mathcal{B}_s$ не благоприятствует предложению $\mathcal{A}$.

В этом случае из пункта 6 T.2 по О.11:

\[
\mathrm{P}\left( \overline{\mathcal{A}}\cdot \mathcal{B}_k\right) = 0
\]

для $k\in \left\{ 1,2,\ldots ,m\right\} $, и 

\[
\mathrm{P}\left( \mathcal{A}\cdot \mathcal{B}_s\right) =0 
\]

для $s\in \left\{ m+1,m+2,\ldots ,n\right\} $.

Следовательно, по О.16:

\[
\mathrm{P}\left( \mathcal{A}\cdot \mathcal{B}_k\right) =\mathrm{P}\left( \mathcal{B}_k\right)
\]

для $k\in \left\{ 1,2,\ldots ,n\right\} $.

По пункту 4 Теоремы 2.1.1:

\[
\mathcal{A}=\left( \mathcal{A}\cdot \left( \mathcal{B}_1 + \mathcal{B}_2 + \ldots  + \mathcal{B}_m + \mathcal{B}_{m+1}\ldots
 + \mathcal{B}_n\right) \right) \mbox{.}
\]

Следовательно, по Т.7:

$\mathrm{P}\left( \mathcal{A}\right) =\mathrm{P}\left( \mathcal{A}\cdot \mathcal{B}_1\right) +\mathrm{P}%
\left( \mathcal{A}\cdot \mathcal{B}_2\right) +\ldots +$

$+\mathrm{P}\left( \mathcal{A}\cdot \mathcal{B}_m\right) +\mathrm{P}\left( \mathcal{A}\cdot
\mathcal{B}_{m+1}\right) +\ldots +\mathrm{P}\left( \mathcal{A}\cdot \mathcal{B}_n\right) =$

$=\mathrm{P}\left( \mathcal{B}_1\right) +\mathrm{P}\left( \mathcal{B}_2\right) +\ldots +\mathrm{P%
}\left( \mathcal{B}_m\right) $.

Следовательно,

\[
\mathrm{P}\left( \mathcal{A}\right) =\frac mn\mbox{.}
\]

\subsection{B-функции и классическая пропозициональная логика}
 
{\bf Теорема 2.2.2} Если предложение $D$ пропозиционально доказуемо, 
то для всех B-функций $\mathfrak{b}$: $\mathfrak{b}\left(^{\circ} D\right) =1 $.

{\bf Доказательство Теоремы 2.2.2: }

Если $D$ есть {\bf A1}, то по Определению 2.1.10, O.10, O.12, O.13:

\[
\mathfrak{b}\left(^{\circ} D\right) =\mathfrak{b} \left(\overline{ ^{\circ}%
 A\cdot \overline{\overline{^{\circ}B\cdot \overline{^{\circ}A}}}}\right) %
\mbox{.} 
\]

По пункту 5 Т.2:

\[
\mathfrak{b}\left(^{\circ} D\right) =1-\mathfrak{b}\left( A\cdot\overline{ %
\overline{^{\circ} B\cdot\overline{^{\circ} A}}}\right) \mbox{.} 
\]

По О.16 и по Теореме 2.1.1:

\[
\begin{array}{c}
\mathfrak{b}\left(^{\circ} D\right) =1-\mathfrak{b}\left(^{\circ} A\right) +%
\mathfrak{b}\left(^{\circ} A\cdot\overline{^{\circ} B\cdot\overline{^{\circ} A}}%
\right) \mbox{,} \\ 
\mathfrak{b}\left(^{\circ} D\right) =1-\mathfrak{b}\left(^{\circ} A\right) +%
\mathfrak{b}\left(^{\circ} A\right) -\mathfrak{b}\left(^{\circ} A\cdot \left(%
^{\circ}B\cdot\overline{^{\circ} A}\right) \right) \mbox{,} \\ 
\mathfrak{b}\left(^{\circ} D\right) =1-\mathfrak{b}\left( \left(^{\circ} A\cdot
^{\circ}B\right) \cdot\overline{^{\circ} A}\right) \mbox{,} \\ 
\mathfrak{b}\left(^{\circ} D\right) =1-\mathfrak{b}\left(^{\circ} A\cdot %
^{\circ}B\right) +\mathfrak{b}\left( \left(^{\circ} A\cdot ^{\circ}B\right) %
\cdot ^{\circ}A\right) \mbox{,} \\ 
\mathfrak{b}\left(^{\circ} D\right) =1-\mathfrak{b}\left(^{\circ} A\cdot ^{\circ}%
B\right) +\mathfrak{b}\left( \left(^{\circ} A\cdot ^{\circ}A\right) \cdot %
^{\circ}B\right) \mbox{,} \\ 
\mathfrak{b}\left(^{\circ} D\right) =1-\mathfrak{b}\left(^{\circ} A\cdot %
^{\circ}B\right) +\mathfrak{b}\left(^{\circ} A\cdot ^{\circ}B\right) \mbox{.}
\end{array}
\]

Для остальных пропозициональных аксиом доказательство аналогично.

Пусть для всех B-функций $\mathfrak{b}$: $\mathfrak{b}(^{\circ}A)=1$ и $%
\mathfrak{b}\left(^{\circ}\left(A\Rightarrow D\right)\right)=1$.

По Определению 2.1.10, O.10, O.12, O.13:

\[
\mathfrak{b}\left(^{\circ}\left( A\Rightarrow D\right)\right) =\mathfrak{b}%
\left(\overline{\left(^{\circ}A\cdot\overline{^{\circ} D}\right)}\right) \mbox{.} 
\]

По пункту 5 Т.2:

\[
\mathfrak{b}\left(^{\circ}\left( A\Rightarrow D\right)\right) =1-\mathfrak{b}%
\left(^{\circ} A\cdot \overline{ ^{\circ}D}\right) \mbox{.} 
\]

Поэтому,

\[
\mathfrak{b}\left(^{\circ} A\cdot\overline{^{\circ} D}\right) =0\mbox{.} 
\]

По О.16:

\[
\mathfrak{b}\left(^{\circ} A\cdot\overline{^{\circ} D}\right) =\mathfrak{b}%
\left(^{\circ}A\right) -\mathfrak{b}\left(^{\circ} A\cdot D\right) \mbox{.} 
\]

Поэтому,

\[
\mathfrak{b}\left(^{\circ} A\cdot ^{\circ}D\right) =\mathfrak{b}\left(^{\circ} %
A\right) =1\mbox{.} 
\]

По О.16 и по Теореме 2.1.1:

\[
\mathfrak{b}\left(^{\circ} A\cdot ^{\circ}D\right) =\mathfrak{b}\left(^{\circ} %
D\right) -{\mathfrak{b}}\left(^{\circ} D\cdot\overline{^{\circ}A}\right) =1\mbox{.} 
\]

Следовательно, для всех B-функций $\mathfrak{b}$:

\[
\mathfrak{b}\left(^{\circ} D\right) =1  _{\bf \Box } 
\]

{\bf Теорема 2.2.3}

1) Если для всех булевых функций $\mathfrak{g}$:

\[
\mathfrak{g}\left( A\right) =1, 
\]

то для всех B-функций $\mathfrak{b}$:

\[
\mathfrak{b}\left(^{\circ} A\right) =1\mbox{.} 
\]

2) Если для всех булевых функций $\mathfrak{g}$:

\[
\mathfrak{g}\left( A\right) =0, 
\]

то для всех B-функций $\mathfrak{b}$:

\[
\mathfrak{b}\left(^{\circ} A\right) =0\mbox{.} 
\]

{\bf Доказательство Теоремы 2.2.3: }

1) Это сразу следует из предыдущей Теоремы и из Теоремы 2.1.3.

2) Если для всех Булевых функций $\mathfrak{g}$: $\mathfrak{g}\left(
A\right) =0$, то по Определению 2.1.6: $\mathfrak{g}\left(\neg A%
\right) =1$. Поэтому, из пункта 1 этой Теоремы: для всех B-функций ${%
\mathfrak{b}}$: $\mathfrak{b}\left(\overline{^{\circ} A}\right) =1$. По 
пункту 5 Т.2: $\mathfrak{b}\left(^{\circ} A\right) =0$ $_{\bf \Box }$

{\bf Теорема 2.2.4} Если функция $\widetilde{\mathfrak{b}}$ определена на множестве 
физических событий, имеет значения в двух-элементном множестве $\left\{ 0;1\right\}$ и подчиняется следующим условиям:

1. $\widetilde{\mathfrak{b}}\left(^{\circ}\left(\neg A\right)\right) = 1-%
\widetilde{\mathfrak{b}}\left(^{\circ} A\right)$;

2. $\widetilde{\mathfrak{b}}\left(^{\circ}\left(A\& B\right)\right) = \widetilde%
{\mathfrak{b}}\left(^{\circ} A\right)\cdot\widetilde{\mathfrak{b}}\left(^{\circ} B\right)$,

то эта функция представляет B-функцию. 

{\bf Доказательство Теоремы 2.2.4: } Если $D$ есть {\bf A1}, то 
$\widetilde{\mathfrak{b}}\left(^{\circ} D\right)=1$ - смотрите Доказательство 
Теоремы 2.2.2.

По О.10 и О.12:

$\widetilde{\mathfrak{b}}\left( ^{\circ }A\cdot ^{\circ }B\right) +\mathfrak{%
\widetilde{\mathfrak{b}}}\left( ^{\circ }A\cdot \overline{^{\circ }B}\right) =%
\widetilde{\mathfrak{b}}\left( ^{\circ }\left( A\&B\right) \right) +\mathfrak{%
\widetilde{\mathfrak{b}}}\left( ^{\circ }\left( A\&\left( \neg B\right) \right)
\right) $.

По условию Теоремы 2.2.4:

$\widetilde{\mathfrak{b}}\left( ^{\circ }A\cdot ^{\circ }B\right) +\mathfrak{%
\widetilde{\mathfrak{b}}}\left( ^{\circ }A\cdot \overline{^{\circ }B}\right) =%
\widetilde{\mathfrak{b}}\left( ^{\circ }A\right) \cdot \mathfrak{\widetilde{%
\mathfrak{b}}}\left( ^{\circ }B\right) +\widetilde{\mathfrak{b}}\left(
^{\circ }A\right) \cdot \left( 1-\widetilde{\mathfrak{b}}\left( ^{\circ
}B\right) \right) $.

Следовательно,

$\widetilde{\mathfrak{b}}\left( ^{\circ }A\cdot ^{\circ }B\right) +%
\widetilde{\mathfrak{b}}\left( ^{\circ }A\cdot \overline{^{\circ }B}\right) =%
\widetilde{\mathfrak{b}}\left( ^{\circ }A\right) {\bf _\Box} $ 

Из Определения 2.1.6 функция $\widetilde{\mathfrak{b}}\left(^{\circ}X\right)$ 
представляет булеву функцию. Следова-тельно, B-функции и булевы функции тесно 
связаны Теоремой 2.2.4. Т.е. B-функции - это функции пропозициональной логики.
Получается, что вероятность это логика еще не произошедших событий.

\subsection{Непротиворечивость логической вероятности}
\subsubsection{Нестандартные числа} 

Рассматриваем множество ${\bf N}$ натуральных чисел.

{\bf Определение 2.7.1.1:} $n${\it -вход} ${\bf S_n}$ для ${\bf N}$ 
определяется рекурсивно: 

1) ${\bf S}_1\stackrel{def}{=}\left\{ 1\right\} $; 

2) ${\bf S}_{\left( n+1\right) }\stackrel{def}{=}{\bf S}_n\cup \left\{ n+1\right\} $. 

{\bf Определение 2.7.1.2: }Если ${\bf S}_n$ есть $n$-вход для ${\bf N}$, и  $%
{\bf A}\subseteq {\bf N}$, то $\left\| {\bf A}\cap {\bf S}_n\right\| $ есть 
количество элементов множества ${\bf A}\cap {\bf S}_n$, и если

\[ 
\varpi _n\left( {\bf A}\right) =\frac{\left\| {\bf A}\cap {\bf S}_n\right\| }%
n\mbox{,} 
\] 

то $\varpi _n\left( {\bf A}\right) $ есть {\it частота} множества $%
{\bf A}$ на $n$-входе ${\bf S}_n$. 

{\bf Теорема 2.7.1.1:} 

1) $\varpi _n({\bf N})=1$; 

2) $\varpi _n(\emptyset )=0$; 

3) $\varpi _n({\bf A})+\varpi _n({\bf N}-{\bf A})=1$; 

4) $\varpi _n({\bf A}\cap {\bf B})+\varpi _n({\bf A}\cap ({\bf N}-{\bf B}%
))=\varpi _n({\bf A})$. 

{\bf Доказательство Теоремы 2.7.1.1:} Из Определений 2.7.1.1 и 2.7.1.2  $_{\bf \Box }$ 

{\bf Определение 2.7.1.3: }Если "$\lim $" есть Коши-Вейерштрасса "limit", то:

\[ 
{\bf \Phi ix}\stackrel{def}{=}\left\{ {\bf A}\subseteq {\bf N}|\lim_{n\rightarrow \infty 
}\varpi _n({\bf A})=1\right\} \mbox{.} 
\] 

{\bf Теорема 2.7.1.2: }${\bf \Phi ix}$ есть фильтр \cite{DVS}, т.е.: 

1) ${\bf N}\in {\bf \Phi ix}$, 

2) $\emptyset \notin {\bf \Phi ix}$, 

3) если ${\bf A}\in {\bf \Phi ix}$, и ${\bf B}\in {\bf \Phi ix}$, то $({\bf A%
}\cap {\bf B})\in {\bf \Phi ix}$ ; 

4) если ${\bf A}\in {\bf \Phi ix}$, и ${\bf A}\subseteq {\bf B}$, то ${\bf B}%
\in {\bf \Phi ix}$. 

{\bf Доказательство Теоремы 2.7.1.2:} Из пункта 3 Теоремы 2.7.1.1: 

\[ 
\lim_{n\rightarrow \infty }\varpi _n({\bf N}-{\bf B})=0\mbox{.} 
\] 

Из пункта 4 Теоремы 2.7.1.1: 

\[ 
\varpi _n({\bf A}\cap ({\bf N}-{\bf B}))\leq \varpi _n({\bf N}-{\bf B})%
\mbox{.} 
\] 

Следовательно, 

\[ 
\lim_{n\rightarrow \infty }\varpi _n\left( {\bf A}\cap ({\bf N}-{\bf B}%
)\right) =0\mbox{.} 
\] 

Следовательно, 

\[ 
\lim_{n\rightarrow \infty }\varpi _n\left( {\bf A}\cap {\bf B}\right) 
=\lim_{n\rightarrow \infty }\varpi _n({\bf A})  _{\bf \Box }
\] 

Далее для нашего предмета приспособим понятия и выводы Robinson нестан- дартного
анализа \cite{DVS}:

{\bf Определение 2.7.1.4:} Последовательности вещественных чисел $\left\langle 
r_n\right\rangle $ и $\left\langle s_n\right\rangle $ {\it Q-эквивалентны} 
(обозначение: $\left\langle r_n\right\rangle \sim \left\langle s_n\right\rangle 
$), если 

\[ 
\left\{ n\in {\bf N}|r_n=s_n\right\} \in {\bf \Phi ix}\mbox{.} 
\] 

{\bf Теорема 2.7.1.3:} Если ${\bf r}$,${\bf s}$,${\bf u}$ - последовательности 
вещественных чисел, то:

1) ${\bf r}\sim {\bf r}$, 

2) если ${\bf r}\sim {\bf s}$, то ${\bf s}\sim {\bf r}$; 

3) если ${\bf r}\sim {\bf s}$, и ${\bf s}\sim {\bf u}$, то ${\bf r}\sim {\bf %
u}$. 

{\bf Доказательство Теоремы 2.7.1.3:} По Определению 2.7.1.4 из Теоремы 2.7.1.2  
$_{\bf \Box }$ 

{\bf Определение 2.7.1.5:} {\it Q-число} есть множество Q-эквивалентных 
последова-тельностей вещественных чисел, т.е. если $\widetilde{a}$ - Q-число, 
и ${\bf r}\in \widetilde{a}$ и ${\bf s}\in \widetilde{a}$, то ${\bf r}\sim 
{\bf s};$ и если ${\bf r}\in \widetilde{a}$, и ${\bf r}\sim {\bf s}$, то $%
{\bf s}\in \widetilde{a}$. 

{\bf Определение 2.7.1.6:} Q-число $\widetilde{a}$ есть {\it стандартное 
Q-число} $a$, если $a$ есть некоторое вещественное число, и существует 
последовательность $\left\langle r_n\right\rangle $, для которой: 
$\left\langle r_n\right\rangle \in \widetilde{a}$, и 

\[ 
\left\{ n\in {\bf N}|r_n=a\right\} \in {\bf \Phi ix}\mbox{.} 
\] 

{\bf Определение 2.7.1.7:} Q-числа $\widetilde{a}$ и $\widetilde{b}$ {\it равны} 
(обозначение: $\widetilde{a}=\widetilde{b}$), если $%
\widetilde{a}\subseteq \widetilde{b}$, и  $\widetilde{b}\subseteq \widetilde{%
a}$. 

{\bf Теорема 2.7.1.4: }Пусть ${\mathfrak f}(x,y,z)$ функция, определенная в
${\bf R}\times {\bf R}\times {\bf R}$, имеющая область значений в 
${\bf R}$ (${\bf R}$ - множество вещественных чисел). 

Пусть $\left\langle y_{1,n}\right\rangle $ , $\left\langle 
y_{2,n}\right\rangle $ , $\left\langle y_{3,n}\right\rangle $ , $%
\left\langle z_{1,n}\right\rangle $ , $\left\langle z_{2,n}\right\rangle $ , 
$\left\langle z_{3,n}\right\rangle $ - последовательности вещественных чисел. 

В этом случае если $\left\langle z_{i,n}\right\rangle \sim \left\langle 
y_{i,n}\right\rangle $, то $\left\langle {\mathfrak f}(y_{1,n},y_{2,n},y_{3,n})%
\right\rangle \sim \left\langle {\mathfrak f}(z_{1,n},z_{2,n},z_{3,n})\right%
\rangle $. 

{\bf Доказательство Теоремы 2.7.1.4:} Обозначим: 

если $k=1$, или $k=2$, или $k=3$, то 

\[ 
{\bf A}_k=\left\{ n\in {\bf N}|y_{k,n}=z_{k,n}\right\} \mbox{.} 
\] 

В этом случае по Определению 2.7.1.4 для всех $k$: 

\[ 
{\bf A}_k\in {\bf \Phi ix}\mbox{.} 
\] 

Т.к. 

\[ 
\left( {\bf A}_1\cap {\bf A}_2\cap {\bf A}_3\right) \subseteq \left\{ n\in 
{\bf N}|{\mathfrak f}(y_{1,n},y_{2,n},y_{3,n})={\mathfrak f}%
(z_{1,n},z_{2,n},z_{3,n})\right\} \mbox{,} 
\] 

то по Теореме 2.7.1.2: 

\[ 
\left\{ n\in {\bf N}|{\mathfrak f}(y_{1,n},y_{2,n},y_{3,n})={\mathfrak f}%
(z_{1,n},z_{2,n},z_{3,n})\right\} \in {\bf \Phi ix}  _{\bf \Box } 
\] 

{\bf Определение 2.7.1.8:} Обозначим: $Q{\bf R}$ - множество Q-чисел. 
~ 

{\bf Определение 2.7.1.9: }Функция $\widetilde{{\mathfrak f}}$, определенная в 
$Q{\bf R}\times Q{\bf R}\times Q{\bf R}$, имеющая область значений в $Q{\bf R}$, 
называется {\it Q-расширением функции} ${\mathfrak f}$, определенной в 
${\bf R}\times {\bf R}\times {\bf R}$, имеющей область значений в ${\bf R}$, 
если выполняются следующие условия: 

Пусть $\left\langle x_n\right\rangle $ ,$\left\langle y_n\right\rangle $ ,$%
\left\langle z_n\right\rangle $ - последовательности вещественных чисел. 
В этом случае: если

$\left\langle x_n\right\rangle \in \widetilde{x}$, $\left\langle 
y_n\right\rangle \in \widetilde{y}$, $\left\langle z_n\right\rangle \in 
\widetilde{z}$, $\widetilde{u}=\widetilde{{\mathfrak f}}\left( \widetilde{x},%
\widetilde{y},\widetilde{z}\right) $, 

то 

$\left\langle {\mathfrak f}\left( x_n,y_n,z_n\right) \right\rangle \in 
\widetilde{u}$. 

{\bf Теорема 2.7.1.5:} Для всех функций ${\mathfrak f}$, определенных в $%
{\bf R}\times {\bf R}\times {\bf R}$, имеющих область значений в ${\bf R}$, 
и для всех вещественных чисел $a$, $b$, $c$, $d$: если $\widetilde{{\mathfrak f}}$ 
- Q-расширение ${\mathfrak f}$; $\widetilde{a}$, $\widetilde{b}$, $%
\widetilde{c}$, $\widetilde{d}$ - стандартные Q-числа $a$, $b$, $c$, $d$, 
то: 

если $d={\mathfrak f}(a,b,c)$, то $\widetilde{d}=\widetilde{{\mathfrak f}}(\widetilde{%
a},\widetilde{b},\widetilde{c})$ и vice versa. ~ 

{\bf Доказательство Теоремы 2.7.1.5:} Если $\left\langle r_n\right\rangle \in 
\widetilde{a}$, $\left\langle s_n\right\rangle \in \widetilde{b}$, $%
\left\langle u_n\right\rangle \in \widetilde{c}$, $\left\langle {\mathfrak t}%
_n\right\rangle \in \widetilde{d}$, то по Определению 2.7.1.6: 

\[ 
\begin{array}{c} 
\left\{ n\in {\bf N}|r_n=a\right\} \in {\bf \Phi ix}\mbox{,} \\ 
\left\{ n\in {\bf N}|s_n=b\right\} \in {\bf \Phi ix}\mbox{,} \\ 
\left\{ n\in {\bf N}|u_n=c\right\} \in {\bf \Phi ix}\mbox{,} \\ 
\left\{ n\in {\bf N}|t_n=d\right\} \in {\bf \Phi ix}\mbox{.} 
\end{array} 
\] 

1) Пусть $d={\mathfrak f}(a,b,c)$. 

В этом случае по Теореме 2.7.1.2: 

\[ 
\left\{ n\in {\bf N}|t_n={\mathfrak f}(r_n,s_n,u_n)\right\} \in {\bf \Phi ix}%
\mbox{.} 
\] 

Следовательно, по Определению 2.7.1.4: 

\[ 
\left\langle t_n\right\rangle \sim \left\langle {\mathfrak f}(r_n,s_n,u_n)\right%
\rangle \mbox{.} 
\] 

Таким образом, по Определению 2.7.1.5: 

\[ 
\left\langle {\mathfrak f}(r_n,s_n,u_n)\right\rangle \in \widetilde{d}\mbox{.} 
\] 

Следовательно, по Определению 2.7.1.9: 

\[ 
\widetilde{d}=\widetilde{{\mathfrak f}}(\widetilde{a},\widetilde{b},\widetilde{c}%
)\mbox{.} 
\] 

2) Пусть $\widetilde{d}=\widetilde{{\mathfrak f}}(\widetilde{a},\widetilde{b},%
\widetilde{c})$. 

В этом случае по Определению 2.7.1.9: 

\[ 
\left\langle {\mathfrak f}(r_n,s_n,u_n)\right\rangle \in \widetilde{d}\mbox{.} 
\] 

Следовательно, по Определению 2.7.1.5: 

\[ 
\left\langle t_n\right\rangle \sim \left\langle {\mathfrak f}(r_n,s_n,u_n)\right%
\rangle \mbox{.} 
\] 

Таким образом, по Определению 2.7.1.4: 

\[ 
\left\{ n\in {\bf N}|t_n={\mathfrak f}(r_n,s_n,u_n)\right\} \in {\bf \Phi ix}%
\mbox{.} 
\] 

Следовательно, по Теореме 2.7.1.2: 

\[ 
\left\{ n\in {\bf N}|t_n={\mathfrak f}(r_n,s_n,u_n),r_n=a,s_n=b,u_n=c,t_n=d%
\right\} \in {\bf \Phi ix}\mbox{.} 
\] 

Следовательно, т.к. это множество не пусто, то 

\[ 
d={\mathfrak f}(a,b,c)  _{\bf \Box } 
\] 

По этой теореме: если $\widetilde{{\mathfrak f}}$ - Q-расширение функции 
${\mathfrak f}$, то выражение "$\widetilde{{\mathfrak f}}(\widetilde{%
x},\widetilde{y},\widetilde{z})$" будет записано как "${\mathfrak f}(%
\widetilde{x},\widetilde{y},\widetilde{z})$", и если $\widetilde{u}$ -  
стандартное Q-число, то выражение "$\widetilde{u}$" будет записано как "$u$". 

{\bf Теорема 2.7.1.6:} Если для всех вещественных чисел $a$, $b$, $c$: 

\[ 
\varphi (a,b,c)=\psi (a,b,c) ,
\] 

то для всех Q-чисел $\widetilde{x}$, $\widetilde{y}$, $\widetilde{z}$: 

\[ 
\varphi (\widetilde{x},\widetilde{y},\widetilde{z})=\psi (\widetilde{x},%
\widetilde{y},\widetilde{z})\mbox{.} 
\] 

{\bf Доказательство Теоремы 2.7.1.6:} Если $\left\langle x_n\right\rangle \in 
\widetilde{x}$, $\left\langle y_n\right\rangle \in \widetilde{y}$, $%
\left\langle z_n\right\rangle \in \widetilde{z}$, $\widetilde{u}=\varphi (%
\widetilde{x},\widetilde{y},\widetilde{z})$, то по Определению 2.7.1.9: $%
\left\langle \varphi (x_n,y_n,z_n)\right\rangle \in \widetilde{u}$. 

Т.к. $\varphi (x_n,y_n,z_n)=\psi (x_n,y_n,z_n)$, то $\left\langle \psi 
(x_n,y_n,z_n)\right\rangle \in \widetilde{u}$. 

Если $\widetilde{v}=\psi (\widetilde{x},\widetilde{y},\widetilde{z})$, то по 
Определению 2.7.1.9: $\left\langle \psi (x_n,y_n,z_n)\right\rangle \in \widetilde{%
v}$, тоже. 

Таким образом, для всех последовательностей $\left\langle t_n\right\rangle $ 
вещественных чисел: если $\left\langle t_n\right\rangle \in \widetilde{u}$, то 
по Определению 2.7.1.5: $\left\langle t_n\right\rangle \sim \left\langle \psi 
(x_n,y_n,z_n)\right\rangle $. 

Следовательно, $\left\langle t_n\right\rangle \in \widetilde{v}$; и если $%
\left\langle t_n\right\rangle \in \widetilde{v}$, то $\left\langle 
t_n\right\rangle \sim \left\langle \varphi (x_n,y_n,z_n)\right\rangle $; 
Следовательно, $\left\langle t_n\right\rangle \in \widetilde{u}$. 

Таким образом, $\widetilde{u}=\widetilde{v}$  $_{\bf \Box }$

{\bf Теорема 2.7.1.7:} Если для всех вещественных чисел $a$, $b$, $c$: 

\[ 
{\mathfrak f}\left( a,\varphi (b,c)\right) =\psi (a,b,c), 
\] 

то для всех Q-чисел $\widetilde{x}$, $\widetilde{y}$, $\widetilde{z}$: 

\[ 
{\mathfrak f}\left( \widetilde{x},\varphi (\widetilde{y},\widetilde{z})\right) 
=\psi (\widetilde{x},\widetilde{y},\widetilde{z})\mbox{.} 
\] 

{\bf Доказательство Теоремы 2.7.1.7:} Пусть $\left\langle w_n\right\rangle \in 
\widetilde{w}$, ${\mathfrak f}(\widetilde{x},\widetilde{w})=\widetilde{u}$, $%
\left\langle x_n\right\rangle \in \widetilde{x}$, $\left\langle 
y_n\right\rangle \in \widetilde{y}$, $\left\langle z_n\right\rangle \in 
\widetilde{z}$, $\varphi (\widetilde{y},\widetilde{z})=\widetilde{w}$, $\psi 
(\widetilde{x},\widetilde{y},\widetilde{z})=\widetilde{v}$. 

По условию этой Теоремы: ${\mathfrak f}(x_n,\varphi (y_n,z_n))=\psi 
(x_n,y_n,z_n)$. 

по Определению 2.7.1.9: $\left\langle \psi (x_n,y_n,z_n)\right\rangle \in 
\widetilde{v}$, $\left\langle \varphi (x_n,y_n)\right\rangle \in \widetilde{w%
}$, $\left\langle {\mathfrak f}(x_n,w_n)\right\rangle \in \widetilde{u}$. 

Для всех последовательностей $\left\langle t_n\right\rangle $ вещественных 
чиселof: 

1) Если $\left\langle t_n\right\rangle \in \widetilde{v}$, то Определению 
2.7.1.5: $\left\langle t_n\right\rangle \sim \left\langle \psi 
(x_n,y_n,z_n)\right\rangle $. 

Следовательно, $\left\langle t_n\right\rangle \sim \left\langle {\mathfrak f}(x_n,\varphi 
(y_n,z_n))\right\rangle $. 

Таким образом, по Определению 2.7.1.4: 

\[ 
\left\{ n\in {\bf N}|t_n={\mathfrak f}(x_n,\varphi \left( y_n,z_n\right) 
)\right\} \in {\bf \Phi ix} ,
\] 

и 

\[ 
\left\{ n\in {\bf N}|w_n=\varphi \left( y_n,z_n\right) \right\} \in {\bf %
\Phi ix}\mbox{.} 
\]
 
Следовательно, по Теореме 2.7.1.2: 

\[ 
\left\{ n\in {\bf N}|t_n={\mathfrak f}(x_n,w_n)\right\} \in {\bf \Phi ix}\mbox{.} 
\] 

Следовательно, по Определению 2.7.1.4: 

\[ 
\left\langle t_n\right\rangle \sim \left\langle {\mathfrak f}(x_n,w_n)\right%
\rangle \mbox{.} 
\] 

Таким образом, по Определению 2.7.1.5: $\left\langle t_n\right\rangle \in \widetilde{u%
}$. 

2) Если $\left\langle t_n\right\rangle \in \widetilde{u}$, то Определению 
2.7.1.5: $\left\langle t_n\right\rangle \sim \left\langle {\mathfrak f}%
(x_n,w_n)\right\rangle $. 

Т.к. $\left\langle w_n\right\rangle \sim \left\langle \varphi 
(y_n,z_n)\right\rangle $, то по Определению 2.7.1.4: 

\[ 
\left\{ n\in {\bf N}|t_n={\mathfrak f}(x_n,w_n)\right\} \in {\bf \Phi ix}\mbox{,} 
\] 

\[ 
\left\{ n\in {\bf N}|w_n=\varphi \left( y_n,z_n\right) \right\} \in {\bf %
\Phi ix}\mbox{.} 
\] 

Таким образом, по Теореме 2.7.1.2: 

\[ 
\left\{ n\in {\bf N}|t_n={\mathfrak f}(x_n,\varphi \left( y_n,z_n\right) 
)\right\} \in {\bf \Phi ix}\mbox{.} 
\] 

Следовательно, по Определению 2.7.1.4: 

\[ 
\left\langle t_n\right\rangle \sim \left\langle {\mathfrak f}(x_n,\varphi 
(y_n,z_n))\right\rangle \mbox{.} 
\] 

Таким образом, 

\[ 
\left\langle t_n\right\rangle \sim \left\langle \psi 
(x_n,y_n,z_n)\right\rangle \mbox{.} 
\] 

Следовательно, по Определению 2.7.1.5: $\left\langle t_n\right\rangle \in \widetilde{v}$. 

Отсюда и из 1) по Определению 2.7.1.7: $\widetilde{u}=\widetilde{v}$  $_{\bf \Box }$ 

{\bf Следствия из Теорем 2.7.1.6 и 2.7.1.7:} \cite{DVS3}: Для всех Q-чисел 
$\widetilde{x}$, $\widetilde{y}$, $\widetilde{z}$: 

${\bf \Phi }${\bf 1:} $(\widetilde{x}+\widetilde{y})=(\widetilde{y}+%
\widetilde{x})$, 

${\bf \Phi }${\bf 2:} $(\widetilde{x}+(\widetilde{y}+\widetilde{z}))=((%
\widetilde{x}+\widetilde{y})+\widetilde{z})$, 

${\bf \Phi }${\bf 3:} $(\widetilde{x}+0)=\widetilde{x}$, 

${\bf \Phi }${\bf 5:} $(\widetilde{x}\cdot \widetilde{y})=(\widetilde{y}%
\cdot \widetilde{x})$, 

${\bf \Phi }${\bf 6:} $(\widetilde{x}\cdot (\widetilde{y}\cdot \widetilde{z}%
))=((\widetilde{x}\cdot \widetilde{y})\cdot \widetilde{z})$, 

${\bf \Phi 7}${\bf : }$(\widetilde{x}\cdot 1)=\widetilde{x}$, 

${\bf \Phi }${\bf 10:} $(\widetilde{x}\cdot (\widetilde{y}+\widetilde{z}))=((%
\widetilde{x}\cdot \widetilde{y})+(\widetilde{x}\cdot \widetilde{z}))$. 

{\bf Теорема 2.7.1.8: }${\bf \Phi }${\bf 4:} Для каждого Q-числа $\widetilde{x}$ 
существует Q-число $\widetilde{y}$, для которого: 

$(\widetilde{x}+\widetilde{y})=0$. 

{\bf Доказательство Теоремы 2.7.1.8: }Если $\left\langle x_n\right\rangle \in 
\widetilde{x}$, то $\widetilde{y}$ есть Q-число, содержащее $%
\left\langle -x_n\right\rangle $  $_{\bf \Box }$ 

{\bf Теорема 2.7.1.9: }${\bf \Phi 9}${\bf :} Неверно, что $0=1$. 

{\bf Доказательство Теоремы 2.7.1.9:} Из Определения 2.7.1.6 и 
Определения 2.7.1.7  $_{\bf \Box }$ 

{\bf Определение 2.7.1.10:} Q-число $\widetilde{x}$ {\it Q-меньше}  
Q-числа $\widetilde{y}$ (обозначение: $\widetilde{x}<\widetilde{y}$), если  
существуют последовательности $\left\langle x_n\right\rangle $ и $\left\langle 
y_n\right\rangle $ вещественных чисел, для которых: $\left\langle 
x_n\right\rangle \in \widetilde{x}$, $\left\langle y_n\right\rangle \in 
\widetilde{y}$ и 

\[ 
\left\{ n\in {\bf N}|x_n<y_n\right\} \in {\bf \Phi ix}\mbox{.} 
\] 

{\bf Теорема 2.7.1.10:} Для всех Q-чисел $\widetilde{x}$, $\widetilde{y}$, $%
\widetilde{z}$: \cite{DVS4} 

${\bf \Omega 1}$: неверно, что $\widetilde{x}<\widetilde{x}$; 

${\bf \Omega 2}$: если $\widetilde{x}<\widetilde{y}$, и $\widetilde{y}<%
\widetilde{z}$, то $\widetilde{x}<\widetilde{z}$; 

${\bf \Omega 4}$: если $\widetilde{x}<\widetilde{y}$, то $(\widetilde{x}+%
\widetilde{z})<(\widetilde{y}+\widetilde{z})$; 

${\bf \Omega 5}$: если $0<\widetilde{z}$, и $\widetilde{x}<\widetilde{y}$, 
то $(\widetilde{x}\cdot \widetilde{z})<(\widetilde{y}\cdot \widetilde{z})$; 

${\bf \Omega 3}^{\prime }$: если $\widetilde{x}<\widetilde{y}$, то неверно, что 
$\widetilde{y}<\widetilde{x}$, или $\widetilde{x}=\widetilde{y}$, и 
vice versa; 

${\bf \Omega 3}^{\prime \prime }$: для всех стандартных Q-чисел $x$, $y$, $z$: 
$x<y$, или $y<x$, или $x=y$. 

{\bf Доказательство Теоремы 2.7.1.10:} Из Определения 2.7.1.10 по  
Теореме 2.7.1.2  $_{\bf \Box }$ 

{\bf Теорема 2.7.1.11: }${\bf \Phi }${\bf 8:} Если $0<|\widetilde{x}|$, то 
существует Q-число $\widetilde{y}$, для которого 
$(\widetilde{x}\cdot \widetilde{y})=1$. 

{\bf Доказательство Теоремы 2.7.1.11:} Если $\left\langle x_n\right\rangle \in 
\widetilde{x}$, то по Определению 2.7.1.10: если 

\[ 
{\bf A}=\left\{ n\in {\bf N}|0<\left| x_n\right| \right\} ,
\] 

то ${\bf A}\in {\bf \Phi ix}$. 

В этом случае: если для последовательности $\left\langle y_n\right\rangle $ : 
если $n\in {\bf A}$ then $y_n=1/x_n$ 

- то 

\[ 
\left\{ n\in {\bf N}|x_n\cdot y_n=1\right\} \in {\bf \Phi ix}  _{\bf \Box } 
\]

Таким образом, Q-числа удовлетворяют всем свойствам вещественных чисел, кроме $%
\Omega $3 \cite{DVS5}. Свойство $\Omega $3 несколько ослаблено 
($\Omega $3' и $\Omega $3''). 

{\bf Определение 2.7.1.11:} Q-число $\widetilde{x}$ называется {\it бесконечно-
малым Q-числом}, если существует последовательность вещественных чисел 
$\left\langle x_n\right\rangle $, для которой: 
$\left\langle x_n\right\rangle \in \widetilde{x}$, и для всех положительных 
вещественных чисел $\varepsilon $: 

\[ 
\left\{ n\in {\bf N}||x_n|<\varepsilon \right\} \in {\bf \Phi ix}\mbox{.} 
\] 

Обозначим множество всех бесконечно-малых Q-чисел как $I$. 

{\bf Определение 2.7.1.12:} Q-числа $\widetilde{x}$ и $\widetilde{y}$ называются  
{\it бесконечно-близкими} (обозна- чение: $\widetilde{x}\approx 
\widetilde{y}$), если $|\widetilde{x}-\widetilde{y}|=0$, или $|\widetilde{x}-%
\widetilde{y}|$ бесконечно мало. 

{\bf Определение 2.7.1.13}: Q-число $\widetilde{x}$ называется {\it бесконечно 
большим}, если суще-ствует последовательность $\left\langle r_n\right\rangle $ 
вещественных чисел, для которой 
$\left\langle r_n\right\rangle \in \widetilde{x}$, и для каждого натурального 
числа $m$: 

\[ 
\left\{ n\in {\bf N}|m<r_n\right\} \in {\bf \Phi ix}\mbox{.} 
\] 

\subsubsection{Модель} 

Определим пропозициональное исчисление как в \cite{MEN63}, но пропозициональные 
формы будем обозначать строчными греческими буквами. 

{\bf Определение C1: }Множество $\Re $ пропозициональных форм есть {\it U-мир}, 
если: 

1) если $\alpha _1,\alpha _2,\ldots ,\alpha _n\in \Re $, и $\alpha _1,\alpha 
_2,\ldots ,\alpha _n\vdash \beta $, то $\beta \in \Re $, 

2) для всех пропозициональных форм $\alpha $: неверно, что $(\alpha \&\left( 
\neg \alpha \right) )\in \Re $, 

3) для каждой пропозициональной формы $\alpha $: $\alpha \in \Re $ или $(\neg 
\alpha )\in \Re $. 

{\bf Определение C2: }Последовательности пропозициональных форм $\left\langle 
\alpha _n\right\rangle $ и $\left\langle \beta _n\right\rangle $ {\it %
Q-эквивалентны} (обозначение: $\left\langle \alpha _n\right\rangle \sim 
\left\langle \beta _n\right\rangle $) if 

\[ 
\left\{ n\in {\bf N}|\alpha _n\equiv \beta _n\right\} \in {\bf \Phi ix}%
\mbox{.} 
\] 

Определяем понятия {\it Q-формы} и {\it Q-расширения функции} для 
пропозициональ-ных форм как в Определениях 2.7.1.5, 2.7.1.9. 

{\bf Определение C3:} Q-форма $\widetilde{\alpha }$ {\it реальна} в  
U-мире $\Re $, если существует последовательность 
$\left\langle \alpha _n\right\rangle $ пропозициональных форм, для которой: 
$\left\langle \alpha _n\right\rangle \in \widetilde{\alpha }$, и 

\[ 
\left\{ n\in {\bf N}|\alpha _n\in \Re \right\} \in {\bf \Phi ix}\mbox{.} 
\] 

{\bf Определение C4: }Множество $\widetilde{\Re }$ Q-форм называется {\it  
Q-расширением U-мира} $\Re $, если $\widetilde{\Re }$ есть множество 
Q-реальных в $\Re $ Q-форм. 

{\bf Определение C5:} Последовательность $\left\langle \widetilde{\Re }%
_k\right\rangle $ Q-расширений U-мира называет-ся {\it S-миром}. 

{\bf Определение C6: }Q-форма $\widetilde{\alpha }$ {\it реальна в S-мире }
$\left\langle \widetilde{\Re }_k\right\rangle $, если 

\[ 
\left\{ k\in {\bf N}|\widetilde{\alpha }\in \widetilde{\Re }_k\right\} \in 
{\bf \Phi ix}\mbox{.} 
\] 

{\bf Определение C7:} Множество ${\bf A}$ (${\bf A}\subseteq {\bf N}$) 
называется {\it регулярным множе-ством}, еали для каждого вещественного 
положительного числа $\varepsilon $ существует натуральное число $n_0$, 
такое, что: для всех натуральных чисел $n$ и $m$, больших или равных $n_0$: 

\[ 
|w_n({\bf A})-w_m({\bf A})|<\varepsilon \mbox{.} 
\] 

{\bf Теорема C1:} Если ${\bf A}$ регулярное множество, и для всех 
вещественных положительных чисел $\varepsilon $: 

\[ 
\left\{ k\in {\bf N}|w_k({\bf A})<\varepsilon \right\} \in {\bf \Phi ix}%
\mbox{,} 
\] 

то 

\[ 
\lim_{k\rightarrow \infty }w_k({\bf A})=0\mbox{.} 
\] 

{\bf Доказательство Теоремы C1:} Допустим, 

\[ 
\lim_{k\rightarrow \infty }w_k({\bf A})\neq 0\mbox{.} 
\] 

То есть существует вещественное число $\varepsilon _0$, такое, что для каждого 
натурального числа $n^{\prime }$ существует натуральное число $n$, для которого: 

\[ 
n>n^{\prime }\mbox{, и }w_n({\bf A})>\varepsilon _0. 
\] 

Пусть $\delta _0$ - некоторое положительное вещественное число, для которого: 
$\varepsilon_0-\delta _0>0$. Т.к. ${\bf A}$ - регулярное множество, то для 
$\delta _0$ существует натуральное число $n_0$, такое, что для всех 
натуральных $n$ и $m$, больших или равных числу $n_0$: 

\[ 
|w_m({\bf A})-w_n({\bf A})|<\delta _0\mbox{.} 
\] 

То есть 

\[ 
w_m({\bf A})>w_n({\bf A})-\delta _0\mbox{.} 
\] 

Т.к. $w_n({\bf A})\geq \varepsilon _0$, то $w_m({\bf A})\geq \varepsilon 
_0-\delta _0$. 

Следовательно, существует натуральное $n_0$, такое, что для всех натуральных 
$m $: если $m\geq n_0$, то $w_m({\bf A})\geq \varepsilon _0-\delta _0$. 

Следовательно, 

\[ 
\left\{ m\in {\bf N}|w_m({\bf A})\geq \varepsilon _0-\delta _0\right\} \in 
{\bf \Phi ix}\mbox{,} 
\] 

и по условию этой Теоремы: 

\[ 
\left\{ k\in {\bf N}|w_k({\bf A})<\varepsilon _0-\delta _0\right\} \in {\bf %
\Phi ix}\mbox{.} 
\] 

Таким образом, 

\[ 
\left\{ k\in {\bf N}|\varepsilon _0-\delta _0<\varepsilon _0-\delta 
_0\right\} \in {\bf \Phi ix}\mbox{.} 
\] 

Т.е. $\emptyset \notin {\bf \Phi ix}$. Это противоречит Теореме 2.7.1.2  $_{\bf \Box }$ 

{\bf Определение C8:} Пусть $\left\langle \widetilde{\Re }_k\right\rangle $ - 
S-мир. 

В этом случае функция ${\mathfrak W}(\widetilde{\beta })$, определенная в множестве 
Q-форм и имеющая знаыения в $Q{\bf R}$, определяется следующим образом: 

если ${\mathfrak W}(\widetilde{\beta })=\widetilde{p}$, то существует 
последовательность $\left\langle p_n\right\rangle $ вещественных чисел, таких, 
что: $\left\langle p_n\right\rangle \in \widetilde{p}$, и 

\[ 
p_n=w_n\left( \left\{ k\in {\bf N}|\widetilde{\beta }\in \widetilde{\Re }%
_k\right\} \right) \mbox{.} 
\] 

{\bf Теорема C2:} Если $\left\{ k\in {\bf N}|\widetilde{\beta }\in \widetilde{%
\Re }_k\right\} $ - регулярное множество, и ${\mathfrak W}(\widetilde{\beta }%
)\approx 1$, то $\widetilde{\beta }$ реальна в $\left\langle \widetilde{%
\Re }_k\right\rangle $. 

{\bf Доказательство Теоремы C2: }Т.к. ${\mathfrak W}(\widetilde{\beta })\approx 1$, 
то по Определениям 2.7.1.12 и 2.7.1.11: для всех положительных вещественных чисел 
$\varepsilon $: 

\[ 
\left\{ n\in {\bf N}|w_n\left( \left\{ k\in {\bf N}|\widetilde{\beta }\in 
\widetilde{\Re }_k\right\} \right) >1-\varepsilon \right\} \in {\bf \Phi ix}%
\mbox{.} 
\] 

Следовательно, по пункту 3 Теоремы 2.7.1.1: для всех положительных веществен-ных 
$\varepsilon$: 

\[ 
\left\{ n\in {\bf N}|\left( {\bf N}-w_n\left( \left\{ k\in {\bf N}|%
\widetilde{\beta }\in \widetilde{\Re }_k\right\} \right) \right) 
<\varepsilon \right\} \in {\bf \Phi ix}\mbox{.} 
\] 

Таким образом, по Теореме C1: 

\[ 
\lim_{n\rightarrow \infty }\left( {\bf N}-w_n\left( \left\{ k\in {\bf N}|%
\widetilde{\beta }\in \widetilde{\Re }_k\right\} \right) \right) =0\mbox{.} 
\] 

Т.е.: 

\[ 
\lim_{n\rightarrow \infty }w_n\left( \left\{ k\in {\bf N}|\widetilde{\beta }%
\in \widetilde{\Re }_k\right\} \right) =1\mbox{.} 
\] 

Следовательно, по Определению 2.7.1.3: 

\[ 
\left\{ k\in {\bf N}|\widetilde{\beta }\in \widetilde{\Re }_k\right\} \in 
{\bf \Phi ix}\mbox{.} 
\] 

И по Определению C6: $\widetilde{\beta }$ реальна в $\left\langle 
\widetilde{\Re }_k\right\rangle $  $_{\bf \Box }$ 

{\bf Теорема C3: }Существует P-функция. 

{\bf Доказательство Теоремы C3:} По Теоремам C2 и 2.7.1.1: ${\mathfrak W}(%
\widetilde{\beta })$ есть P-функция в $\left\langle \widetilde{\Re }%
_k\right\rangle $ $_{\bf \Box }$ 

\section{Квантовая теория}
\subsection{События и уравнения движения}

  Обозначим: 

\begin{eqnarray*}
&&\mathbf{x}\stackrel{def}{=}\left( x_1,x_2,...,x_\mu\right) \mbox{,}\\
&&\underline{x}\stackrel{def}{=}\left\langle t,\mathbf{x}\right\rangle\mbox{,}\\
&&\int d^{\mu +1}\underline{x}\stackrel{def}{=}\int dt\int dx_1\int
dx_2\cdots \int dx_\mu \mbox{,}\\
&&\int d^\mu \mathbf{y}\stackrel{def}{=}\int\limits_{-\infty }^\infty
dy_1\int\limits_{-\infty }^\infty dy_2\cdots \int\limits_{-\infty }^\infty
dy_\mu \mbox{.}
\end{eqnarray*}  

  $A(D)$ означает $\left(A\left( t,\mathbf{x}\right)\&\ll\left( t,%
\mathbf{x}\right)\in D\gg\right)$.

  Пусть ${\rm P}$ - логическая функция вероятности.

  Я называю {\it абсолютной плотностью} вероятности события $A$ функцию 
$p_A\left(\underline{x}\right)$ такую, что для любой области $D$: если 
$D\subseteq R^{\mu +1}$, то

\[
\int\limits_Dd^{\mu +1}\underline{x}\cdot p_A\left( \underline{x}\right) =\mathrm{P}%
\left( A\left( D\right) \right) \mbox{.}
\]

  Если $J$ - якобиан преобразований

\begin{eqnarray}
&&t\rightarrow t^{\prime }=\frac{t-vx_k}{\sqrt{1-v^2}}\mbox{,}\nonumber\\
&&x_k\rightarrow x_k^{\prime }=\frac{x_k-vt}{\sqrt{1-v^2}}\mbox{,}\label{lr}\\
&&x_j\rightarrow x_j^{\prime }=x_j \mbox{ для }j\neq k\mbox{,}\nonumber
\end{eqnarray}

то 

\[
J=\frac{\partial \left( t^{\prime },x^{\prime }\right) }{\partial \left(
t,x\right) }=1\mbox{.}
\]

Поэтому абсолютная плотность вероятности инвариантна относительно 
пре-образований Лоренца.

  Если

\[
\rho _A\left( t,\mathbf{x}\right) =\frac{p_A\left( t,\mathbf{x}\right) }{%
\int d^\mu \mathbf{y}\cdot p_A\left( t,\mathbf{y}\right) }\mbox{,}
\]

то $\rho _A\left( t,\mathbf{x}\right)$ представляет {\it плотность распределения} 
вероятности $A$ в момент $t$.

В преобразованиях (\ref{lr}):

\[
\rho _A\left( t,\mathbf{x}\right) \rightarrow \rho _A^{\prime }\left( t,%
\mathbf{x}\right) =\frac{p_A\left( t,\mathbf{x}\right) }{\int d^\mu \mathbf{y%
}\cdot p_A\left( t+v\left( y_k-x_k\right) ,\mathbf{y}\right) }\mbox{.}
\]

Следовательно, $\rho_A$ не инвариантна относительно этих преобразований. 

Далее я рассматриваю события $A\left(\underline{x}\right)$, для которых $\rho_A$ 
представляет нулевую компоненту некоторого $3+1$-векторного поля $\underline{j}_A$ 

($\underline{j}_A=\left(%
 j_{A,0},\mathbf{j}_A\right) =\left( j_{A,0},j_{A,1},j_{A,2},\ldots,j_{A,\mu} \right) $). 

То есть найдутся вещественные функции $j_{A,k}\left(\underline{x}\right)$ такие, 
что:

\[
\rho _A=j_{A,0}
\]

и в преобразованиях (\ref{lr}):

\begin{eqnarray*}
&&j_{A,0}\rightarrow j_{A,0}^{\prime }=\frac{j_{A,0}-vj_{A,k}}{\sqrt{1-v^2}}\mbox{,}\\
&&j_{A,k}\rightarrow j_{A,k}^{\prime }=\frac{j_{A,k}-vj_{A,0}}{\sqrt{1-v^2}}\mbox{,}\\
&&j_{A,s}\rightarrow j_{A,s}^{\prime }=j_{A,s}\mbox{ для }s\neq k \mbox{.}
\end{eqnarray*}

Обозначим:

\[
1_2\stackrel{def}{=}\left[ 
\begin{array}{cc}
1 & 0 \\ 
0 & 1
\end{array}
\right] \mbox{, }0_2\stackrel{def}{=}\left[ 
\begin{array}{cc}
0 & 0 \\ 
0 & 0
\end{array}
\right] \mbox{, }\beta^{[0]}\stackrel{def}{=}-\left[  
\begin{array}{cc}
1_2 & 0_2 \\ 
0_2 & 1_2
\end{array}
\right] \mbox{.}
\]

матрицы Паули:

\[
\sigma _1=\left( 
\begin{array}{cc}
0 & 1 \\ 
1 & 0
\end{array}
\right) \mbox{, }\sigma _2=\left( 
\begin{array}{cc}
0 & -\mathrm{i} \\ 
\mathrm{i} & 0
\end{array}
\right) \mbox{, }\sigma _3=\left( 
\begin{array}{cc}
1 & 0 \\ 
0 & -1
\end{array}
\right) \mbox{.} 
\]

Множество $\widetilde{C}$ комплексных $n\times n$ матриц называется {\it %
Клиффордовым мно-жеством ранга $n$} \cite{Md} если выполняются следующие условия:

если $\alpha _k\in \widetilde{C}$ и $\alpha _r\in \widetilde{C}$, то $%
\alpha _k\alpha _r+\alpha _r\alpha _k=2\delta _{k,r}$;

если $\alpha _k\alpha _r+\alpha _r\alpha _k=2\delta _{k,r}$ для всех элементов $%
\alpha _r$ множества $\widetilde{C}$, то $\alpha _k\in \widetilde{C}$.

Далее до специальной отметки пусть $\mu =3$.

Если $n=4$, то Клиффордово множество либо содержит $3$ матрицы ({\it %
Клиффор-дова тройка}), либо - $5$ матриц ({\it Клиффордова пентада}).

Существует только шесть Клиффордовых пентад \cite{Md}: одна {\it легкая пентада} $%
\beta $:

\begin{equation}
\beta ^{\left[ 1\right] }\stackrel{def}{=}\left[ 
\begin{array}{cc}
\sigma _1 & 0_2 \\ 
0_2 & -\sigma _1
\end{array}
\right] \mbox{, }\beta ^{\left[ 2\right] }\stackrel{def}{=}\left[ 
\begin{array}{cc}
\sigma _2 & 0_2 \\ 
0_2 & -\sigma _2
\end{array}
\right] \mbox{, }\beta ^{\left[ 3\right] }\stackrel{def}{=}\left[ 
\begin{array}{cc}
\sigma _3 & 0_2 \\ 
0_2 & -\sigma _3
\end{array}
\right] \mbox{,}  \label{lghr}
\end{equation}

\begin{equation}
\gamma ^{\left[ 0\right] }\stackrel{def}{=}\left[ 
\begin{array}{cc}
0_2 & 1_2 \\ 
1_2 & 0_2
\end{array}
\right] \mbox{, }  \label{lghr1}
\end{equation}

\begin{equation}
\beta ^{\left[ 4\right] }\stackrel{def}{=}\mathrm{i}\cdot \left[ 
\begin{array}{cc}
0_2 & 1_2 \\ 
-1_2 & 0_2
\end{array}
\right] \mbox{;}  \label{lghr2}
\end{equation}

три {\it цветных} пентады:

{\it красная} пентада $\zeta $:

\[
\zeta ^{[1]}=\left[ 
\begin{array}{cc}
\sigma _1 & 0_2 \\ 
0_2 & -\sigma _1
\end{array}
\right] ,\zeta ^{[2]}=\left[ 
\begin{array}{cc}
\sigma _2 & 0_2 \\ 
0_2 & \sigma _2
\end{array}
\right] ,\zeta ^{[3]}=\left[ 
\begin{array}{cc}
-\sigma _3 & 0_2 \\ 
0_2 & -\sigma _3
\end{array}
\right] , 
\]

\[
\gamma _\zeta ^{[0]}=\left[ 
\begin{array}{cc}
0_2 & -\sigma _1 \\ 
-\sigma _1 & 0_2
\end{array}
\right] \mbox{, }\zeta ^{[4]}=-{\rm i} \left[ 
\begin{array}{cc}
0_2 & \sigma _1 \\ 
-\sigma _1 & 0_2
\end{array}
\right] ; 
\]

{\it зеленая} пентада $\eta $:

\[
\eta ^{[1]}=\left[ 
\begin{array}{cc}
-\sigma _1 & 0_2 \\ 
0_2 & -\sigma _1
\end{array}
\right] ,\eta ^{[2]}=\left[ 
\begin{array}{cc}
\sigma _2 & 0_2 \\ 
0_2 & -\sigma _2
\end{array}
\right] ,\eta ^{[3]}=\left[ 
\begin{array}{cc}
-\sigma _3 & 0_2 \\ 
0_2 & -\sigma _3
\end{array}
\right] , 
\]

\[
\gamma _\eta ^{[0]}=\left[ 
\begin{array}{cc}
0_2 & -\sigma _2 \\ 
-\sigma _2 & 0_2
\end{array}
\right] \mbox{, }\eta ^{[4]}={\rm i} \left[ 
\begin{array}{cc}
0_2 & \sigma _2 \\ 
-\sigma _2 & 0_2
\end{array}
\right] ; 
\]

{\it синяя} пентада $\theta $:

\[
\theta ^{[1]}=\left[ 
\begin{array}{cc}
-\sigma _1 & 0_2 \\ 
0_2 & -\sigma _1
\end{array}
\right] ,\theta ^{[2]}=\left[ 
\begin{array}{cc}
\sigma _2 & 0_2 \\ 
0_2 & \sigma _2
\end{array}
\right] ,\theta ^{[3]}=\left[ 
\begin{array}{cc}
\sigma _3 & 0_2 \\ 
0_2 & -\sigma _3
\end{array}
\right] , 
\]

\[
\gamma _\theta ^{[0]}=\left[ 
\begin{array}{cc}
0_2 & -\sigma _3 \\ 
-\sigma _3 & 0_2
\end{array}
\right] ,\theta ^{[4]}=-{\rm i} \left[ 
\begin{array}{cc}
0_2 & \sigma _3 \\ 
-\sigma _3 & 0_2
\end{array}
\right] ; 
\]

две {\it вкусовые} пентады:

{\it сладкая} пентада $\underline{\Delta }$:

\[
\underline{\Delta }^{[1]}=\left[ 
\begin{array}{cc}
0_2 & -\sigma _1 \\ 
-\sigma _1 & 0_2
\end{array}
\right] ,\underline{\Delta }^{[2]}=\left[ 
\begin{array}{cc}
0_2 & -\sigma _2 \\ 
-\sigma _2 & 0_2
\end{array}
\right] ,\underline{\Delta }^{[3]}=\left[ 
\begin{array}{cc}
0_2 & -\sigma _3 \\ 
-\sigma _3 & 0_2
\end{array}
\right] , 
\]

\[
\underline{\Delta }^{[0]}=\left[ 
\begin{array}{cc}
-1_2 & 0_2 \\ 
0_2 & 1_2
\end{array}
\right] ,\underline{\Delta }^{[4]}={\rm i} \left[ 
\begin{array}{cc}
0_2 & 1_2 \\ 
-1_2 & 0_2
\end{array}
\right] ; 
\]

{\it горькая} пентада $\underline{\Gamma}$:

\[
\underline{\Gamma }^{[1]}={\rm i} \left[ 
\begin{array}{cc}
0_2 & -\sigma _1 \\ 
\sigma _1 & 0_2
\end{array}
\right] ,\underline{\Gamma }^{[2]}={\rm i} \left[ 
\begin{array}{cc}
0_2 & -\sigma _2 \\ 
\sigma _2 & 0_2
\end{array}
\right] ,\underline{\Gamma }^{[3]}={\rm i} \left[ 
\begin{array}{cc}
0_2 & -\sigma _3 \\ 
\sigma _3 & 0_2
\end{array}
\right] , 
\]

\[
\underline{\Gamma }^{[0]}=\left[ 
\begin{array}{cc}
-1_2 & 0_2 \\ 
0_2 & 1_2
\end{array}
\right] ,\underline{\Gamma }^{[4]}=\left[ 
\begin{array}{cc}
0_2 & 1_2

 \\ 
1_2 & 0_2
\end{array}
\right] \mbox{.} 
\]

Так как следующая система четырех вещественных уравнений с восемью вещественными 
неизвестными: $b^2$, при $b>0$, $\stackrel{*}{\alpha }$, $\stackrel{*}{\beta }$, $%
\stackrel{*}{\chi }$, $\stackrel{*}{\theta }$, $\stackrel{*}{\gamma }$, $%
\stackrel{*}{\upsilon }$, $\stackrel{*}{\lambda }$

\begin{equation}
\left\{ 
\begin{array}{c}
b^2 = -\rho_A \mbox{,} \\ 
b^2\left( 
\begin{array}{c}
\cos ^2\left( \stackrel{*}{\alpha }\right) \sin \left( 2\stackrel{*}{\beta }%
\right) \cos \left( \stackrel{*}{\theta }-\stackrel{*}{\gamma }\right) - \\ 
-\sin ^2\left( \stackrel{*}{\alpha }\right) \sin \left( 2\stackrel{*}{\chi }%
\right) \cos \left( \stackrel{*}{\upsilon }-\stackrel{*}{\lambda }\right) 
\end{array}
\right) = -j_{A,1}\mbox{,} \\ 
b^2\left( 
\begin{array}{c}
\cos ^2\left( \stackrel{*}{\alpha }\right) \sin \left( 2\stackrel{*}{\beta }%
\right) \sin \left( \stackrel{*}{\theta }-\stackrel{*}{\gamma }\right) - \\ 
-\sin ^2\left( \stackrel{*}{\alpha }\right) \sin \left( 2\stackrel{*}{\chi }%
\right) \sin \left( \stackrel{*}{\upsilon }-\stackrel{*}{\lambda }\right) 
\end{array}
\right) = -j_{A,2}\mbox{,} \\ 
b^2\left( \cos ^2\left( \stackrel{*}{\alpha }\right) \cos
\left( 2\stackrel{*}{\beta }\right) -\sin ^2\left( \stackrel{*}{\alpha }%
\right) \cos \left( 2\stackrel{*}{\chi }\right) \right) = -j_{A,3}\mbox{.}
\end{array}
\right|\label{abc} 
\end{equation}

имеет решения для любых $\rho_A$ и $j_{A,k}$, то при

\begin{eqnarray*}
&&\varphi _1=b\exp \left({\rm i}\stackrel{*}{\gamma }\right) \cos \left( \stackrel{%
*}{\beta }\right) \cos \left( \stackrel{*}{\alpha }\right)\mbox{,}\\
&&\varphi _2=b\exp \left({\rm i}\stackrel{*}{\theta }\right) \sin \left( \stackrel{%
*}{\beta }\right) \cos \left( \stackrel{*}{\alpha }\right)\mbox{,}\\
&&\varphi _3=b\exp \left({\rm i}\stackrel{*}{\lambda }\right) \cos \left( 
\stackrel{*}{\chi }\right) \sin \left( \stackrel{*}{\alpha }\right)\mbox{,}\\
&&\varphi _4=b\exp \left({\rm i}\stackrel{*}{\upsilon }\right) \sin \left( 
\stackrel{*}{\chi }\right) \sin \left( \stackrel{*}{\alpha }\right)
\end{eqnarray*}

\begin{equation}
j_{A,\alpha} \stackrel{def}{=}-\sum_{k=1}^4\sum_{s=1}^4\varphi _s^{*}\beta
_{s,k}^{\left[ \alpha \right] }\varphi _k  \label{j}
\end{equation}

с $\alpha \in \left\{ 0,1,2,3\right\} $.

Если 3-вектор $\mathbf{u}_A$ определяется как

\begin{equation}
\mathbf{j}_A\stackrel{def}{=}\rho _A\mathbf{u}_A \mbox{,}  \label{v2}
\end{equation}

то $\mathbf{u}_A$ представляет {\it локальную скорость распространения вероятности
собы-тия} $A$.

Обозначим:

\begin{eqnarray*}
&&\partial _k\stackrel{Def}{=}\partial /\partial x_k\mbox{;}\\
&&\partial _t\stackrel{Def}{=}\partial _0\stackrel{Def}{=}\partial /\partial_ t\mbox{;}\\
&&\partial _k^{\prime }\stackrel{Def}{=}\partial /\partial x_k^{\prime }\mbox{;}\\
&&\sum_{\mathbf{k}}\stackrel{Def}{=}\sum_{k_1=-\infty }^\infty
\sum_{k_2=-\infty }^\infty \sum_{k_3=-\infty }^\infty \mbox{.} 
\end{eqnarray*}

Я рассматриваю события, подчиняющиеся следующим условиям: существует крошечное 
вещественное положительное число $h$ такое, что если $\left| x_r\right| \geq
\frac \pi h$ ($r\in \left\{ 1,2,3\right\} $), то

\[
\varphi _j\left( t,\mathbf{x}\right) =0\mbox{.} 
\]

Пусть $\left( V\right) $ - область, для которой: $\mathbf{x}\in \left(
V\right) $, если и только если $\left| x_r\right| \leq \frac \pi h$ для $r\in
\left\{ 1,2,3\right\} $. То есть:

\[
\int_{\left( V\right) }d^3\mathbf{x}=\int_{-\frac \pi h}^{\frac \pi
h}dx_1\int_{-\frac \pi h}^{\frac \pi h}dx_2\int_{-\frac \pi h}^{\frac \pi
h}dx_3\mbox{.} 
\]

Пусть $j\in \left\{ 1,2,3,4\right\} $, $k\in \left\{ 1,2,3,4\right\} $

Если 

\[
\varsigma _{w,\mathbf{p}}\left( t,\mathbf{x}\right) \stackrel{def}{=}%
\exp \left( \mathrm{i}h\left( wt-\mathbf{px}\right) \right) \mbox{,}
\]

то ряд Фурье для $\varphi _j\left( t,\mathbf{x}\right) $ имеет следующую форму: 

\[
\varphi _j\left( t,\mathbf{x}\right) =\sum_{w,\mathbf{p}}c_{j,w,\mathbf{p}%
}\varsigma _{w,\mathbf{p}}\left( t,\mathbf{x}\right)\mbox{.} 
\]

Обозначим: $\varphi _{j,w,\mathbf{p}}\left( t,\mathbf{x}\right) \stackrel{def}{=}%
c_{j,w,\mathbf{p}}\varsigma _{w,\mathbf{p}}\left( t,\mathbf{x}\right) $.

Пусть $\left\langle t,\mathbf{x}\right\rangle $ - какая нибудь пространственно-
временная точка.

Обозначим 

\[
A_k\stackrel{def}{=}\varphi _{k,w,\mathbf{p}}|_{\left\langle t,\mathbf{x}\right\rangle }
\]

- значение функции $\varphi _{k,w,\mathbf{p}}$ в этой точке,

а 

\[
C_j\stackrel{def}{=}\left( \partial _t\varphi _{j,w,\mathbf{p}}-\sum_{s=1}^4\sum_{\alpha
=1}^3\beta _{j,s}^{\left[ \alpha \right] }\partial _\alpha \varphi _{s,w,%
\mathbf{p}}\right) |_{\left\langle t,\mathbf{x}\right\rangle }
\]

 - значение Функции $\left(\partial _t\varphi _{j,w,\mathbf{p}%
}-\sum_{s=1}^4\sum_{\alpha =1}^3\beta _{j,s}^{\left[ \alpha \right]
}\partial _\alpha \varphi _{s,w,\mathbf{p}}\right)$.

Здесь $A_k$ и $C_j$ - комплексные числа. Поэтому следующая система уравнений:

\begin{equation}
\left\{ 
\begin{array}{c}
\sum_{k=1}^4z_{j,k,w,\mathbf{p}}A_k=C_j\mbox{,} \\ 
z_{j,k,w,\mathbf{p}}^{*}=-z_{k,j,w,\mathbf{p}}
\end{array}
\right|  \label{sys}
\end{equation}

представляет систему из 20 алгебраических комплексных уравнений с 16 комплексными 
неизвестными $z_{k,j,w,\mathbf{p}}$. Эта система может быть преобразована в 
систему из 8 линейных вещественных уравнений с 16 вещественными неизвестными 
$x_{s,k}\stackrel{def}{=}\mathrm{Re}\left( z_{s,k,w,\mathbf{p}}\right) $ для 
$s<k$ и $y_{s,k}\stackrel{def}{=}\mathrm{Im}\left(z_{s,k,w,\mathbf{p}}\right) $ 
для $s\leq k$:

\[
\left\{ 
\begin{array}{c}
-y_{1,1}b_1+x_{1,2}a_2-y_{1,2}b_2+\allowbreak
x_{1,3}a_3-y_{1,3}b_3+x_{1,4}a_4-y_{1,4}b_4\\=-hwb_1-hp_3b_1-\allowbreak
hp_1b_2+hp_2a_2\mbox{,} \\ 
y_{1,1}a_1+x_{1,2}b_2+y_{1,2}a_2+x_{1,3}b_3+y_{1,3}a_3+x_{1,4}b_4+%
\allowbreak y_{1,4}a_4\\=hwa_1+hp_3a_1+hp_1a_2+hp_2b_2\mbox{,} \\ 
-x_{1,2}a_1-y_{1,2}b_1-y_{2,2}b_2+\allowbreak
x_{2,3}a_3-y_{2,3}b_3+x_{2,4}a_4-y_{2,4}b_4\\=-hwb_2-hp_1b_1-hp_2a_1+hp_3b_2%
\mbox{,} \\ 
-x_{1,2}b_1+y_{1,2}a_1+y_{2,2}a_2+x_{2,3}b_3+y_{2,3}a_3+x_{2,4}b_4+%
\allowbreak y_{2,4}a_4\\=hwa_2+hp_1a_1-\allowbreak hp_2b_1-hp_3a_2\mbox{,} \\ 
-x_{1,3}a_1-y_{1,3}b_1-x_{2,3}a_2-y_{2,3}b_2-y_{3,3}b_3+x_{3,4}a_4-y_{3,4}b_4\\=-hwb_3+hp_3b_3+\allowbreak hp_1b_4-hp_2a_4%
\mbox{,} \\ 
-x_{1,3}b_1+y_{1,3}a_1-x_{2,3}b_2+\allowbreak
y_{2,3}a_2+y_{3,3}a_3+x_{3,4}b_4+\allowbreak
y_{3,4}a_4\\=hwa_3-hp_3a_3-hp_1a_4-hp_2b_4\mbox{,} \\ 
-x_{1,4}a_1-y_{1,4}b_1-x_{2,4}a_2-y_{2,4}b_2-x_{3,4}a_3-y_{3,4}b_3-y_{4,4}b_4\\=-hwb_4+hp_1b_3+hp_2a_3-hp_3b_4%
\mbox{,} \\ 
-x_{1,4}b_1+y_{1,4}a_1-x_{2,4}b_2+\allowbreak
y_{2,4}a_2-x_{3,4}b_3+y_{3,4}a_3+\allowbreak
y_{4,4}a_4\\=hwa_4-hp_1a_3+\allowbreak hp_2b_3+hp_3a_4\mbox{;}
\end{array}
\right| 
\]

(здесь $a_k=\mathrm{Re}A_k$ и $b_k=\mathrm{Im}A_k$.)

 Эта система имеет решения по теореме Кронекера-Капелли. Поэтому в каждой точке 
$\left\langle t,\mathbf{x}\right\rangle $ существует такое комплексное число 
$z_{j,k,w,\mathbf{p}}|_{\left\langle t,\mathbf{x}\right\rangle }$.

Пусть $\kappa _{w,\mathbf{p}}$ - линейные операторы на линейном пространстве, 
натянутом на базис из функций $\varsigma _{w,\mathbf{p}}\left( t,\mathbf{x}%
\right) $, такие, что

\[
\kappa _{w,\mathbf{p}}\varsigma _{w^{\prime },\mathbf{p}^{\prime }}\stackrel{%
def}{=}\left\{ 
\begin{array}{c}
\varsigma _{w^{\prime },\mathbf{p}^{\prime }}\mbox{, if }w=w^{\prime }%
\mbox{, }\mathbf{p}=\mathbf{p}^{\prime }\mbox{;} \\ 
0\mbox{, if }w\neq w^{\prime }\mbox{или/и }\mathbf{p}\neq \mathbf{p}%
^{\prime }
\end{array}
\mbox{.}\right| 
\]

Пусть $Q_{j,k}$ - оператор такой, что в каждой точке $\left\langle t,\mathbf{%
x}\right\rangle $:

\[
Q_{j,k}|_{\left\langle t,\mathbf{x}\right\rangle }\stackrel{def}{=}\sum_{w,%
\mathbf{p}}\left( z_{j,k,w,\mathbf{p}}|_{\left\langle t,\mathbf{x}%
\right\rangle }\right) \kappa _{w,\mathbf{p}} 
\]

Следовательно, для каждой функции $\varphi_j $ существует оператор $Q_{j,k}$ 
такой, что зависимость $\varphi_j $ от $t$ описывается следующим 
дифференциальным уравнением\footnote{%
Эта система уравнений похожа на уравнение Дирака с массовой матрицей 
\cite{VVD}, \cite{Barut}, \cite{Wilson}. Я выбрал такую форму этой системы для 
того, чтобы выразить поведение  $\rho _A \left( t,\mathbf{x}%
\right) $ спинорами и элементами Клиффордова множества.}:

\begin{equation}
\partial _t\varphi _j=\sum_{k=1}^4\left( \beta _{j,k}^{\left[ 1\right]
}\partial _1+\beta _{j,k}^{\left[ 2\right] }\partial _2+\beta _{j,k}^{\left[
3\right] }\partial _3+Q_{j,k}\right) \varphi _k\mbox{.}  \label{ham}
\end{equation}

и $Q_{j,k}^{*}=\sum_{w,\mathbf{p}}\left( z_{j,k,w,\mathbf{p}%
}^{*}|_{\left\langle t,\mathbf{x}\right\rangle }\right) \kappa _{w,\mathbf{p}%
}=-Q_{k,j}$.

В этом случае, если

\[
\widehat{H}_{j,k}\stackrel{def}{=}\mathrm{i}\left( \beta _{j,k}^{\left[
1\right] }\partial _1+\beta _{j,k}^{\left[ 2\right] }\partial _2+\beta
_{j,k}^{\left[ 3\right] }\partial _3+Q_{j,k}\right)\mbox{,} 
\]

то $\widehat{H}$ называется {\it гамильтонианом} движения с уравнением (\ref{ham}).

Пусть $\mathbf{H}$ - некоторое гильбертово пространство, на элементах которого 
определены линейные операторы $\psi_s\left( \mathbf{x}\right) $ со следующими 
свойствами:

1. $\mathbf{H}$ содержит элемент $\Phi _0$ такой, что:

\[
\Phi _0^{\dagger }\Phi _0=1\mbox{,} 
\]

и

\[
\psi _s\Phi _0=0\mbox{, }\Phi _0^{\dagger }\psi _s^{\dagger }=0\mbox{;} 
\]

2.7.1.

\[
\psi _s\left( \mathbf{x}\right) \psi _s\left( \mathbf{x}\right) =0\mbox{,} 
\]

и

\[
\psi _s^{\dagger }\left( \mathbf{x}\right) \psi _s^{\dagger }\left( \mathbf{x%
}\right) =0\mbox{;} 
\]

3.

\begin{equation}
\begin{array}{c}
\left\{ \psi _{s^{\prime }}^{\dagger }\left( \mathbf{y}\right) ,\psi
_s\left( \mathbf{x}\right) \right\} \stackrel{Def}{=} \\ 
\stackrel{Def}{=}\psi _{s^{\prime }}^{\dagger }\left( \mathbf{y}\right) \psi
_s\left( \mathbf{x}\right) +\psi _s\left( \mathbf{x}\right) \psi _{s^{\prime
}}^{\dagger }\left( \mathbf{y}\right) =\delta \left( \mathbf{y}-\mathbf{x}%
\right) \delta _{s^{\prime },s}
\end{array}
\label{dddd}
\end{equation}

Пусть:

\begin{equation}
\Psi \left( t,\mathbf{x}\right) \stackrel{def}{=}\sum_{s=1}^4\varphi
_s\left( t,\mathbf{x}\right) \psi _s^{\dagger }\left( \mathbf{x}\right) \Phi
_0  \label{Sat}
\end{equation}

Из (\ref{dddd}):

\[
\Psi ^{\dagger }\left( t,\mathbf{x}^{\prime }\right) \Psi \left( t,\mathbf{x}%
\right) =\sum_{s=1}^4\varphi _s^{*}\left( t,\mathbf{x}^{\prime }\right)
\varphi _s\left( t,\mathbf{x}\right) \delta \left( \mathbf{x}^{\prime }-%
\mathbf{x}\right) \mbox{.} 
\]

То есть из (\ref{j}):

\[
\int dx^{\prime }\cdot \Psi ^{\dagger }\left( t,\mathbf{x}^{\prime }\right)
\Psi \left( t,\mathbf{x}\right) =\rho _A \left( t,\mathbf{x}\right) \mbox{.}
\]

Оператор $\mathcal{H}$, определенный как:

\begin{equation}
\mathcal{H}\left( t,\mathbf{x}\right) \stackrel{def}{=}\sum_{s=1}^4\psi
_s^{\dagger }\left( \mathbf{x}\right) \sum_{k=1}^4\widehat{H}_{s,k}\left( t,%
\mathbf{x}\right) \psi _k\left( \mathbf{x}\right)  \label{hmm}
\end{equation}

называется {\it плотностью гамильтониана} $\widehat{H}$.

Из (\ref{Sat}):

\[
-\mathrm{i}\int d^3\mathbf{x}\cdot \mathcal{H}\left( t,\mathbf{x}\right)
\Psi \left( t,\mathbf{x}_0\right) =\partial _t\Psi \left( t,\mathbf{x}%
_0\right) \mbox{.} 
\]

Следовательно, плотность гамильтониана определяет изменение по времени 
вероятности события $A$ в пространственной точке $\mathbf{x}_0$.

Я называю оператор $\psi ^{\dagger }\left( \mathbf{x}\right) $ {\it 
оператором рождения}, а $\psi \left( \mathbf{x}\right) $ - {\it оператором 
уничтожения вероятности} события $A$ в точке  $\mathbf{x}$. Оператор 
$\psi ^{\dagger }\left( \mathbf{x}\right) $ не является оператором рождения 
частицы в точке $\mathbf{x}$, но этот оператор изменяет вероятность события $A$ 
в этой точке. Аналогично - для $\psi \left( \mathbf{x}\right) $.

Матричная форма формулы (\ref{ham}) имеет следующий вид:

\begin{equation}
\partial _t\varphi =\left( \beta ^{\left[ 1\right] }\partial _1+\beta
^{\left[ 2\right] }\partial _2+\beta ^{\left[ 3\right] }\partial _3+\widehat{%
Q}\right) \varphi \mbox{,}  \label{ham1}
\end{equation}

с

\[
\varphi =\left[ 
\begin{array}{c}
\varphi _1 \\ 
\varphi _2 \\ 
\varphi _3 \\ 
\varphi _4
\end{array}
\right] 
\]

и

\[
\widehat{Q}=\left[ 
\begin{array}{cccc}
\mathrm{i}\vartheta _{1,1} & \mathrm{i}\vartheta _{1,2}-\varpi _{1,2} & 
\mathrm{i}\vartheta _{1,3}-\varpi _{1,3} & \mathrm{i}\vartheta _{1,4}-\varpi
_{1,4} \\ 
\mathrm{i}\vartheta _{1,2}+\varpi _{1,2} & \mathrm{i}\vartheta _{2,2} & 
\mathrm{i}\vartheta _{2,3}-\varpi _{2,3} & \mathrm{i}\vartheta _{2,4}-\varpi
_{2,4} \\ 
\mathrm{i}\vartheta _{1,3}+\varpi _{1,3} & \mathrm{i}\vartheta _{2,3}+\varpi
_{2,3} & \mathrm{i}\vartheta _{3,3} & \mathrm{i}\vartheta _{3,4}-\varpi
_{3,4} \\ 
\mathrm{i}\vartheta _{1,4}+\varpi _{1,4} & \mathrm{i}\vartheta _{2,4}+\varpi
_{2,4} & \mathrm{i}\vartheta _{3,4}+\varpi _{3,4} & \mathrm{i}\vartheta
_{4,4}
\end{array}
\right] 
\]

с $\varpi _{s,k}={\rm Re}\left( Q_{s,k}\right) $ и $\vartheta _{s,k}=%
{\rm Im}\left(Q_{s,k}\right) $.

Пусть $\Theta _0$, $\Theta _3$, $\Upsilon _0$ и $\Upsilon _3$ - решение 
следующей системы уравнений:

\[
\left\{ 
\begin{array}{c}
{-\Theta _0+\Theta _3-\Upsilon _0+\Upsilon _3}{}{=\vartheta _{1,1}}\mbox{;}
\\ 
{-\Theta _0-\Theta _3-\Upsilon _0-\Upsilon _3}{}{=\vartheta _{2,2}}\mbox{;}
\\ 
{-\Theta _0-\Theta _3+\Upsilon _0+\Upsilon _3}{}{=\vartheta _{3,3}}\mbox{;}
\\ 
{-\Theta _0+\Theta _3+\Upsilon _0-\Upsilon _3}{}{=\vartheta _{4,4}}
\end{array}
\right| \mbox{,} 
\]

ф $\Theta _1$, $\Upsilon _1$, $\Theta _2$, $\Upsilon _2$, ${M_0}$, ${M_4}$%
, ${M_{1,0}}$, ${M_{1,4}}$, ${M_{2,0}}$, ${M_{2,4}}$, ${M_{3,0}}$, ${M_{3,4}}
$ - решения следующих систем:

\[
\left\{ 
\begin{array}{c}
{\ \Theta _1+\Upsilon _1}{}{=\vartheta _{1,2}}\mbox{;} \\ 
{-\Theta _1+\Upsilon _1}{}{=\vartheta _{3,4}}\mbox{;}
\end{array}
\right| 
\]

\[
\left\{ 
\begin{array}{c}
{-\Theta _2-\Upsilon _2}{}{=\varpi _{1,2}}\mbox{;} \\ 
{\Theta _2-\Upsilon _2}{}{=\varpi _{3,4}}\mbox{;}
\end{array}
\right| 
\]

\[
\left\{ 
\begin{array}{c}
{M_0+M_{3,0}}{}{=\vartheta _{1,3}}\mbox{;} \\ 
{M_0-M_{3,0}}{}{=\vartheta _{2,4}}\mbox{;}
\end{array}
\right| 
\]

\[
\left\{ 
\begin{array}{c}
{M_4+M_{3,4}}{}{=\varpi _{1,3}}\mbox{;} \\ 
{M_4-M_{3,4}}{}{=\varpi _{2,4}}\mbox{;}
\end{array}
\right| 
\]

\[
\left\{ 
\begin{array}{c}
{M_{1,0}-M_{2,4}}{}{=\vartheta _{1,4}}\mbox{;} \\ 
{M_{1,0}+M_{2,4}}{}{=\vartheta _{2,3}}\mbox{;}
\end{array}
\right| 
\]

\[
\left\{ 
\begin{array}{c}
{M_{1,4}-M_{2,0}}{}{=\varpi _{1,4}}\mbox{;} \\ 
{M_{1,4}+M_{2,0}}{}{=\varpi _{2,3}}
\end{array}
\right|\mbox{.} 
\]

Из (\ref{ham1}):

\begin{eqnarray}
&&\left( \partial _t+\mathrm{i}\Theta _0+\mathrm{i}\Upsilon _0\gamma
^{\left[ 5\right] }\right) \varphi =  \nonumber \\
&=&\left( 
\begin{array}{c}
\sum_{k=1}^3\beta ^{\left[ k\right] }\left( \partial _k+\mathrm{i}\Theta _k+%
\mathrm{i}\Upsilon _k\gamma ^{\left[ 5\right] }\right) +\mathrm{i}M_0\gamma
^{\left[ 0\right] }+\mathrm{i}M_4\beta ^{\left[ 4\right] } \\ 
-\mathrm{i}M_{1,0}\gamma _\zeta ^{\left[ 0\right] }-\mathrm{i}M_{1,4}\zeta
^{\left[ 4\right] }- \\ 
-\mathrm{i}M_{2,0}\gamma _\eta ^{\left[ 0\right] }-\mathrm{i}M_{2,4}\eta
^{\left[ 4\right] }- \\ 
-\mathrm{i}M_{3,0}\gamma _\theta ^{\left[ 0\right] }-\mathrm{i}M_{3,4}\theta
^{\left[ 4\right] }
\end{array}
\right) \varphi \label{ham0}
\end{eqnarray}

с

\[
\gamma ^{\left[ 5\right] }\stackrel{def}{=}\left[ 
\begin{array}{cc}
1_2 & 0_2 \\ 
0_2 & -1_2
\end{array}
\right] \mbox{.}
\]

Здесь слагаемые

\[
\begin{array}{c}
-\mathrm{i}M_{1,0}\gamma _\zeta ^{\left[ 0\right] }-\mathrm{i}M_{1,4}\zeta
^{\left[ 4\right] }- \\ 
-\mathrm{i}M_{2,0}\gamma _\eta ^{\left[ 0\right] }-\mathrm{i}M_{2,4}\eta
^{\left[ 4\right] }- \\ 
-\mathrm{i}M_{3,0}\gamma _\theta ^{\left[ 0\right] }-\mathrm{i}M_{3,4}\theta
^{\left[ 4\right] }
\end{array}
\]

содержат элементы цветных пентад, а

\[
\sum_{k=1}^3\beta ^{\left[ k\right] }\left( \partial _k+\mathrm{i}\Theta _k+%
\mathrm{i}\Upsilon _k\gamma ^{\left[ 5\right] }\right) +\mathrm{i}M_0\gamma
^{\left[ 0\right] }+\mathrm{i}M_4\beta ^{\left[ 4\right] } 
\]

содержит только элементы легкой пентады. Я называю сумму

\begin{equation}
\widehat{H}_l\stackrel{def}{=}\sum_{k=1}^3\beta ^{\left[ k\right] }\left( 
\mathrm{i}\partial _k-\Theta _k-\Upsilon _k\gamma ^{\left[ 5\right] }\right)
-M_0\gamma ^{\left[ 0\right] }-M_4\beta ^{\left[ 4\right] } \label{oh}
\end{equation}

{\it лептоннным (3 н) гамильтонианом}.

\subsection{Вращения системы $x_5Ox_4$ и $B$-бозонн}

Если определить (\ref{j}):

\begin{center}
$\varphi ^{\dagger }\gamma ^{\left[ 0\right] }\varphi \stackrel{def}{=}%
-j_{A,0}$ и $\varphi ^{\dagger }\beta ^{\left[ 4\right] }\varphi \stackrel{def}{=}%
-j_{A,4}$,
\end{center}

и (\ref{v2}):

\begin{equation}
\rho _A u_{A,4}\stackrel{def}{=}j_{A,4}\mbox{ и }\rho_A u_{A,5}\stackrel{def}%
{=}j_{A,5}\mbox{,}\label{5u} 
\end{equation}

то

\begin{eqnarray*}
&&-u_{A,5}=\sin 2\stackrel{*}{\alpha }\left( \sin \stackrel{*}{\beta }\sin 
\stackrel{*}{\chi }\cos \left( -\stackrel{*}{\theta }+\stackrel{*}{\upsilon }%
\right) +\cos \stackrel{*}{\beta }\cos \stackrel{*}{\chi }\cos \left( 
\stackrel{*}{\gamma }-\stackrel{*}{\lambda }\right) \right)\mbox{,}\\
&&-u_{A,4}=\sin 2\stackrel{*}{\alpha }\left( -\sin \stackrel{*}{\beta }\sin 
\stackrel{*}{\chi }\sin \left( -\stackrel{*}{\theta }+\stackrel{*}{\upsilon }%
\right) +\cos \stackrel{*}{\beta }\cos \stackrel{*}{\chi }\sin \left( 
\stackrel{*}{\gamma }-\stackrel{*}{\lambda }\right) \right)\mbox{.}
\end{eqnarray*}

Поэтому из (\ref{abc}):

\[
u_{A,1}^2+u_{A,2}^2+u_{A,3}^2+u_{A,4}^2+u_{A,5}^2=1\mbox{.} 
\]

Следовательно, только все пять элементов клиффордовой пентады дают полный набор 
компонент скорости. Повидимому стоит к нашим трем пространст-венным координатам 
$x_1,x_2,x_3$ добавить еще две квазипространственные координа-ты $x_5$ и $x_4$. 
Эти дополнительные координаты могут быть выбраны такими, чтобы 

\[
-\frac \pi h\leq x_5\leq \frac \pi h,-\frac \pi h\leq x_4\leq \frac \pi h%
\mbox{.} 
\]

$x_4$ и $x_5$ не являются координатами каких-нибудь событий. Поэтому наши 
приборы не обнаруживают их как пространственные координаты. 

Пусть:

\begin{eqnarray}
&&\widetilde{\varphi }\left( t,x_1,x_2,x_3,x_5,x_4\right) \stackrel{def}{=}%
\varphi \left( t,x_1,x_2,x_3\right) \cdot  \nonumber \\
&&\cdot \left( \exp \left( -\mathrm{i}\left( x_5M_0\left(
t,x_1,x_2,x_3\right) +x_4M_4\left( t,x_1,x_2,x_3\right) \right) \right)
\right) \mbox{.}  \nonumber
\end{eqnarray}

В этом случае уравнение движения с лептоннным гамильтонианом (\ref{oh}) имеет
следующий вид:

\begin{equation}
\left( \sum_{\mu =0}^3\beta ^{\left[ \mu \right] }\left( \mathrm{i}\partial
_\mu -\Theta _\mu -\Upsilon _\mu \gamma ^{\left[ 5\right] }\right) +\gamma
^{\left[ 0\right] }\mathrm{i}\partial _5+\beta ^{\left[ 4\right] }\mathrm{i}%
\partial _4\right) \widetilde{\varphi }=0.  \label{gkk}
\end{equation}

Пусть $g_1$ - некоторое положительное вещественное число, и для $\mu \in \left\{
0,1,2,3\right\} $: $F_\mu $ и $B_\mu $ представляет решение следующей системы 
уравнений:

\[
\left\{ 
\begin{array}{c}
{-0.5g_1B_\mu +F_\mu }{}{=-\Theta _\mu -\Upsilon _\mu ,}\mbox{;} \\ 
{-g_1B_\mu +F_\mu }{}{=-\Theta _\mu +\Upsilon _\mu }\mbox{.}
\end{array}
\right| 
\]

\textit{Матрицу заряда} определяем следующим образом:

\[
Y\stackrel{Def}{=}-\left[ 
\begin{array}{cc}
1_2 & 0_2 \\ 
0_2 & 2\cdot 1_2
\end{array}
\right] \mbox{.} 
\]

Следовательно, из (\ref{gkk}):

\begin{equation}
\left( \sum_{\mu =0}^3\beta ^{\left[ \mu \right] }\left( \mathrm{i}\partial
_\mu +F_\mu +0.5g_1YB_\mu \right) +\gamma ^{\left[ 0\right] }\mathrm{i}%
\partial _5+\beta ^{\left[ 4\right] }\mathrm{i}\partial _4\right) \widetilde{%
\varphi }=0\mbox{.}  \label{gkB}
\end{equation}

Пусть $\chi \left( t,x_1,x_2,x_3\right) $ - вещественная функция, и:

\begin{equation}
\widetilde{U}\left( \chi \right) \stackrel{def}{=}\left[ 
\begin{array}{cc}
\exp \left( \mathrm{i}\frac \chi 2\right) 1_2 & 0_2 \\ 
0_2 & \exp \left( \mathrm{i}\chi \right) 1_2
\end{array}
\right] \mbox{.}  \label{ux}
\end{equation}

Так как

\[
\partial _\mu \widetilde{U}=-\mathrm{i}\frac{\partial _\mu \chi }2Y%
\widetilde{U} 
\]

и

\begin{eqnarray*}
&&\widetilde{U}^{\dagger }\gamma ^{\left[ 0\right] }\widetilde{U}=\gamma
^{\left[ 0\right] }\cos \frac \chi 2+\beta ^{\left[ 4\right] }\sin \frac
\chi 2\mbox{,} \\ 
&&\widetilde{U}^{\dagger }\beta ^{\left[ 4\right] }\widetilde{U}=\beta
^{\left[ 4\right] }\cos \frac \chi 2-\gamma ^{\left[ 0\right] }\sin \frac
\chi 2\mbox{,} \\ 
&&\widetilde{U}^{\dagger }\widetilde{U}=1_4\mbox{,} \\ 
&&\widetilde{U}^{\dagger }Y\widetilde{U}=Y\mbox{,} \\ 
&&\beta ^{\left[ k\right] }\widetilde{U}=\widetilde{U}\beta ^{\left[ k\right] }
\end{eqnarray*}

для $k\in \left\{ 1,2,3\right\} $,

то уравнение движения (\ref{gkB}) инвариантно относительно следующего 
преобра-зования (повороты $x_4Ox_5$):

\begin{eqnarray}
&&x_4\rightarrow x_4^{\prime }=x_4\cos \frac \chi 2-x_5\sin \frac \chi 2%
\mbox{;}\nonumber \\ 
&&x_5\rightarrow x_5^{\prime }=x_5\cos \frac \chi 2+x_4\sin \frac \chi 2%
\mbox{;}\nonumber \\ 
&&x_\mu \rightarrow x_\mu ^{\prime }=x_\mu \mbox{ для }\mu \in \left\{
0,1,2,3\right\} \mbox{;}\nonumber \\ 
&&Y\rightarrow Y^{\prime }=\widetilde{U}^{\dagger }Y\widetilde{U}=Y\mbox{;}\label{T} \\ 
&&\widetilde{\varphi }\rightarrow \widetilde{\varphi }^{\prime }=\widetilde{U}%
\widetilde{\varphi }\mbox{,}\nonumber \\ 
&&B_\mu \rightarrow B_\mu ^{\prime }=B_\mu -\frac 1{g_1}\partial _\mu \chi %
\mbox{,}\nonumber \\ 
&&F_\mu \rightarrow F_\mu ^{\prime }=F_\mu \mbox{.}\nonumber
\end{eqnarray}

Следовательно, $B_\mu $ подобно $B$-бозонному полю Стандартной Модели. Я называю 
это поле $B$-{\it бозоннным}.

\subsection{Массы}

Пусть $\epsilon _\mu $ ($\mu \in \left\{ 1,2,3,4\right\} $) - базис, в котором 
легкая пентада имеет форму (\ref{lghr}).

Спинорные функции типа

\[
\frac h{2\pi }\exp \left( -\mathrm{i}h\left( sx_4+nx_5\right) \right)
\epsilon _k 
\]

с целыми $n$ и $s$ образуют ортнормированный базиc некоторого линейного 
пространства $\Im $ со следующим скалярным произведением:

\begin{equation}
\widetilde{\varphi }*\widetilde{\chi }\stackrel{def}{=}\int_{-\frac \pi
h}^{\frac \pi h}dx_5\int_{-\frac \pi h}^{\frac \pi h}dx_4\cdot \left( 
\widetilde{\varphi }^{\dagger }\cdot \widetilde{\chi }\right) \mbox{.}
\label{sp}
\end{equation}

В этом случае из (\ref{j}):

\begin{equation}
\widetilde{\varphi }*\beta ^{\left[ \mu \right] }\widetilde{\varphi }=-j_{A,\mu}
\label{jax}
\end{equation}

для $\mu \in \left\{ 0,1,2,3\right\} $.

Гамильтониан называем {\it планковым гамильтонианом}, если существуют фун-кции 
 $N_\vartheta \left(t,x_1,x_2,x_3\right)$ и $N_\varpi \left( t,x_1,x_2,x_3%
\right) $, имеющие область значений в множестве целых чисел, для которых:

\[
M_0=N_\vartheta h\mbox{ и }M_4=N_\varpi h\mbox{.} 
\]

В этом случае ряд Фурье для $\widetilde{\varphi }$ имеет следующий вид:

\[
\begin{array}{c}
\widetilde{\varphi }\left( t,x_1,x_2,x_3,x_5,x_4\right) = \\ 
=\varphi \left( t,x_1,x_2,x_3\right) \cdot \\ 
\cdot \sum_{n,s}\delta _{-n,N_\vartheta \left( t,\mathbf{x}\right) }\delta
_{-s,N_\varpi \left( t,\mathbf{x}\right) }\exp \left( -\mathrm{i}h\left(
nx_5+sx_4\right) \right)\mbox{,}
\end{array}
\]

где:

\begin{eqnarray*}
\delta _{-n,N_\vartheta } &=&\frac h{2\pi }\int_{-\frac \pi h}^{\frac \pi
h}\exp \left( \mathrm{i}h\left( nx_5\right) \right) \exp \left( \mathrm{i}%
N_\vartheta hx_5\right) dx_5=\frac{\sin \left( \pi \left( n+N_\vartheta
\right) \right) }{\pi \left( n+N_\vartheta \right) }\mbox{,} \\
\delta _{-s,N_\varpi } &=&\frac h{2\pi }\int_{-\frac \pi h}^{\frac \pi
h}\exp \left( \mathrm{i}h\left( sx_4\right) \right) \exp \left( \mathrm{i}%
N_\varpi hx_4\right) dx_4=\frac{\sin \left( \pi \left( s+N_\varpi \right)
\right) }{\pi \left( s+N_\varpi \right) }
\end{eqnarray*}

с целыми $n$ и $s$.

Если обозначить:

\[
\phi \left( t,\mathbf{x},-n,-s\right) \stackrel{Def}{=}\varphi \left( t,%
\mathbf{x}\right) \delta _{n,N_\vartheta \left( t,\mathbf{x}\right) }\delta
_{s,N_\varpi \left( t,\mathbf{x}\right) \mbox{,} } 
\]

то

\begin{equation}
\begin{array}{c}
\widetilde{\varphi }\left( t,\mathbf{x},x_5,x_4\right) = \\ 
=\sum_{n,s}\phi \left( t,\mathbf{x},n,s\right) \exp \left( -\mathrm{i}%
h\left( nx_5+sx_4\right) \right) \mbox{.}
\end{array}
\label{lt}
\end{equation}

Целые числа $n$ и $s$ называются \textit{массовыми числами}.

Из свойств функции $\delta $: в каждой точке $\left\langle t,\mathbf{x}%
\right\rangle $: либо

\[
\widetilde{\varphi }\left( t,\mathbf{x},x_5,x_4\right) =0 \mbox{,}
\]

либо целые числа $n_0$ и $s_0$ существуют, для которых:

\begin{equation}
\begin{array}{c}
\widetilde{\varphi }\left( t,\mathbf{x},x_5,x_4\right) = \\ 
=\phi \left( t,\mathbf{x},n_0,s_0\right) \exp \left( -\mathrm{i}h\left(
n_0x_5+s_0x_4\right) \right) \mbox{.}
\end{array}
\label{dlt}
\end{equation}

Здесь если

\[
m_0\stackrel{Def}{=}\sqrt{n_0^2+s_0^2} 
\]

то

\[
m\stackrel{Def}{=}hm_0 
\]

называется \textit{массой} функции $\widetilde{\varphi }$.

То есть для каждой точки пространства-времени: либо эта точка пустая,
либо в этой точке помещается единственная масса.

Уравнение движения (\ref{gkB}) при преобразовании (\ref{T}) принимает
следующую форму:

\[
\begin{array}{c}
\sum_{n^{\prime },s^{\prime }}\left( \sum_{\mu =0}^3\beta ^{\left[ \mu
\right] }\left( \mathrm{i}\partial _\mu +F_\mu +0.5g_1YB_\mu \gamma ^{\left[
5\right] }\right) +\gamma ^{\left[ 0\right] }\mathrm{i}\partial _5^{\prime
}+\beta ^{\left[ 4\right] }\mathrm{i}\partial _4^{\prime }\right) \cdot \\ 
\cdot \exp \left( -\mathrm{i}h\left( n^{\prime }x_5+s^{\prime }x_4\right)
\right) \widetilde{U}\phi =0
\end{array}
\]

с:

\[
\begin{array}{c}
n^{\prime }=n\cos \frac \chi 2-s\sin \frac \chi 2\mbox{,} \\ 
s^{\prime }=n\sin \frac \chi 2+s\cos \frac \chi 2\mbox{.}
\end{array}
\]

Но $s$ и $n$ являются целыми числами, и $s^{\prime }$ и $n^{\prime }$
должны оставаться тоже целыми.

Тройка $\left\langle \lambda ;n,s\right\rangle $ целых чисел называется
\textit{пифагоровой тройкой} \cite{Pf}, если

\[
\lambda ^2=n^2+s^2\mbox{.} 
\]

Пусть $\varepsilon $ - маленькое положительное вещественное число. 
Целое число $\lambda $ называется \textit{число-отец с точностью} 
$\varepsilon $ если для каждого вещественного числа $\chi $ и для
каждой пифагоровой тройки $\left\langle\lambda ;n,s\right\rangle $: 
существует пифагорова тройка 
$\left\langle \lambda;n^{\prime },s^{\prime }\right\rangle $,
для которой:

\[
\begin{array}{c}
\left| -s\sin \frac \chi 2+n\cos \frac \chi 2-n^{\prime }\right|
<\varepsilon \mbox{,} \\ 
\left| s\cos \frac \chi 2+n\sin \frac \chi 2-s^{\prime }\right| <\varepsilon %
\mbox{.}
\end{array}
\]

\textit{Для любого} $\varepsilon $\textit{: существует бесконечно много 
чисел-отцов с точностью }$\varepsilon $.

\subsection{Одномассовые состояния, частицы и \\античастицы}

Пусть (\ref{lt}):

\[
\widetilde{\varphi }\left( t,\mathbf{x},x_5,x_4\right) =\exp \left( -\mathrm{%
i}hnx_5\right) \sum_{k=1}^4\phi _k\left( t,\mathbf{x},n,0\right) \epsilon _k%
\mbox{.} 
\]

В этом случае гамильтониан имеет следующий вид (из \ref{gkB}):

\[
\widehat{H}=\sum_{k=1}^3\beta ^{\left[ k\right] }\mathrm{i}\partial
_k+hn\gamma ^{\left[ 0\right] }+\widehat{G} 
\]

с

\[
\widehat{G}\stackrel{Def}{=}\sum_{\mu =0}^3\beta ^{\left[ \mu \right]
}\left( F_\mu +0.5g_1YB_\mu \right) \mbox{.} 
\]

Если

\begin{equation}
\widehat{H}_0\stackrel{Def}{=}\sum_{k=1}^3\beta ^{\left[ k\right] }\mathrm{i}%
\partial _k+hn\gamma ^{\left[ 0\right] } \mbox{,}  \label{hmm0}
\end{equation}

то функции

\[
u_1\left( \mathbf{k}\right) \exp \left( -\mathrm{i}h\mathbf{kx}\right) 
\mbox{
и }u_2\left( \mathbf{k}\right) \exp \left( -\mathrm{i}h\mathbf{kx}\right) 
\]

с

\[
u_1\left( \mathbf{k}\right) \stackrel{Def}{=}\frac 1{2\sqrt{\omega\left( \mathbf{k%
}\right) \left( \omega\left( \mathbf{k}\right) +n\right) }}\left[ 
\begin{array}{c}
\omega\left( \mathbf{k}\right) +n+k_3 \\ 
k_1+\mathrm{i}k_2 \\ 
\omega\left( \mathbf{k}\right) +n-k_3 \\ 
-k_1-\mathrm{i}k_2
\end{array}
\right] 
\]

и

\[
u_2\left( \mathbf{k}\right) \stackrel{Def}{=}\frac 1{2\sqrt{\omega\left( \mathbf{k%
}\right) \left( \omega\left( \mathbf{k}\right) +n\right) }}\left[ 
\begin{array}{c}
k_1-\mathrm{i}k_2 \\ 
\omega\left( \mathbf{k}\right) +n-k_3 \\ 
-k_1+\mathrm{i}k_2 \\ 
\omega\left( \mathbf{k}\right) +n+k_3
\end{array}
\right] 
\]

являются собственными векторами для $\widehat{H}_0$ с собственным значением
$\omega\left( \mathbf{k}\right) \stackrel{Def}{=}\sqrt{\mathbf{k}^2+n^2}$, 
а функции

\[
u_3\left( \mathbf{k}\right) \exp \left( -\mathrm{i}h\mathbf{kx}\right) 
\mbox{
и }u_4\left( \mathbf{k}\right) \exp \left( -\mathrm{i}h\mathbf{kx}\right) 
\]

с

\[
u_3\left( \mathbf{k}\right) \stackrel{Def}{=}\frac 1{2\sqrt{\omega\left( \mathbf{k%
}\right) \left( \omega\left( \mathbf{k}\right) +n\right) }}\left[ 
\begin{array}{c}
-\omega\left( \mathbf{k}\right) -n+k_3 \\ 
k_1+\mathrm{i}k_2 \\ 
\omega\left( \mathbf{k}\right) +n+k_3 \\ 
k_1+\mathrm{i}k_2
\end{array}
\right] 
\]

и

\[
u_4\left( \mathbf{k}\right) \stackrel{Def}{=}\frac 1{2\sqrt{\omega\left( \mathbf{k%
}\right) \left( \omega\left( \mathbf{k}\right) +n\right) }}\left[ 
\begin{array}{c}
k_1-\mathrm{i}k_2 \\ 
-\omega\left( \mathbf{k}\right) -n-k_3 \\ 
k_1-\mathrm{i}k_2 \\ 
\omega\left( \mathbf{k}\right) +n-k_3
\end{array}
\right] 
\]

- собственными векторами для $\widehat{H}_0$ с собственными значениями 
$-\omega\left( \mathbf{k}\right) $.

Здесь $u_\mu \left( \mathbf{k}\right) $ формируют ортонормированный базис
в пространстве, натянутом на векторы $\epsilon _\mu $.

Пусть:

\[
b_{r,\mathbf{k}}\stackrel{Def}{=}\left( \frac h{2\pi }\right)
^3\sum_{j^{\prime }=1}^4\int_{\left( V\right) }d^3\mathbf{x}^{\prime }\cdot
e^{\mathrm{i}h\mathbf{kx}^{\prime }}u_{r,j^{\prime }}^{*}\left( \mathbf{k}%
\right) \psi _{j^{\prime }}\left( \mathbf{x}^{\prime }\right) \mbox{.} 
\]

В этом случае так как

\[
\sum_{r=1}^4u_{r,j}^{*}\left( \mathbf{k}\right) u_{r,j^{\prime }}\left( 
\mathbf{k}\right) =\delta _{j,j^{\prime }} \mbox{,} 
\]

то

\begin{equation}
\psi _j\left( \mathbf{x}\right) =\sum_{\mathbf{k}}e^{-\mathrm{i}h\mathbf{kx}%
}\sum_{r=1}^4b_{r,\mathbf{k}}u_{r,j}\left( \mathbf{k}\right)  \label{bb}
\end{equation}

и

\begin{equation}
\begin{array}{c}
\left\{ b_{s,\mathbf{k}^{\prime }}^{\dagger },b_{r,\mathbf{k}}\right\}
=\left( \frac h{2\pi }\right) ^3\delta _{s,r}\delta _{\mathbf{k},\mathbf{k}%
^{\prime }}\mbox{,} \\ 
\left\{ b_{s,\mathbf{k}^{\prime }}^{\dagger },b_{r,\mathbf{k}}^{\dagger
}\right\} =0=\left\{ b_{s,\mathbf{k}^{\prime }},b_{r,\mathbf{k}}\right\} %
\mbox{,} \\ 
b_{r,\mathbf{k}}\Phi _0=0\mbox{.}
\end{array}
\label{bubu}
\end{equation}

Плотность гамильтониана (\ref{hmm}) для $\widehat{H}_0$ имеет вид:

\[
\mathcal{H}_0\left( \mathbf{x}\right) =\sum_{j=1}^4\psi _j^{\dagger }\left( 
\mathbf{x}\right) \sum_{k=1}^4\widehat{H}_{0,j,k}\psi _k\left( \mathbf{x}%
\right) \mbox{.} 
\]

Следовательно, из (\ref{bb}):

\[
\int_{\left( V\right) }d^3\mathbf{x}\cdot \mathcal{H}_0\left( \mathbf{x}%
\right) =\left( \frac{2\pi }h\right) ^3\sum_{\mathbf{k}}h\omega\left( \mathbf{k}%
\right) \cdot \left( \sum_{r=1}^2b_{r,\mathbf{k}}^{\dagger }b_{r,\mathbf{k}%
}-\sum_{r=3}^4b_{r,\mathbf{k}}^{\dagger }b_{r,\mathbf{k}}\right) \mbox{.}
\]

Пусть преобразование Фурье для $\varphi $ имеет вид:

\[
\varphi _j\left( t,\mathbf{x}\right) =\sum_{\mathbf{p}}\sum_{r=1}^4c_r\left(
t,\mathbf{p}\right) u_{r,j}\left( \mathbf{p}\right) e^{-\mathrm{i}h\mathbf{px%
}} 
\]

с

\[
c_r\left( t,\mathbf{p}\right) \stackrel{Def}{=}\left( \frac h{2\pi }\right)
^3\sum_{j^{\prime }=1}^4\int_{\left( V\right) }d^3\mathbf{x}^{\prime }\cdot
u_{r,j^{\prime }}^{*}\left( \mathbf{p}\right) \varphi _{j^{\prime }}\left( t,%
\mathbf{x}^{\prime }\right) e^{\mathrm{i}h\mathbf{px}^{\prime }} 
\]

Я называю функцию $\varphi _j\left( t,\mathbf{x}\right) $ \textit{обычной},
если существует вещественное число $K$, для которого:

если $\left| p_1\right| >K$ или/и $\left| p_2\right| >K$ или/и $\left|
p_3\right| >K$, то $c_r\left( t,\mathbf{p}\right) =0$.

В этом случае обозначаем:

\[
\sum_{\mathbf{p\in \Xi }}\stackrel{Def}{=}\sum_{p_1=-L}^L\sum_{p_2=-L}^L%
\sum_{p_3=-L}^L \mbox{.}
\]

Если $\varphi _j\left( t,\mathbf{x}\right) $ - обычные функции, то

\[
\varphi _j\left( t,\mathbf{x}\right) =\sum_{\mathbf{p\in \Xi }%
}\sum_{r=1}^4c_r\left( t,\mathbf{p}\right) u_{r,j}\left( \mathbf{p}\right)
e^{-\mathrm{i}h\mathbf{px}}\mbox{.} 
\]

Следовательно, из (\ref{Sat}):

\[
\Psi \left( t,\mathbf{x}\right) =\sum_{\mathbf{p}}\sum_{r=1}^4\sum_{\mathbf{k%
}}\sum_{r^{\prime }=1}^4c_r\left( t,\mathbf{p}\right) e^{\mathrm{i}h\left( 
\mathbf{k}-\mathbf{p}\right) \mathbf{x}}\sum_{j=1}^4u_{r^{\prime
},j}^{*}\left( \mathbf{k}\right) u_{r,j}\left( \mathbf{p}\right)
b_{r^{\prime },\mathbf{k}}^{\dagger }\Phi _0 
\]

и

\[
\int_{\left( V\right) }d^3\mathbf{x}\cdot \Psi \left( t,\mathbf{x}\right)
=\left( \frac{2\pi }h\right) ^3\sum_{\mathbf{p}}\sum_{r=1}^4c_r\left( t,%
\mathbf{p}\right) b_{r,\mathbf{p}}^{\dagger }\Phi _0 \mbox{.}
\]

Если обозначить:

\[
\widetilde{\Psi }\left( t,\mathbf{p}\right) \stackrel{Def}{=}\left( \frac{%
2\pi }h\right) ^3\sum_{r=1}^4c_r\left( t,\mathbf{p}\right) b_{r,\mathbf{p}%
}^{\dagger }\Phi _0 \mbox{,}
\]

то

\[
\int_{\left( V\right) }d^3\mathbf{x}\cdot \Psi \left( t,\mathbf{x}\right)
=\sum_{\mathbf{p}}\widetilde{\Psi }\left( t,\mathbf{p}\right) 
\]

и

\[
H_0\widetilde{\Psi }\left( t,\mathbf{p}\right) =\left( \frac{2\pi }h\right)
^3\sum_{\mathbf{k}}h\omega\left( \mathbf{k}\right) \cdot \left(
\sum_{r=1}^2c_r\left( t,\mathbf{k}\right) b_{r,\mathbf{k}}^{\dagger }\Phi
_0-\sum_{r=3}^4c_r\left( t,\mathbf{k}\right) b_{r,\mathbf{k}}^{\dagger }\Phi
_0\right) \mbox{.}
\]

На множестве обычных функций $H_0$ эквивалентен оператору:

\[
\stackrel{\Xi }{H}_0\stackrel{Def}{=}\left( \frac{2\pi }h\right) ^3\sum_{%
\mathbf{k\in \Xi }}h\omega\left( \mathbf{k}\right) \cdot \left( \sum_{r=1}^2b_{r,%
\mathbf{k}}^{\dagger }b_{r,\mathbf{k}}-\sum_{r=3}^4b_{r,\mathbf{k}}^{\dagger
}b_{r,\mathbf{k}}\right) \mbox{.} 
\]

Так как (из (\ref{bubu}))

\[
b_{r,\mathbf{k}}^{\dagger }b_{r,\mathbf{k}}=\left( \frac h{2\pi }\right)
^3-b_{r,\mathbf{k}}b_{r,\mathbf{k}}^{\dagger } \mbox{,}
\]

то

\begin{equation}
\stackrel{\Xi }{H}_0=\left( \frac{2\pi }h\right) ^3\sum_{\mathbf{k\in \Xi }%
}h\omega\left( \mathbf{k}\right) \left( \sum_{r=1}^2b_{r,\mathbf{k}}^{\dagger
}b_{r,\mathbf{k}}+\sum_{r=3}^4b_{r,\mathbf{k}}b_{r,\mathbf{k}}^{\dagger
}\right) -h\sum_{\mathbf{k\in \Xi }}\omega\left( \mathbf{k}\right) \mbox{.}
\label{uW}
\end{equation}

Пусть:

\begin{equation}
\begin{array}{c}
v_{\left( 1\right) }\left( \mathbf{k}\right) \stackrel{Def}{=}\gamma
^{\left[ 0\right] }u_3\left( \mathbf{k}\right) \mbox{,} \\ 
v_{\left( 2\right) }\left( \mathbf{k}\right) \stackrel{Def}{=}\gamma
^{\left[ 0\right] }u_4\left( \mathbf{k}\right) \mbox{,} \\ 
u_{\left( 1\right) }\left( \mathbf{k}\right) \stackrel{Def}{=}u_1\left( 
\mathbf{k}\right) \mbox{,} \\ 
u_{\left( 2\right) }\left( \mathbf{k}\right) \stackrel{Def}{=}u_2\left( 
\mathbf{k}\right) \mbox{.}
\end{array}
\label{xa}
\end{equation}

и пусть:

\[
\begin{array}{c}
d_1\left( \mathbf{k}\right) \stackrel{Def}{=}-b_3^{\dagger }\left( -\mathbf{k%
}\right) \mbox{,} \\ 
d_2\left( \mathbf{k}\right) \stackrel{Def}{=}-b_4^{\dagger }\left( -\mathbf{k%
}\right) \mbox{.}
\end{array}
\]

В этом случае:

\begin{center}
\[
\psi _j\left( \mathbf{x}\right) =\sum_{\mathbf{k}}\sum_{\alpha =1}^2\left(
e^{-\mathrm{i}h\mathbf{kx}}b_{\alpha ,\mathbf{k}}u_{\left( \alpha \right)
,j}\left( \mathbf{k}\right) +e^{\mathrm{i}h\mathbf{kx}}d_{\alpha ,\mathbf{k}%
}^{\dagger }v_{\left( \alpha \right) ,j}\left( \mathbf{k}\right) \right) 
\]
\end{center}

и из (\ref{uW}) Wick-упорядоченный гамильтониан имеет следующую форму: 

\[
:\stackrel{\Xi }{H}_0:=\left( \frac{2\pi }h\right) ^3h\sum_{\mathbf{k\in \Xi 
}}\omega\left( \mathbf{k}\right) \sum_{\alpha =1}^2\left( b_{\alpha ,\mathbf{k}%
}^{\dagger }b_{\alpha ,\mathbf{k}}+d_{\alpha ,\mathbf{k}}^{\dagger
}d_{\alpha ,\mathbf{k}}\right) \mbox{.} 
\]

Здесь $b_{\alpha ,\mathbf{k}}^{\dagger }$ - \textit{оператор рождения},
а $b_{\alpha ,\mathbf{k}}$ - \textit{оперетор уничтожения} $n$-%
\textit{лептонна} с \textit{импульсом} $\mathbf{k}$ и \textit{спиновым
индексом} $\alpha $; $d_{\alpha ,\mathbf{k}}^{\dagger }$ - \textit{оператор
рождения}, а $d_{\alpha ,\mathbf{k}}$ - \textit{оперетор уничтожения}
{\it анти}-$n$-\textit{лептонна} с \textit{импульсом} $\mathbf{k}$ и 
\textit{спиновым индексом} $\alpha $.

Функции:

\[
u_{\left( 1\right) }\left( \mathbf{k}\right) \exp \left( -\mathrm{i}h\mathbf{%
kx}\right) \mbox{
и }u_{\left( 2\right) }\left( \mathbf{k}\right) \exp \left( -\mathrm{i}h%
\mathbf{kx}\right) 
\]

называются \textit{базисными $n$-лептонными функциями} с импульсом $\mathbf{k}$, и

\[
v_{\left( 1\right) }\left( \mathbf{k}\right) \exp \left( \mathrm{i}h\mathbf{%
kx}\right) \mbox{ и}v_{\left( 2\right) }\left( \mathbf{k}\right) \exp \left( \mathrm{i}h%
\mathbf{kx}\right) 
\]

называются \textit{базисными анти-$n$-лептонными функциями} с импульсом 
$\mathbf{k}$.

\subsection{Двухмассовое состояние \cite{DVB}, \cite{AV}}

Рассматриваем подпространство $\Im _{\jmath }$ пространства $\Im $, натянутое на следующий подбазис:

\[
\jmath =\left\langle 
\begin{array}{c}
\frac h{2\pi }\exp \left( -\mathrm{i}h\left( s_0x_4\right) \right) \epsilon
_1,\frac h{2\pi }\exp \left( -\mathrm{i}h\left( s_0x_4\right) \right)
\epsilon _2, \\ 
\frac h{2\pi }\exp \left( -\mathrm{i}h\left( s_0x_4\right) \right) \epsilon
_3,\frac h{2\pi }\exp \left( -\mathrm{i}h\left( s_0x_4\right) \right)
\epsilon _4, \\ 
\frac h{2\pi }\exp \left( -\mathrm{i}h\left( n_0x_5\right) \right) \epsilon
_1,\frac h{2\pi }\exp \left( -\mathrm{i}h\left( n_0x_5\right) \right)
\epsilon _2, \\ 
\frac h{2\pi }\exp \left( -\mathrm{i}h\left( n_0x_5\right) \right) \epsilon
_3,\frac h{2\pi }\exp \left( -\mathrm{i}h\left( n_0x_5\right) \right)
\epsilon _4
\end{array}
\right\rangle 
\]

с некоторыми натуральными $s_0$ и $n_0$. 

Пусть $U$ - какое-нибудь линейное преобразование в пространстве $\Im _{\jmath }$ такое, что для любого $\widetilde{\varphi }$ : если $\widetilde{\varphi }%
\in \Im _{\jmath }$, то

\begin{equation}
-\left( U\widetilde{\varphi }\right) ^{\dagger }*\beta ^{\left[ \mu \right]
}\left( U\widetilde{\varphi }\right) =j_{A,\mu}  \label{uni}
\end{equation}

для $\mu \in \left\{ 0,1,2,3\right\} $ (\ref{jax}).

В этом случае:

\begin{center}
$U^{\dagger }\beta ^{\left[\mu\right] }U=\beta ^{\left[
\mu\right] }$. \footnote{Я применяю следующее правило действий: если 
$V_{k,s}$ и $\beta$ - $4\times 4$-матрицы, и 

\begin{center}
$V\stackrel{def}{=}\left[ 
\begin{array}{cc}
V_{1,1} & V_{1,2} \\ 
V_{2,1} & V_{2,2}
\end{array}
\right] $,
\end{center}

то 

\begin{center}
$\beta V\stackrel{def}{=}\left[ 
\begin{array}{cc}
\beta V_{1,1} & \beta V_{1,2} \\ 
\beta V_{2,1} & \beta V_{2,2}
\end{array}
\right] $
\end{center}

и

\begin{center}
$ V\beta\stackrel{def}{=}\left[ 
\begin{array}{cc}
V_{1,1} \beta & V_{1,2} \beta \\ 
V_{2,1} \beta & V_{2,2} \beta
\end{array}
\right] $.
\end{center}

}
 
\end{center}

Для каждого такого преобразования $U$ существуют вещественные функции $\chi \left( t,%
\mathbf{x}\right) $, $\alpha \left( t,\mathbf{x}\right) $, $a\left( t,%
\mathbf{x}\right) $, $b\left( t,\mathbf{x}\right) $, $c\left( t,\mathbf{x}%
\right) $, $q\left( t,\mathbf{x}\right) $, $u\left( t,\mathbf{x}\right) $, $%
v\left( t,\mathbf{x}\right) $, $k\left( t,\mathbf{x}\right) $, $s\left( t,%
\mathbf{x}\right) $ такие, что

\[
U=\exp \left( \mathrm{i}\alpha \right) \widetilde{U}\left(\chi\right)
U^{\left( -\right) }U^{\left( +\right) }\mbox{;} 
\]

здесь $\widetilde{U}\left( \chi \right) $ определяется формулой (\ref{ux}), а 
$U^{\left( -\right) }$ и $U^{\left( +\right) }$ имеют следующую матричную форму 
в базисе $\jmath $:

\[
U^{\left( -\right)}\left( t,\mathbf{x}\right)\stackrel = \rm{S}\left( a\left(%
t,\mathbf{x}\right),b\left( t,\mathbf{x}\right),c\left( t,\mathbf{x}\right),q%
\left( t,\mathbf{x}\right)\right) 
\]

с

\[
a^2\left( t,\mathbf{x}\right)+b^2\left( t,\mathbf{x}\right)+c^2\left( t,\mathbf%
{x}\right)+q^2\left( t,\mathbf{x}\right)=1 
\]

и с

\[
\mathrm{S}\left( a,b,c,q\right) \stackrel{def}{=}\left[ 
\begin{array}{cccc}
\left( a+\mathrm{i}b\right) 1_2 & 0_2 & \left( c+\mathrm{i}q\right) 1_2 & 0_2
\\ 
0_2 & 1_2 & 0_2 & 0_2 \\ 
\left( -c+\mathrm{i}q\right) 1_2 & 0_2 & \left( a-\mathrm{i}b\right) 1_2 & 
0_2 \\ 
0_2 & 0_2 & 0_2 & 1_2
\end{array}
\right] \mbox{,} 
\]

и

\begin{equation}
U^{\left( +\right) }\left( t,\mathbf{x}\right)\stackrel =\rm{R}\left( u\left(%
 t,\mathbf{x}\right),v\left( t,\mathbf{x}\right),k\left( t,\mathbf{x}\right),%
s\left( t,\mathbf{x}\right)\right)
\label{upls}
\end{equation}

с

\[
u^2\left( t,\mathbf{x}\right)+v^2\left( t,\mathbf{x}\right)+k^2\left( t,\mathbf{x}\right)%
+s^2\left( t,\mathbf{x}\right)=1 
\]

и с

\begin{equation}
\mathrm{R}\left( u,v,k,s\right) \stackrel{def}{=}\left[ 
\begin{array}{cccc}
1_2 & 0_2 & 0_2 & 0_2 \\ 
0_2 & \left( u+\mathrm{i}v\right) 1_2 & 0_2 & \left( k+\mathrm{i}s\right) 1_2
\\ 
0_2 & 0_2 & 1_2 & 0_2 \\ 
0_2 & \left( -k+\mathrm{i}s\right) 1_2 & 0_2 & \left( u-\mathrm{i}v\right)
1_2
\end{array}
\right] \mbox{.}
\end{equation}

$U^{\left( +\right) }$ соответствует антилептоннам, т.к. $\mathrm{R}=\mathrm{S}\gamma ^{\left[
5\right] }$ (\ref{xa}).

Рассматриваем $U^{\left( -\right) }$.

Пусть:

\[
\ell _{\circ }\stackrel{def}{=}\imath _{\circ }\left( a,b,q,c\right) 
\mbox{,
}\ell _{*}\stackrel{def}{=}\imath _{*}\left( a,b,q,c\right) 
\]

с

\[
\imath _{\circ }\left( a,b,q,c\right) \stackrel{def}{=}\frac 1{2\sqrt{\left(
1-a^2\right) }}\left[ 
\begin{array}{cc}
\left( b+\sqrt{\left( 1-a^2\right) }\right) 1_4 & \left( q-\mathrm{i}%
c\right) 1_4 \\ 
\left( q+\mathrm{i}c\right) 1_4 & \left( \sqrt{\left( 1-a^2\right) }%
-b\right) 1_4
\end{array}
\right] 
\]

и

\[
\imath _{*}\left( a,b,q,c\right) \stackrel{def}{=}\frac 1{2\sqrt{\left(
1-a^2\right) }}\left[ 
\begin{array}{cc}
\left( \sqrt{\left( 1-a^2\right) }-b\right) 1_4 & \left( -q+\mathrm{i}%
c\right) 1_4 \\ 
\left( -q-\mathrm{i}c\right) 1_4 & \left( b+\sqrt{\left( 1-a^2\right) }%
\right) 1_4
\end{array}
\right] \mbox{.} 
\]

Эти операторы подчиняются следующим условиям:

\[
\begin{array}{c}
\ell _{\circ }\ell _{\circ }=\ell _{\circ }\mbox{, }\ell _{*}\ell _{*}=\ell
_{*}\mbox{;} \\ 
\ell _{\circ }\ell _{*}=0=\ell _{*}\ell _{\circ }\mbox{,} \\ 
\left( \ell _{\circ }-\ell _{*}\right) \left( \ell _{\circ }-\ell
_{*}\right) =1_8\mbox{,} \\ 
\ell _{\circ }+\ell _{*}=1_8\mbox{,}
\end{array}
\]

\[
\begin{array}{c}
\ell _{\circ }\gamma ^{\left[ 0\right] }=\gamma
^{\left[ 0\right] }\ell _{\circ }\mbox{, }\ell _{*}\gamma
^{\left[ 0\right] }=\gamma ^{\left[ 0\right] }\ell _{*}\mbox{,}
\\ 
\ell _{\circ }\beta ^{\left[ 4\right] }=\beta
^{\left[ 4\right] }\ell _{\circ }\mbox{, }\ell _{*}\beta
^{\left[ 4\right] }=\beta ^{\left[ 4\right] }\ell _{*}
\end{array}
\]

и

\begin{equation}
\begin{array}{c}
U^{\left( -\right) \dagger }\gamma ^{\left[ 0\right] }U^{\left(
-\right) }=a\gamma ^{\left[ 0\right] }-\left( \ell _{\circ
}-\ell _{*}\right) \sqrt{1-a^2}\beta ^{\left[ 4\right] }\mbox{,}
\\ 
U^{\left( -\right) \dagger }\beta ^{\left[ 4\right] }U^{\left(
-\right) }=a\beta ^{\left[ 4\right] }+\left( \ell _{\circ }-\ell
_{*}\right) \sqrt{1-a^2}\gamma ^{\left[ 0\right] }\mbox{.}
\end{array}
\label{gaa}
\end{equation}

Из (\ref{gkB}): лептонное уравнение движения имеет вид:

\[
\left( \sum_{\mu =0}^3\beta ^{\left[ \mu \right] }\left( \mathrm{%
i}\partial _\mu +F_\mu +0.5g_1 Y B_\mu \right) +\gamma
^{\left[ 0\right] }\mathrm{i}\partial _5+\beta ^{\left[ 4\right]
}\mathrm{i}\partial _4\right) U^{\left( -\right) \dagger }U^{\left(
-\right) }\widetilde{\varphi }=0\mbox{.} 
\]

Если

\begin{equation}
\partial _kU^{\left( -\right) \dagger }=U^{\left( -\right) \dagger }\partial
_k\label{aa} 
\end{equation}

для $k\in \left\{ 0,1,2,3,4,5\right\} $, то

\[
\left( 
\begin{array}{c}
U^{\left( -\right) \dagger }\mathrm{i}\sum_{\mu =0}^3\beta
^{\left[ \mu \right] }\left( \mathrm{i}\partial _\mu +F_\mu +0.5g_1%
Y B_\mu \right) \\ 
+\gamma ^{\left[ 0\right] }U^{\left( -\right) \dagger }\mathrm{i}%
\partial _5+\beta ^{\left[ 4\right] }U^{\left( -\right) \dagger }%
\mathrm{i}\partial _4
\end{array}
\right) U^{\left( -\right) }\widetilde{\varphi }=0\mbox{.} 
\]

Следовательно, из (\ref{gaa}):

\[
U^{\left( -\right) \dagger }\left( 
\begin{array}{c}
\sum_{\mu =0}^3\beta ^{\left[ \mu \right] }\left( \mathrm{i}%
\partial _\mu +F_\mu +0.5g_1 Y B_\mu \right) \\ 
+\gamma ^{\left[ 0\right] }\mathrm{i}\left( a\partial _5-\left(
\ell _{\circ }-\ell _{*}\right) \sqrt{1-a^2}\partial _4\right) \\ 
+\beta ^{\left[ 4\right] }\mathrm{i}\left( \sqrt{1-a^2}\left(
\ell _{\circ }-\ell _{*}\right) \partial _5+a\partial _4\right)
\end{array}
\right) U^{\left( -\right) }\widetilde{\varphi }=0\mbox{.} 
\]

Таким образом, если обозначить:

\[
\begin{array}{c}
x_4^{\prime }=\left( \ell _{\circ }+\ell _{*}\right) ax_4+\left( \ell
_{\circ }-\ell _{*}\right) \sqrt{1-a^2}x_5 \\ 
x_5^{\prime }=\left( \ell _{\circ }+\ell _{*}\right) ax_5-\left( \ell
_{\circ }-\ell _{*}\right) \sqrt{1-a^2}x_4 \mbox{,}
\end{array}
\]

то

\begin{equation}
\left( \sum_{\mu =0}^3\beta ^{\left[ \mu \right] }\left( \mathrm{%
i}\partial _\mu +F_\mu +0.5g_1 Y B_\mu \right) +\left( %
\gamma ^{\left[ 0\right] }\mathrm{i}\partial _5^{\prime }+\beta
^{\left[ 4\right] }\mathrm{i}\partial _4^{\prime }\right) \right) 
\widetilde{\varphi }^{\prime }=0\mbox{.}  \label{me8}
\end{equation}

с

\[
\widetilde{\varphi }^{\prime }=U^{\left( -\right) }\widetilde{\varphi }%
\mbox{.} 
\]

То есть лептонный гамильтониан инвариантен относительно следующего гло-бального 
преобразования:

\begin{eqnarray}
&&\widetilde{\varphi }\rightarrow \widetilde{\varphi }^{\prime }=U^{\left(
-\right) }\widetilde{\varphi }\mbox{,}  \nonumber \\
&&x_4\rightarrow x_4^{\prime }=\left( \ell _{\circ }+\ell _{*}\right)
ax_4+\left( \ell _{\circ }-\ell _{*}\right) \sqrt{1-a^2}x_5\mbox{,}
\label{glb} \\
&&x_5\rightarrow x_5^{\prime }=\left( \ell _{\circ }+\ell _{*}\right)
ax_5-\left( \ell _{\circ }-\ell _{*}\right) \sqrt{1-a^2}x_4\mbox{,} 
\nonumber \\
&&x_\mu \rightarrow x_\mu ^{\prime }=x_\mu \mbox{.}  \nonumber
\end{eqnarray}

\subsubsection{Нейтринно}

Пусть:

$
\begin{array}{c}
\widetilde{\varphi }\left( t,\mathbf{x},x_5,x_4\right) = \\ 
=\exp \left( -\mathrm{i}hs_0x_4\right) \sum_{r=1}^4\phi _{4,r}\left( t,%
\mathbf{x},0,s_0\right) \epsilon _r+\exp \left( -\mathrm{i}hn_0x_5\right)
\sum_{r=1}^4\phi _{5,r}\left( t,\mathbf{x},n_0,0\right) \epsilon _r
\end{array}
$

и

\begin{center}
$\widehat{H}_{0,4}\stackrel{Def.}{=}\sum_{r=1}^3\beta ^{\left[ r\right] }%
\mathrm{i}\partial _r+h\left( n_0\gamma ^{\left[ 0\right] }\kappa
_{n_0,0}^{\circ }+s_0\beta ^{\left[ 4\right] }\kappa _{0,s_0}^{\circ
}\right) $.
\end{center}

8-векторы в базисе $\jmath$:

\[
\underline{u}_1\left( \mathbf{k}\right) \stackrel{Def}{=}\frac 1{2\sqrt{%
\omega \left( \mathbf{k}\right) \left( \omega \left( \mathbf{k}\right)
+n_0\right) }}\left[ 
\begin{array}{c}
0 \\ 
0 \\ 
0 \\ 
0 \\ 
\omega \left( \mathbf{k}\right) +n_0+k_3 \\ 
k_1+\mathrm{i}k_2 \\ 
\omega \left( \mathbf{k}\right) +n_0-k_3 \\ 
-k_1-\mathrm{i}k_2
\end{array}
\right] 
\]

и

\[
\underline{u}_2\left( \mathbf{k}\right) \stackrel{Def}{=}\frac 1{2\sqrt{%
\omega \left( \mathbf{k}\right) \left( \omega \left( \mathbf{k}\right)
+n_0\right) }}\left[ 
\begin{array}{c}
0 \\ 
0 \\ 
0 \\ 
0 \\ 
k_1-\mathrm{i}k_2 \\ 
\omega \left( \mathbf{k}\right) +n_0-k_3 \\ 
-k_1+\mathrm{i}k_2 \\ 
\omega \left( \mathbf{k}\right) +n_0+k_3
\end{array}
\right] 
\]

соответствуют собственным векторам для $\widehat{H}_{0,4}$ с
собственным значением $\omega \left( \mathbf{k}\right) =\sqrt{\mathbf{k}^2+n_0^2}$, 
а 8-векторы

\[
\underline{u}_3\left( \mathbf{k}\right) \stackrel{Def}{=}\frac 1{2\sqrt{%
\omega \left( \mathbf{k}\right) \left( \omega \left( \mathbf{k}\right)
+n_0\right) }}\left[ 
\begin{array}{c}
0 \\ 
0 \\ 
0 \\ 
0 \\ 
-\omega \left( \mathbf{k}\right) -n_0+k_3 \\ 
k_1+\mathrm{i}k_2 \\ 
\omega \left( \mathbf{k}\right) +n_0+k_3 \\ 
k_1+\mathrm{i}k_2
\end{array}
\right] 
\]

и

\[
\underline{u}_4\left( \mathbf{k}\right) \stackrel{Def}{=}\frac 1{2\sqrt{%
\omega \left( \mathbf{k}\right) \left( \omega \left( \mathbf{k}\right)
+n_0\right) }}\left[ 
\begin{array}{c}
0 \\ 
0 \\ 
0 \\ 
0 \\ 
k_1-\mathrm{i}k_2 \\ 
-\omega \left( \mathbf{k}\right) -n_0-k_3 \\ 
k_1-\mathrm{i}k_2 \\ 
\omega \left( \mathbf{k}\right) +n_0-k_3
\end{array}
\right] 
\]

соответствуют собственным векторам для $\widehat{H}_{0,4}$ с
собственным значением $-\omega \left( \mathbf{k}\right) $.

Пусть

\[
\begin{array}{c}
\widehat{H}_{0,4}^{\prime }\stackrel{Def}{=}U^{\left( -\right) }%
\widehat{H}_{0,4}U^{\left( -\right) \dagger }\mbox{,} \\ 
\underline{u}_\mu ^{\prime }\left( \mathbf{k}\right) \stackrel{Def}{=}%
U^{\left( -\right) }\underline{u}_\mu \left( \mathbf{k}\right) \mbox{.}
\end{array}
\]

То есть

\[
\underline{u}_1^{\prime }\left( \mathbf{k}\right) =\frac 1{2\sqrt{\omega
\left( \mathbf{k}\right) \left( \omega \left( \mathbf{k}\right) +n\right) }%
}\left[ 
\begin{array}{c}
\left( c+\mathrm{i}q\right) \left( \omega \left( \mathbf{k}\right)
+n_0+k_3\right)  \\ 
\left( c+\mathrm{i}q\right) \left( k_1+\mathrm{i}k_2\right)  \\ 
0 \\ 
0 \\ 
\left( a-\mathrm{i}b\right) \left( \omega \left( \mathbf{k}\right)
+n_0+k_3\right)  \\ 
\left( a-\mathrm{i}b\right) \left( k_1+\mathrm{i}k_2\right)  \\ 
\omega \left( \mathbf{k}\right) +n_0-k_3 \\ 
-k_1-\mathrm{i}k_2
\end{array}
\right] \mbox{,}
\]

\[
\underline{u}_2^{\prime }\left( \mathbf{k}\right) =\frac 1{2\sqrt{\omega
\left( \mathbf{k}\right) \left( \omega \left( \mathbf{k}\right) +n_0\right) }%
}\left[ 
\begin{array}{c}
\left( c+\mathrm{i}q\right) \left( k_1-\mathrm{i}k_2\right)  \\ 
\left( c+\mathrm{i}q\right) \left( \omega \left( \mathbf{k}\right)
+n_0-k_3\right)  \\ 
0 \\ 
0 \\ 
\left( a-\mathrm{i}b\right) \left( k_1-\mathrm{i}k_2\right)  \\ 
\left( a-\mathrm{i}b\right) \left( \omega \left( \mathbf{k}\right)
+n_0-k_3\right)  \\ 
-k_1+\mathrm{i}k_2 \\ 
\omega \left( \mathbf{k}\right) +n_0+k_3
\end{array}
\right] \mbox{,}
\]

\[
\underline{u}_3^{\prime }\left( \mathbf{k}\right) =\frac 1{2\sqrt{\omega
\left( \mathbf{k}\right) \left( \omega \left( \mathbf{k}\right) +n_0\right) }%
}\left[ 
\begin{array}{c}
-\left( c+\mathrm{i}q\right) \left( \omega \left( \mathbf{k}\right)
+n_0-k_3\right)  \\ 
\left( c+\mathrm{i}q\right) \left( k_1+\mathrm{i}k_2\right)  \\ 
0 \\ 
0 \\ 
-\left( a-\mathrm{i}b\right) \left( \omega \left( \mathbf{k}\right)
+n_0-k_3\right)  \\ 
\left( a-\mathrm{i}b\right) \left( k_1+\mathrm{i}k_2\right)  \\ 
\omega \left( \mathbf{k}\right) +n_0+k_3 \\ 
k_1+\mathrm{i}k_2
\end{array}
\right] \mbox{,}
\]

\[
\underline{u}_4^{\prime }\left( \mathbf{k}\right) =\frac 1{2\sqrt{\omega
\left( \mathbf{k}\right) \left( \omega \left( \mathbf{k}\right) +n_0\right) }%
}\left[ 
\begin{array}{c}
\left( c+\mathrm{i}q\right) \left( k_1-\mathrm{i}k_2\right)  \\ 
-\left( c+\mathrm{i}q\right) \left( \omega \left( \mathbf{k}\right)
+n_0+k_3\right)  \\ 
0 \\ 
0 \\ 
\left( a-\mathrm{i}b\right) \left( k_1-\mathrm{i}k_2\right)  \\ 
-\left( a-\mathrm{i}b\right) \left( \omega \left( \mathbf{k}\right)
+n_0+k_3\right)  \\ 
k_1-\mathrm{i}k_2 \\ 
\omega \left( \mathbf{k}\right) +n_0-k_3
\end{array}
\right] \mbox{.}
\]

Здесь $\underline{u}_1^{\prime }\left( \mathbf{k}\right) $ и $\underline{u}%
_2^{\prime }\left( \mathbf{k}\right) $ соответствуют собственным векторам
для $\widehat{H}_{0,4}^{\prime }$ с собствен-ным значением 
$\omega \left( \mathbf{k}\right) =\sqrt{\mathbf{k}^2+n_0^2}$, и 
$\underline{u}_3^{\prime}\left( \mathbf{k}\right) $ и $\underline{u}_4^{\prime }\left( \mathbf{k}%
\right) $ соответствует собственным векторам для $\widehat{H}_{0,4}$
с собственным значением $-\omega \left( \mathbf{k}\right) $.

Пусть в (\ref{xa}):

\[
\begin{array}{c}
\underline{v}_{\left( 1\right) }\left( \mathbf{k}\right) \stackrel{Def}{=}%
\underline{\gamma ^{\left[ 0\right] }}\underline{u}_3^{\prime }\left( 
\mathbf{k}\right) \mbox{,} \\ 
\underline{v}_{\left( 2\right) }\left( \mathbf{k}\right) \stackrel{Def}{=}%
\underline{\gamma ^{\left[ 0\right] }}\underline{u}_4^{\prime }\left( 
\mathbf{k}\right) \mbox{,} \\ 
\underline{u}_{\left( 1\right) }\left( \mathbf{k}\right) \stackrel{Def}{=}%
\underline{u}_1^{\prime }\left( \mathbf{k}\right) \mbox{,} \\ 
\underline{u}_{\left( 2\right) }\left( \mathbf{k}\right) \stackrel{Def}{=}%
\underline{u}_2^{\prime }\left( \mathbf{k}\right) \mbox{.}
\end{array}
\]

Поэтому,

\[
\underline{v}_{\left( 1\right) }\left( \mathbf{k}\right) =\frac 1{2\sqrt{%
\omega \left( \mathbf{k}\right) \left( \omega \left( \mathbf{k}\right)
+n_0\right) }}\left[ 
\begin{array}{c}
0 \\ 
0 \\ 
-\left( c+\mathrm{i}q\right) \left( \omega \left( \mathbf{k}\right)
+n_0-k_3\right)  \\ 
\left( c+\mathrm{i}q\right) \left( k_1+\mathrm{i}k_2\right)  \\ 
\omega \left( \mathbf{k}\right) +n_0+k_3 \\ 
k_1+\mathrm{i}k_2 \\ 
-\left( a-\mathrm{i}b\right) \left( \omega \left( \mathbf{k}\right)
+n_0-k_3\right)  \\ 
\left( a-\mathrm{i}b\right) \left( k_1+\mathrm{i}k_2\right) 
\end{array}
\right] 
\]

и

\[
\underline{v}_{\left( 2\right) }\left( \mathbf{k}\right) =\frac 1{2\sqrt{%
\omega \left( \mathbf{k}\right) \left( \omega \left( \mathbf{k}\right)
+n_0\right) }}\left[ 
\begin{array}{c}
0 \\ 
0 \\ 
\left( c+\mathrm{i}q\right) \left( k_1-\mathrm{i}k_2\right)  \\ 
-\left( c+\mathrm{i}q\right) \left( \omega \left( \mathbf{k}\right)
+n_0+k_3\right)  \\ 
k_1-\mathrm{i}k_2 \\ 
\omega \left( \mathbf{k}\right) +n_0-k_3 \\ 
\left( a-\mathrm{i}b\right) \left( k_1-\mathrm{i}k_2\right)  \\ 
-\left( a-\mathrm{i}b\right) \left( \omega \left( \mathbf{k}\right)
+n_0+k_3\right) 
\end{array}
\right] \mbox{.}
\]

$\underline{u}_{\left( \alpha \right) }^{\prime }\left( \mathbf{k}\right) $ 
назывюется \textit{би-}$n_0$-\textit{лептонными}, а  $\underline{v}_{\left(
\alpha \right) }\left( \mathbf{k}\right) $ - \textit{би-анти-}$n_0$\textit{%
-лептонными} базисными векторами с импульсом $\mathbf{k}$ и спиновым индексом 
$\alpha $.

Следовательно, би-анти-$n_0$-лептонные базисные векторы являются результатом 
действия оперетора  $U^{\left( +\right) }$ (\ref{upls}).

Векторы

\begin{eqnarray*}
l_{n_0,\left( 1\right) }\left( \mathbf{k}\right)  &=&\left[ 
\begin{array}{c}
\left( a-\mathrm{i}b\right) \left( \omega \left( \mathbf{k}\right)
+n_0+k_3\right)  \\ 
\left( a-\mathrm{i}b\right) \left( k_1+\mathrm{i}k_2\right)  \\ 
\omega \left( \mathbf{k}\right) +n_0-k_3 \\ 
-k_1-\mathrm{i}k_2
\end{array}
\right] \mbox{ и } \\
l_{n_0,\left( 2\right) }\left( \mathbf{k}\right)  &=&\left[ 
\begin{array}{c}
\left( a-\mathrm{i}b\right) \left( k_1-\mathrm{i}k_2\right)  \\ 
\left( a-\mathrm{i}b\right) \left( \omega \left( \mathbf{k}\right)
+n_0-k_3\right)  \\ 
-k_1+\mathrm{i}k_2 \\ 
\omega \left( \mathbf{k}\right) +n_0+k_3
\end{array}
\right] 
\end{eqnarray*}

называются \textit{лептонными компонентами} би-$n_0$-лептонных базисных
векторов, а векторы

\[
\nu _{n_0,\left( 1\right) }\left( \mathbf{k}\right) =\left[ 
\begin{array}{c}
\omega \left( \mathbf{k}\right) +n_0+k_3 \\ 
k_1+\mathrm{i}k_2 \\ 
0 \\ 
0
\end{array}
\right] \mbox{ и }\nu _{n_0,\left( 2\right) }\left( \mathbf{k}\right)
=\left[ 
\begin{array}{c}
k_1-\mathrm{i}k_2 \\ 
\omega \left( \mathbf{k}\right) +n_0-k_3 \\ 
0 \\ 
0
\end{array}
\right] 
\]

называются \textit{нейтринными компонентами} би-$n_0$-лептонных базисных 
век-торов.

Векторы

\begin{eqnarray*}
\overline{l}_{n_0,\left( 1\right) }\left( \mathbf{k}\right)  &=&\left[ 
\begin{array}{c}
\omega \left( \mathbf{k}\right) +n_0+k_3 \\ 
k_1+\mathrm{i}k_2 \\ 
-\left( a-\mathrm{i}b\right) \left( \omega \left( \mathbf{k}\right)
+n_0-k_3\right)  \\ 
\left( a-\mathrm{i}b\right) \left( k_1+\mathrm{i}k_2\right) 
\end{array}
\right] \mbox{ и } \\
\overline{l}_{n_0,\left( 2\right) }\left( \mathbf{k}\right)  &=&\left[ 
\begin{array}{c}
k_1-\mathrm{i}k_2 \\ 
\omega \left( \mathbf{k}\right) +n_0-k_3 \\ 
\left( a-\mathrm{i}b\right) \left( k_1-\mathrm{i}k_2\right)  \\ 
-\left( a-\mathrm{i}b\right) \left( \omega \left( \mathbf{k}\right)
+n_0+k_3\right) 
\end{array}
\right] 
\end{eqnarray*}

называются \textit{лептонными компонентами} анти-би-$n_0$-лептонных базисных
век-торов, а векторы

\[
\overline{\nu }_{n_0,\left( 1\right) }\left( \mathbf{k}\right) =\left[ 
\begin{array}{c}
0 \\ 
0 \\ 
-\left( \omega \left( \mathbf{k}\right) +n_0-k_3\right)  \\ 
k_1+\mathrm{i}k_2
\end{array}
\right] \mbox{ и }\overline{\nu }_{n_0,\left( 2\right) }\left( \mathbf{k}%
\right) =\left[ 
\begin{array}{c}
0 \\ 
0 \\ 
k_1-\mathrm{i}k_2 \\ 
-\left( \omega \left( \mathbf{k}\right) +n_0+k_3\right) 
\end{array}
\right] 
\]

называются \textit{нейтринными компонентами} анти-би-$n_0$-лептонных базисных
векторов.

\subsubsection{Электрослабые преобразования}

Пусть (\ref{aa}) не выполняется для $k\in \left\{ 0,1,2,3\right\} $, и:

\begin{equation}
\ \ K\stackrel{def}{=}\sum_{\mu =0}^3\beta ^{\left[ \mu \right] }\left(
F_\mu +0.5g_1YB_\mu \right) \mbox{.}  \label{kdf}
\end{equation}

В таком случае из (\ref{gkB}) уравнение движения получает такую форму:

\begin{equation}
\left( K+\sum_{\mu =0}^3\beta ^{\left[ \mu \right] }\mathrm{i}\partial _\mu
+\gamma ^{\left[ 0\right] }\mathrm{i}\partial _5+\beta ^{\left[ 4\right] }%
\mathrm{i}\partial _4\right) \widetilde{\varphi }=0\mbox{.}  \label{me81}
\end{equation}

Поэтому для слуедующего преобразования:

\begin{eqnarray}
&&\widetilde{\varphi }\rightarrow \widetilde{\varphi }^{\prime }\stackrel{def%
}{=}U^{\left( -\right) }\widetilde{\varphi }\mbox{,}  \nonumber \\
&&x_4\rightarrow x_4^{\prime }\stackrel{def}{=}\left( \ell _{\circ }+\ell
_{*}\right) ax_4+\left( \ell _{\circ }-\ell _{*}\right) \sqrt{1-a^2}x_5%
\mbox{,}  \nonumber \\
&&x_5\rightarrow x_5^{\prime }\stackrel{def}{=}\left( \ell _{\circ }+\ell
_{*}\right) ax_5-\left( \ell _{\circ }-\ell _{*}\right) \sqrt{1-a^2}x_4%
\mbox{,}  \label{gll} \\
&&x_\mu \rightarrow x_\mu ^{\prime }\stackrel{def}{=}x_\mu \mbox{, для }\mu
\in \left\{ 0,1,2,3\right\} \mbox{,}  \nonumber \\
&&K\rightarrow K^{\prime }  \nonumber
\end{eqnarray}

с

\[
\begin{array}{c}
\partial _4U^{\left( -\right) }=U^{\left( -\right) }\partial _4\mbox{ и }%
\partial _5U^{\left( -\right) }=U^{\left( -\right) }\partial _5
\end{array}
\]

это уравнение имеет такую форму:

\begin{equation}
\left( 
\begin{array}{c}
U^{\left( -\right) \dagger }K^{\prime }U^{\left( -\right) }+ \\ 
+\sum_{\mu =0}^3\beta ^{\left[ \mu \right] }\mathrm{i}\left( \partial _\mu
+U^{\left( -\right) \dagger }\left( \partial _\mu U^{\left( -\right)
}\right) \right) +\gamma ^{\left[ 0\right] }\mathrm{i}\partial _5+\beta
^{\left[ 4\right] }\mathrm{i}\partial _4
\end{array}
\right) \widetilde{\varphi }=0\mbox{.}  \label{me82}
\end{equation}

Следовательно, если

\begin{equation}
K^{\prime }=K-\mathrm{i}\sum_{\mu =0}^3\beta ^{\left[ \mu \right] }\left(
\partial _\mu U^{\left( -\right) }\right) U^{\left( -\right) \dagger }\mbox{,}
\label{ksht}
\end{equation}

то уравнение (\ref{me81}) инвариантно для локальных преобразований (%
\ref{gll}).

Пусть $g_2$ - некоторое положительное число.

Если обозначить ($a,b,c,q$ из $U^{\left( -\right) }$):

\[
\begin{array}{c}
W_\mu ^{0,}\stackrel{def}{=}-\frac 2{g_2q}\left( 
\begin{array}{c}
q\left( \partial _\mu a\right) b-q\left( \partial _\mu b\right) a+\left(
\partial _\mu c\right) q^2+ \\ 
+a\left( \partial _\mu a\right) c+b\left( \partial _\mu b\right) c+c^2\left(
\partial _\mu c\right)
\end{array}
\right) \\ 
W_\mu ^{1,}\stackrel{def}{=}-\frac 2{g_2q}\left( 
\begin{array}{c}
\left( \partial _\mu a\right) a^2-bq\left( \partial _\mu c\right) +a\left(
\partial _\mu b\right) b+ \\ 
+a\left( \partial _\mu c\right) c+q^2\left( \partial _\mu a\right) +c\left(
\partial _\mu b\right) q
\end{array}
\right) \\ 
W_\mu ^{2,}\stackrel{def}{=}-\frac 2{g_2q}\left( 
\begin{array}{c}
q\left( \partial _\mu a\right) c-a\left( \partial _\mu a\right) b-b^2\left(
\partial _\mu b\right) - \\ 
-c\left( \partial _\mu c\right) b-\left( \partial _\mu b\right) q^2-\left(
\partial _\mu c\right) qa
\end{array}
\right)
\end{array}
\]

и

\[
W_\mu \stackrel{def}{=}\left[ 
\begin{array}{cccc}
W_\mu ^{0,}1_2 & 0_2 & \left( W_\mu ^{1,}-\mathrm{i}W_\mu ^{2,}\right) 1_2 & 
0_2 \\ 
0_2 & 0_2 & 0_2 & 0_2 \\ 
\left( W_\mu ^{1,}+\mathrm{i}W_\mu ^{2,}\right) 1_2 & 0_2 & -W_\mu ^{0,}1_2
& 0_2 \\ 
0_2 & 0_2 & 0_2 & 0_2
\end{array}
\right]\mbox{,} 
\]

то

\begin{equation}
-\mathrm{i}\left( \partial _\mu U^{\left( -\right) }\right) U^{\left(
-\right) \dagger }=\frac 12g_2W_\mu \mbox{,}  \label{w}
\end{equation}

и из (\ref{w}), (\ref{kdf}), (\ref{ksht}), (\ref{me81}):

\begin{equation}
\left( 
\begin{array}{c}
\sum_{\mu =0}^3\beta ^{\left[ \mu \right] }\mathrm{i}\left( \partial _\mu -%
\mathrm{i}0.5g_1B_\mu Y-\mathrm{i}\frac 12g_2W_\mu -\mathrm{i}F_\mu \right)
\\ 
+\gamma ^{\left[ 0\right] }\mathrm{i}\partial _5^{\prime }+\beta ^{\left[
4\right] }\mathrm{i}\partial _4^{\prime }
\end{array}
\right) \widetilde{\varphi }^{\prime }=0\mbox{.}  \label{hW}
\end{equation}

Пусть

\[
U^{\prime }\stackrel{def}{=}\mathrm{S}\left( a^{\prime },b^{\prime
},c^{\prime },q^{\prime }\right) \mbox{.} 
\]

В этом случае если

\[
U^{\prime \prime }\stackrel{def}{=}U^{\prime }U^{\left( -\right) }\mbox{,}
\]

то существуют вещественные функции $a^{\prime \prime }\left( t,\mathbf{x}%
\right) $, $b^{\prime \prime }\left( t,\mathbf{x}\right) $, $c^{\prime
\prime }\left( t,\mathbf{x}\right) $, $q^{\prime \prime }\left( t,\mathbf{x}%
\right) $ такие, что $U^{\prime \prime }$ имеет подобную форму:

\[
U^{\prime \prime }\stackrel{def}{=}\mathrm{S}\left( a^{\prime \prime
},b^{\prime \prime },c^{\prime \prime },q^{\prime \prime }\right) \mbox{.} 
\]

Если

\[
\ell _{\circ }^{\prime \prime }\stackrel{def}{=}\imath _{\circ }\left(
a^{\prime \prime },b^{\prime \prime },q^{\prime \prime },c^{\prime \prime
}\right) \mbox{, }\ell _{*}^{\prime \prime }\stackrel{def}{=}\imath
_{*}\left( a^{\prime \prime },b^{\prime \prime },q^{\prime \prime
},c^{\prime \prime }\right) \mbox{,} 
\]

и

\begin{eqnarray}
&&\widetilde{\varphi }\rightarrow \widetilde{\varphi }^{\prime \prime }%
\stackrel{def}{=}U^{\prime \prime }\widetilde{\varphi }\mbox{,}  \nonumber \\
&&x_4\rightarrow x_4^{\prime \prime }\stackrel{def}{=}\left( \ell _{\circ
}^{\prime \prime }+\ell _{*}^{\prime \prime }\right) a^{\prime \prime
}x_4+\left( \ell _{\circ }^{\prime \prime }-\ell _{*}^{\prime \prime
}\right) \sqrt{1-a^{\prime \prime 2}}x_5\mbox{,}  \nonumber \\
&&x_5\rightarrow x_5^{\prime \prime }\stackrel{def}{=}\left( \ell _{\circ
}^{\prime \prime }+\ell _{*}^{\prime \prime }\right) a^{\prime \prime
}x_5-\left( \ell _{\circ }^{\prime \prime }-\ell _{*}^{\prime \prime
}\right) \sqrt{1-a^{\prime \prime 2}}x_4\mbox{,}  \label{tt2} \\
&&x_\mu \rightarrow x_\mu ^{\prime \prime }\stackrel{def}{=}x_\mu 
\mbox{,
для }\mu \in \left\{ 0,1,2,3\right\} \mbox{,}  \nonumber \\
&&K\rightarrow K^{\prime \prime }\stackrel{def}{=}\sum_{\mu =0}^3\beta
^{\left[ \mu \right] }\left( F_\mu +0.5g_1YB_\mu +\frac 12g_2W_\mu ^{\prime
\prime }\right)\mbox{,}  \nonumber
\end{eqnarray}

то из (\ref{w}):

\[
W_\mu ^{\prime \prime }=-\frac{2i}{g_2}\left( \partial _\mu \left( U^{\prime
}U^{\left( -\right) }\right) \right) \left( U^{\prime }U^{\left( -\right)
}\right) ^{\dagger }\mbox{.} 
\]

Поэтому:

\[
W_\mu ^{\prime \prime }=-\frac{2i}{g_2}\left( \partial _\mu U^{\prime
}\right) U^{\prime \dagger }-\frac{2i}{g_2}U^{\prime }\left( \partial _\mu
U^{\left( -\right) }\right) U^{\left( -\right) \dagger }U^{\prime \dagger }%
\mbox{,} 
\]

т.е. из (\ref{w}):

\begin{equation}
W_\mu ^{\prime \prime }=U^{\prime }W_\mu U^{\prime \dagger }-\frac{2i}{g_2}%
\left( \partial _\mu U^{\prime }\right) U^{\prime \dagger }  \label{w00}
\end{equation}

как в Стандартной Модели.

Уравнение движения SU(2) Янг-Миллса поля без материи (например в \cite{Sd} 
или в \cite{Rd} ) имеет следующую форму:

\[
\partial ^\nu \mathbf{W}_{\mu \nu }=-g_2\mathbf{W}^\nu \times \mathbf{W}%
_{\mu \nu } 
\]

с:

\[
\mathbf{W}_{\mu \nu }=\partial _\mu \mathbf{W}_\nu -\partial _\nu \mathbf{W}%
_\mu +g_2\mathbf{W}_\mu \times \mathbf{W}_\nu 
\]

и

\[
\mathbf{W}_\mu =\left[ 
\begin{array}{c}
W_\mu ^{0,} \\ 
W_\mu ^{1,} \\ 
W_\mu ^{2,}
\end{array}
\right] \mbox{.} 
\]

Поэтому уравнение движения для $W_\mu ^{0,}$ имеет такой вид:

\begin{equation}
\begin{array}{c}
\partial ^\nu \partial _\nu W_\mu ^{0,}=g_2^2\left( W^{2,\nu }W_\nu
^{2,}+W^{1,\nu }W_\nu ^{1,}\right) W_\mu ^{0,}- \\ 
-g_2^2\left( W^{1,\nu }W_\mu ^{1,}+W^{2,\nu }W_\mu ^{2,}\right) W_\nu ^{0,}+
\\ 
+g_2\partial ^\nu \left( W_\mu ^{1,}W_\nu ^{2,}-W_\mu ^{2,}W_\nu
^{1,}\right) + \\ 
+g_2\left( W^{1,\nu }\partial _\mu W_\nu ^{2,}-W^{1,\nu }\partial _\nu W_\mu
^{2,}-W^{2,\nu }\partial _\mu W_\nu ^{1,}+W^{2,\nu }\partial _\nu W_\mu
^{1,}\right) + \\ 
+\partial ^\nu \partial _\mu W_\nu ^{0,}
\end{array}
\label{b}
\end{equation}

с $g_{0,0}=1$, $g_{1,1}=g_{2,2}=g_{3,3}=-1$ (т.е.: $W^\nu W_\nu
=W^0W_0-W^1W_1-W^2W_2-W^3W_3$. В калибровке с $W_0=0$: $W^\nu W_\nu
=-\left( W^1W_1+W^2W_2+W^3W_3\right) $ ).

$W_\mu ^{1,}$ и $W_\mu ^{2,}$ подчиняются таким же уравнениям.

Это уравнение может быть преобразовано к следующему виду:

\[
\begin{array}{c}
\partial ^\nu \partial _\nu W_\mu ^{0,}=\left[ g_2^2\left( W^{2,\nu }W_\nu
^{2,}+W^{1,\nu }W_\nu ^{1,}+W^{0,\nu }W_\nu ^{0,}\right) \right] \cdot W_\mu
^{0,}- \\ 
-g_2^2\left( W^{1,\nu }W_\mu ^{1,}+W^{2,\nu }W_\mu ^{2,}+W^{0,\nu }W_\mu
^{0,}\right) W_\nu ^{0,}+ \\ 
+g_2\partial ^\nu \left( W_\mu ^{1,}W_\nu ^{2,}-W_\mu ^{2,}W_\nu
^{1,}\right) + \\ 
+g_2\left( W^{1,\nu }\partial _\mu W_\nu ^{2,}-W^{1,\nu }\partial _\nu W_\mu
^{2,}-W^{2,\nu }\partial _\mu W_\nu ^{1,}+W^{2,\nu }\partial _\nu W_\mu
^{1,}\right) + \\ 
+\partial ^\nu \partial _\mu W_\nu ^{0,}\mbox{.}
\end{array}
\]

Оно похоже на уравнение Клейна-Гордона поля $W_\mu ^{0,}$ с массой

\begin{equation}
g_2\left[ -\left( W^{2,\nu }W_\nu ^{2,}+W^{1,\nu }W_\nu ^{1,}+W^{0,\nu
}W_\nu ^{0,}\right) \right] ^{\frac 12}.  \label{z10}
\end{equation}

и с дополнительными членами взаимодействия поля $W_\mu ^{0,}$ с другими 
компонен-тами поля $\mathbf{W}$.

"Масса" (\ref{z10}) инвариантна относительно поворотов и преобразований Лоренца:

\[
\left\{ 
\begin{array}{c}
W_r^{k,\prime }=W_r^{k,}\cos \lambda -W_s^{k,}\sin \lambda \mbox{,} \\ 
W_s^{k,\prime }=W_r^{k,}\sin \lambda +W_s^{k,}\cos \lambda \mbox{;}
\end{array}
\right| 
\]

\[
\left\{ 
\begin{array}{c}
W_0^{k,\prime }=W_0^{k,}\cosh \lambda -W_s^{k,}\sinh \lambda \mbox{,} \\ 
W_s^{k,\prime }=W_s^{k,}\cosh \lambda -W_0^{k,}\sinh \lambda
\end{array}
\right| 
\]

с вещественным $\lambda $ и с $r\in \left\{ 1,2,3\right\} $ и $s\in
\left\{ 1,2,3\right\} $, и (\ref{z10}) инвариантно относительно глобальных 
изоспиновых преобразований $U^{\left(+ \right)}$:

\[
W_\nu ^{\prime }\rightarrow W_\nu ^{\prime \prime }=U^{\prime }W_\nu
U^{\prime \dagger }\mbox{,} 
\]

но не инвариантно относительно локальных преобразований (\ref{w00})

Уравнение (\ref{b}) можно упростить:

\[
\begin{array}{c}
\sum_\nu g_{\nu ,\nu }\partial ^\nu \partial _\nu W_\mu ^{0,}=\left[
g_2^2\sum_{\nu \neq \mu }g_{\nu ,\nu }\left( \left( W_\nu ^{2,}\right)
^2+\left( W_\nu ^{1,}\right) ^2\right) \right] \cdot W_\mu ^{0,}- \\ 
-g_2^2\sum_{\nu \neq \mu }g_{\nu ,\nu }\left( W^{1,\nu }W_\mu ^{1,}+W^{2,\nu
}W_\mu ^{2,}\right) W_\nu ^{0,}- \\ 
+g_2\sum_\nu g_{\nu ,\nu }\partial ^\nu \left( W_\mu ^{1,}W_\nu ^{2,}-W_\mu
^{2,}W_\nu ^{1,}\right) + \\ 
+g_2\sum_\nu g_{\nu ,\nu }\left( W^{1,\nu }\partial _\mu W_\nu
^{2,}-W^{1,\nu }\partial _\nu W_\mu ^{2,}-W^{2,\nu }\partial _\mu W_\nu
^{1,}+W^{2,\nu }\partial _\nu W_\mu ^{1,}\right) + \\ 
+\partial _\mu \sum_\nu g_{\nu ,\nu }\partial ^\nu W_\nu ^{0,}\mbox{.}
\end{array}
\]

(здесь нет суммирования по повторяющимся индексам "$_\nu ^\nu $"; 
суммирование выражается символом "$\sum $" ) .

В этом уравнении форма

\[
g_2\left[ -\sum_{\nu \neq \mu }g_{\nu ,\nu }\left( \left( W_\nu ^{2,}\right)
^2+\left( W_\nu ^{1,}\right) ^2\right) \right] ^{\frac 12} 
\]

может меняться в пространстве, но она не содержит $W_\mu ^{0,}$ и локально 
действует как масса, т.е. не позволяет частицам этого поля вести себя подобно 
безмассовым частицам.

Пусть

\[
\begin{array}{c}
\alpha \stackrel{def}{=}\arctan \frac{g_1}{g_2}\mbox{,} \\ 
Z_\mu \stackrel{def}{=}\left( W_\mu ^{0,}\cos \alpha -B_\mu \sin \alpha
\right) \mbox{,} \\ 
A_\mu \stackrel{def}{=}\left( B_\mu \cos \alpha +W_\mu ^{0,}\sin \alpha
\right) \mbox{.}
\end{array}
\]

В этом случае:

\[
\begin{array}{c}
\sum_\nu g_{\nu ,\nu }\partial _\nu \partial _\nu W_\mu ^{0,}=\cos \alpha
\cdot \sum_\nu g_{\nu ,\nu }\partial _\nu \partial _\nu Z_\mu +\sin \alpha
\cdot \sum_\nu g_{\nu ,\nu }\partial _\nu \partial _\nu A_\mu \mbox{.}
\end{array}
\]

Если

\[
\sum_\nu g_{\nu ,\nu }\partial _\nu \partial _\nu A_\mu =0\mbox{,}
\]

то

\[
m_Z=\frac{m_W}{\cos \alpha } 
\]

с $m_W$ из (\ref{z10}). Это почти как в Стандартной Модели.





\subsection{Повороты декартовой системы координат и кваррки}

Пусть $\alpha$ - вещественное число, и

\begin{eqnarray}
&&x_1^{\prime }\stackrel{def}{=}x_1\cos \left( \alpha \right) -x_2\sin \left( \alpha \right) %
\mbox{;}\nonumber \\ 
&&x_2^{\prime }\stackrel{def}{=}x_1\sin \left( \alpha \right) +x_2\cos \left( \alpha \right) %
\mbox{;}\label{rot} \\ 
&&x_3^{\prime }\stackrel{def}{=}x_3\mbox{.}\nonumber
\end{eqnarray}

Следовательно для любой функции $\varphi $:

\begin{eqnarray}
&&\partial _1^{\prime }\varphi =\left( \partial _1\varphi \cdot \cos \alpha
-\partial _2\varphi \cdot \sin \alpha \right) \mbox{;}\nonumber \\ 
&&\partial _2^{\prime }\varphi =\left( \partial _2\varphi \cdot \cos \alpha
+\partial _1\varphi \cdot \sin \alpha \right) \mbox{;} \label{d}\\ 
&&\partial _3^{\prime }\varphi =\partial _3\varphi \mbox{.}
\end{eqnarray}

Так как $\mathbf{j}_A$ - 3-вектор, то из (\ref{j}):

\begin{eqnarray*}
&&j_{A,1}^{\prime }=-\varphi ^{\dagger }\left( \beta ^{\left[ 1\right] }\cos
\left( \alpha \right) -\beta ^{\left[ 2\right] }\sin \left( \alpha \right)
\right) \varphi \mbox{;} \\ 
&&j_{A,2}^{\prime }=-\varphi ^{\dagger }\left( \beta ^{\left[ 1\right] }\sin
\left( \alpha \right) +\beta ^{\left[ 2\right] }\cos \left( \alpha \right)
\right) \varphi \mbox{;} \\ 
&&j_{A,3}^{\prime }=-\varphi ^{\dagger }\beta ^{\left[ 3\right] }\varphi \mbox{.}
\end{eqnarray*}

Следовательно, если для $\varphi ^{\prime }$:

\[
\begin{array}{c}
j_{A,1}^{\prime }=-\varphi ^{\prime \dagger }\beta ^{\left[ 1\right] }\varphi
^{\prime }\mbox{;} \\ 
j_{A,2}^{\prime }=-\varphi ^{\prime \dagger }\beta ^{\left[ 2\right] }\varphi
^{\prime }\mbox{;} \\ 
j_{A,3}^{\prime }=-\varphi ^{\prime \dagger }\beta ^{\left[ 3\right] }\varphi
^{\prime }\mbox{,}
\end{array}
\]

и

\[
\varphi ^{\prime }\stackrel{def}{=}U_{1,2}\left( \alpha \right) \varphi \mbox{,}
\]

то

\begin{eqnarray}
&&U_{1,2}^{\dagger }\left( \alpha \right) \beta ^{\left[ 1\right]
}U_{1,2}\left( \alpha \right) =\beta ^{\left[ 1\right] }\cos \alpha -\beta
^{\left[ 2\right] }\sin \alpha \mbox{;}\nonumber \\ 
&&U_{1,2}^{\dagger }\left( \alpha \right) \beta ^{\left[ 2\right]
}U_{1,2}\left( \alpha \right) =\beta ^{\left[ 2\right] }\cos \alpha +\beta
^{\left[ 1\right] }\sin \alpha \mbox{;}\label{con1} \\ 
&&U_{1,2}^{\dagger }\left( \alpha \right) \beta ^{\left[ 3\right]
}U_{1,2}\left( \alpha \right) =\beta ^{\left[ 3\right] }\mbox{;}\nonumber
\end{eqnarray}

из (\ref{j}): так как

\[
\rho_A =\varphi ^{\dagger }\varphi =\varphi ^{\prime \dagger }\varphi ^{\prime
} \mbox{,} 
\]

то

\begin{equation}
U_{1,2}^{\dagger }\left( \alpha \right) U_{1,2}\left( \alpha \right) =1_4%
\mbox{.}  \label{con2}
\end{equation}

Если

\[
U_{1,2}\left( \alpha \right) =\cos \frac \alpha 2\cdot 1_4-\sin \frac \alpha
2\cdot \beta ^{\left[ 1\right] }\beta ^{\left[ 2\right] } \mbox{,}
\]

то $U_{1,2}\left( \alpha \right) $ подчиняется всем этим условиям ((\ref{con1}), (\ref{con2})). 
Более того:

\begin{equation}
\begin{array}{c}
U_{1,2}^{\dagger }\left( \alpha \right) \beta ^{\left[ 4\right]
}U_{1,2}\left( \alpha \right) =\beta ^{\left[ 4\right] }\mbox{;} \\ 
U_{1,2}^{\dagger }\left( \alpha \right) \gamma ^{\left[ 0\right]
}U_{1,2}\left( \alpha \right) =\gamma ^{\left[ 0\right] }\mbox{,}
\end{array}
\label{con3}
\end{equation}

\[
U_{1,2}^{\dagger }\left( \alpha \right) \gamma ^{\left[ 5\right]
}U_{1,2}\left( \alpha \right) =\gamma ^{\left[ 5\right] }\mbox{.} 
\]

Пусть $\widehat{H}_l^{\prime }$ - результат замены $\beta ^{\left[
k\right] }$ на $\beta ^{\left[ k\right] \prime }=U_{1,2}^{\dagger }\left(
\alpha \right) \beta ^{\left[ k\right] }U_{1,2}\left( \alpha \right) $ и $%
\partial _k$ - на $\partial _k^{\prime }=\frac \partial {\partial x_k^{\prime
}}$ в $\widehat{H}_l$.

Из (\ref{d}), (\ref{con1}), (\ref{con2}) и (\ref{con3}):

\[
\widehat{H}_l^{\prime }=\mathrm{i}\left( 
\begin{array}{c}
\beta ^{\left[ 1\right] }\left( 
\begin{array}{c}
\partial _1+\mathrm{i}\left( \Theta _1^{\prime }\cos \left( \alpha \right)
+\Theta _2^{\prime }\sin \left( \alpha \right) \right) + \\ 
+\mathrm{i}\left( \Upsilon _1^{\prime }\cos \left( \alpha \right) +\Upsilon
_2^{\prime }\sin \left( \alpha \right) \right) \gamma ^{\left[ 5\right] }
\end{array}
\right) + \\ 
+\beta ^{\left[ 2\right] }\left( 
\begin{array}{c}
\partial _2+\mathrm{i}\left( -\Theta _1^{\prime }\sin \left( \alpha \right)
+\Theta _2^{\prime }\cos \left( \alpha \right) \right) + \\ 
+\mathrm{i}\left( -\Upsilon _1^{\prime }\sin \left( \alpha \right) +\mathrm{i%
}_2^{\prime }\cos \left( \alpha \right) \right) \gamma ^{\left[ 5\right] }
\end{array}
\right) + \\ 
+\beta ^{\left[ 3\right] }\left( \partial _3+\mathrm{i}\Theta _3^{\prime }+%
\mathrm{i}\Upsilon _3^{\prime }\gamma ^{\left[ 5\right] }\right) + \\ 
+\mathrm{i}M_0^{\prime }\gamma ^{\left[ 0\right] }+\mathrm{i}M_4\beta
^{\left[ 4\right] }\mbox{.}
\end{array}
\right) 
\]

Следовательно, если

\[
\begin{array}{c}
\Theta _0^{\prime }=\Theta _0\mbox{,} \\ 
\Theta _1^{\prime }=\Theta _1\cos \left( \alpha \right) -\Theta _2\sin
\left( \alpha \right) \mbox{,} \\ 
\Theta _2^{\prime }=\Theta _1\sin \left( \alpha \right) +\Theta _2\cos
\left( \alpha \right) \mbox{,} \\ 
\Theta _3^{\prime }=\Theta _3 \mbox{,}
\end{array}
\]

и такая же формула для $\left\langle \Upsilon _0,\Upsilon
_1,\Upsilon _2,\Upsilon _3\right\rangle $, то $\widehat{H}_l^{\prime }=%
\widehat{H}_l$ при поворотах декартовой системы (\ref{rot}).

Но:

\begin{equation}
\begin{array}{c}
U_{1,2}^{\dagger }\left( \alpha \right) \zeta ^{\left[ 1\right]
}U_{1,2}\left( \alpha \right) =\zeta ^{\left[ 1\right] }\cos \alpha -\eta
^{\left[ 2\right] }\sin \alpha \mbox{;} \\ 
U_{1,2}^{\dagger }\left( \alpha \right) \eta ^{\left[ 2\right]
}U_{1,2}\left( \alpha \right) =\eta ^{\left[ 2\right] }\cos \alpha +\zeta
^{\left[ 1\right] }\sin \alpha \mbox{;}
\end{array}
\label{conk}
\end{equation}

\begin{equation}
\begin{array}{c}
U_{1,2}^{\dagger }\left( \alpha \right) \zeta ^{\left[ 2\right]
}U_{1,2}\left( \alpha \right) =\zeta ^{\left[ 2\right] }\cos \alpha -\eta
^{\left[ 1\right] }\sin \alpha \mbox{;} \\ 
U_{1,2}^{\dagger }\left( \alpha \right) \eta ^{\left[ 1\right]
}U_{1,2}\left( \alpha \right) =\eta ^{\left[ 1\right] }\cos \alpha +\zeta
^{\left[ 2\right] }\sin \alpha \mbox{;}
\end{array}
\label{conk1}
\end{equation}

\[
\begin{array}{c}
U_{1,2}^{\dagger }\left( \alpha \right) \zeta ^{\left[ 3\right]
}U_{1,2}\left( \alpha \right) =\zeta ^{\left[ 3\right] }\mbox{;} \\ 
U_{1,2}^{\dagger }\left( \alpha \right) \eta ^{\left[ 3\right]
}U_{1,2}\left( \alpha \right) =\eta ^{\left[ 3\right] }\mbox{;}
\end{array}
\]

\begin{equation}
\begin{array}{c}
U_{1,2}^{\dagger }\left( \alpha \right) \gamma _\zeta ^{\left[ 0\right]
}U_{1,2}\left( \alpha \right) =\gamma _\zeta ^{\left[ 0\right] }\cos \alpha
-\gamma _\eta ^{\left[ 0\right] }\sin \alpha \mbox{;} \\ 
U_{1,2}^{\dagger }\left( \alpha \right) \gamma _\eta ^{\left[ 0\right]
}U_{1,2}\left( \alpha \right) =\gamma _\eta ^{\left[ 0\right] }\cos \alpha
+\gamma _\zeta ^{\left[ 0\right] }\sin \alpha \mbox{;} \\ 
U_{1,2}^{\dagger }\left( \alpha \right) \gamma _\theta ^{\left[ 0\right]
}U_{1,2}\left( \alpha \right) =\gamma _\theta ^{\left[ 0\right] }\mbox{;}
\end{array}
\label{conk2}
\end{equation}

\begin{equation}
\begin{array}{c}
U_{1,2}^{\dagger }\left( \alpha \right) \zeta ^{\left[ 4\right]
}U_{1,2}\left( \alpha \right) =\zeta ^{\left[ 4\right] }\cos \alpha +\eta
^{\left[ 4\right] }\sin \alpha \mbox{;} \\ 
U_{1,2}^{\dagger }\left( \alpha \right) \eta ^{\left[ 4\right]
}U_{1,2}\left( \alpha \right) =\eta ^{\left[ 4\right] }\cos \alpha -\zeta
^{\left[ 4\right] }\sin \alpha \mbox{;} \\ 
U_{1,2}^{\dagger }\left( \alpha \right) \theta ^{\left[ 4\right]
}U_{1,2}\left( \alpha \right) =\theta ^{\left[ 4\right] }
\end{array}
\label{conk3}
\end{equation}

и

\[
\begin{array}{c}
U_{1,2}^{\dagger }\left( \alpha \right) \theta ^{\left[ 1\right]
}U_{1,2}\left( \alpha \right) =\theta ^{\left[ 1\right] }\cos \alpha +\theta
^{\left[ 2\right] }\sin \alpha \mbox{;} \\ 
U_{1,2}^{\dagger }\left( \alpha \right) \theta ^{\left[ 2\right]
}U_{1,2}\left( \alpha \right) =\theta ^{\left[ 2\right] }\cos \alpha -\theta
^{\left[ 1\right] }\sin \alpha \mbox{;} \\ 
U_{1,2}^{\dagger }\left( \alpha \right) \theta ^{\left[ 3\right]
}U_{1,2}\left( \alpha \right) =\theta ^{\left[ 3\right] }\mbox{.}
\end{array}
\]

Следовательно из (\ref{conk}), (\ref{conk1}), (\ref{conk2}), (\ref{conk3}):

$\widehat{H}\left( \zeta \right) $ перемешивается с $\widehat{H}\left( \eta
\right) $ при таком повороте. Для других поворотов декартовой системы: 
лептонные гамильтонианы преобразуются в лептонные, а цветные перемешиваются 
между собой.

Поэтому цветная тройка элементов не может быть разделена в пространстве.
Эти частицы должны быть локализованы в одном и том же месте (конфайнмент?). 

Каждая цветная пентада содержит по два массовых элемента. Следовательно, 
каждое семейство содержит два сорта цветных частиц трех цветов. Я называю эти 
частицы {\it кваррками}.

\subsection{Вкусовые пентады}

Я называю $4\times 4$ матрицы типа

\[
\left[ 
\begin{array}{cc}
\vartheta  & 0_2 \\ 
0_2 & \upsilon 
\end{array}
\right] 
\]

2-диагональными, а 

\[
\left[ 
\begin{array}{cc}
0_2 & \vartheta  \\ 
\upsilon  & 0_2
\end{array}
\right] 
\]

- 2-антидиагональными.

Таким образом, в лептоннном уравнении движения три 2-диагональных элемента 
легкой пентады определяет 3-мерное пространство событий, а два 
2-антидиагональных элемента образуют 2-мерное пространство электрослабых 
взаимодействий. Аналогично для цветных пентад.

Для сладкой пентады вектор локальной скорости имеет следующие компоненты:

\[
\rho u_0^{\underline{\Delta }}\stackrel{def}{=}\varphi ^{\dagger }\underline{%
\Delta }^{[0]}\varphi =-\cos \left( 2\cdot \stackrel{*}{\alpha }\right) , 
\]

\[
\rho u_1^{\underline{\Delta }}\stackrel{def}{=}\varphi ^{\dagger }\underline{%
\Delta }^{[1]}\varphi =-\sin \left( 2\cdot \stackrel{*}{\alpha }\right)
\cdot \left[ 
\begin{array}{c}
\cos \left( \stackrel{*}{\beta }\right) \cdot \sin \left( \stackrel{*}{\chi }%
\right) \cos \left( \stackrel{*}{\gamma }-\stackrel{*}{\upsilon }\right)  \\ 
+\sin \left( \stackrel{*}{\beta }\right) \cdot \cos \left( \stackrel{*}{\chi 
}\right) \cos \left( \stackrel{*}{\theta }-\stackrel{*}{\lambda }\right) 
\end{array}
\right] ,
\]

\[
\rho u_2^{\underline{\Delta }}\stackrel{def}{=}\varphi ^{\dagger }\underline{%
\Delta }^{[2]}\varphi =-\sin \left( 2\cdot \stackrel{*}{\alpha }\right)
\cdot \left[ 
\begin{array}{c}
-\cos \left( \stackrel{*}{\beta }\right) \cdot \sin \left( \stackrel{*}{\chi 
}\right) \sin \left( \stackrel{*}{\gamma }-\stackrel{*}{\upsilon }\right) 
\\ 
+\sin \left( \stackrel{*}{\beta }\right) \cdot \cos \left( \stackrel{*}{\chi 
}\right) \sin \left( \stackrel{*}{\theta }-\stackrel{*}{\lambda }\right) 
\end{array}
\right] ,
\]

\[
\rho u_3^{\underline{\Delta }}\stackrel{def}{=}\varphi ^{\dagger }\underline{%
\Delta }^{[3]}\varphi =-\sin \left( 2\cdot \stackrel{*}{\alpha }\right)
\cdot \left[ 
\begin{array}{c}
\cos \left( \stackrel{*}{\beta }\right) \cdot \cos \left( \stackrel{*}{\chi }%
\right) \cos \left( \stackrel{*}{\gamma }-\stackrel{*}{\lambda }\right)  \\ 
-\sin \left( \stackrel{*}{\beta }\right) \cdot \sin \left( \stackrel{*}{\chi 
}\right) \cos \left( \stackrel{*}{\theta }-\stackrel{*}{\upsilon }\right) 
\end{array}
\right] ,
\]

\[
\rho u_4^{\underline{\Delta }}\stackrel{def}{=}\varphi ^{\dagger }\underline{%
\Delta }^{[4]}\varphi =-\sin \left( 2\cdot \stackrel{*}{\alpha }\right)
\cdot \left[ 
\begin{array}{c}
-\cos \left( \stackrel{*}{\beta }\right) \cdot \cos \left( \stackrel{*}{\chi 
}\right) \sin \left( \stackrel{*}{\gamma }-\stackrel{*}{\lambda }\right)  \\ 
-\sin \left( \stackrel{*}{\beta }\right) \cdot \sin \left( \stackrel{*}{\chi 
}\right) \sin \left( \stackrel{*}{\theta }-\stackrel{*}{\upsilon }\right) 
\end{array}
\right] .
\]

Следовательно, здесь 2-антидиогональные матрицы $\underline{\Delta }^{[1]}$ и $%
\underline{\Delta }^{[2]}$ определяют 2-мерное пространство ($u_1^{\underline{%
\Delta }}$, $u_2^{\underline{\Delta }}$), в котором действует преобразование 
изоспина. 2-антидиагональные матрицы $\underline{\Delta }^{[3]}$ и $%
\underline{\Delta }^{[4]}$ определяют такое же пространство ($u_3^{\underline{\Delta }}$, $%
u_4^{\underline{\Delta }}$). Эта пентада содержит единственную 2-диагональную матрицу, 
определяющую 1-мерное пространство ($u_0^{\underline{\Delta }}$%
) для размещения событий.

Подобно сладкой пентаде горькая пентада с четырьмя 2-антидиагональными матрицами 
и с единственной 2-диагональной матрицей определяет два 2-мерных пространства 
для изоспиновых преобразований и одно 1-мерное пространство для размещения 
событий.

\subsection{Два события}

Пусть

\[
\int_{D_1}d^3\mathbf{x}\int_{D_2}d^3\underline{\mathbf{y}}\cdot \rho \left(
t,\mathbf{x},\mathbf{y}\right) \stackrel{Def}{=}\mathbf{P}\left( A_1\left(
t,D_1\right) \&A_2\left( t,D_2\right) \right) 
\]

Существуют комплексные функции $\varphi _{s_1,s_2}\left( t,\mathbf{x},\mathbf{y}\right) $
($s_k\in \left\{ 1,2,3,4\right\} $), для которых:

\[
\rho \left( t,\mathbf{x},\mathbf{\ y}\right)
=4\sum_{s_1=1}^4\sum_{s_2=1}^4\varphi _{s_1,s_2}^{*}\left( t,\mathbf{x},%
\mathbf{y}\right) \varphi _{s_1,s_2}\left( t,\mathbf{x},\mathbf{y}\right) %
\mbox{.} 
\]

Если

\[
\begin{array}{c}
\Psi \left( t,\mathbf{x},\mathbf{y}\right) \stackrel{Def}{=} \\ 
=\sum_{s_1=1}^4\sum_{s_2=1}^4\varphi _{s_1,s_2}\left( t,\mathbf{x},\mathbf{y}%
\right) \left( \psi _{s_1}^{ \dagger }\left( \mathbf{x}%
\right) \psi _{s_2}^{ \dagger }\left( \mathbf{y}\right) -\psi
_{s_2}^{ \dagger }\left( \mathbf{y}\right) \psi
_{s_1}^{ \dagger }\left( \mathbf{x}\right) \right) \Phi _0
\end{array}\mbox{,}
\]

то

\[
\begin{array}{c}
\Psi ^{\dagger }\left( t,\mathbf{x},\mathbf{y}\right) \Psi \left( t,\mathbf{x%
},\mathbf{y}\right) = \\ 
=4\sum_{s_1=1}^4\sum_{s_2=1}^4\varphi _{s_1,s_2}^{*}\left(
t,\mathbf{x}^{\prime },\mathbf{y}^{\prime }\right) \varphi
_{s_1,s_2}\left( t,\mathbf{x},\mathbf{y}\right) \cdot
\delta \left( \mathbf{x}-\mathbf{x}^{\prime }\right) \delta \left( \mathbf{y}%
-\mathbf{y}^{\prime }\right) \mbox{.}
\end{array}
\]

Подобно (\ref{sys}): система с неизвестными комплексными функциями 
$Q_{s_1,k_1}^{\left( 1\right) }$, $Q_{s_2,k_2}^{ }$, 
$Q_{s_1,k_1;s_2,k_2}^{\left( 1,2\right) }$:

\[
\left\{ 
\begin{array}{c}
\sum_{k_1=1}^4Q_{s_1,k_1}^{\left( 1\right) }\varphi
_{k_1,s_2}+\sum_{k_2=1}^4Q_{s_2,k_2}^{\left( 2\right) }\varphi _{s_1,k_2}+
\\ 
+\sum_{k_1=1}^4\sum_{k_2=1}^4Q_{s_1,k_1;s_2,k_2}^{\left( 1,2\right) }\varphi
_{k_1,k_2}= \\ 
=\partial _t\varphi _{s_1,s_2}-\sum_{r=1}^3\left( \sum_{k_1=1}^4\beta
_{s_1,k_1}^{\left[ r\right] }\frac \partial {\partial x_r}\varphi
_{k_1,s_2}+\sum_{k_2=1}^4\beta _{s_2,k_2}^{\left[ r\right] }\frac \partial
{\partial y_r}\varphi _{s_1,k_2}\right) ; \\ 
Q_{k_1,s_1}^{\left( 1\right) *}=-Q_{s_1,k_1}^{\left( 1\right) }; \\ 
Q_{k_2,s_2}^{\left( 2\right) *}=-Q_{s_2,k_2}^{\left( 2\right) }; \\ 
Q_{k_1,s_1;s_2,k_2}^{\left( 1,2\right) *}=-Q_{s_1,k_1;s_2,k_2}^{\left(
1,2\right) }; \\ 
Q_{s_1,k_1;k_2,s_2}^{\left( 1,2\right) *}=-Q_{s_1,k_1;s_2,k_2}^{\left(
1,2\right) }
\end{array}
\right| 
\]

имеет решения.

et cetera...

\subsection{Размерность физического пространства}

Теперь пусть размерность пространства событий $\mu$ - любое натуральное число, 
необязательно равное 3.

По \cite{ZH}: для каждого натурального числа $z$ существует клиффордово 
множество ранга $2^z$.

Для каждого вектора плотности вероятности $\left\langle \rho \left( t,%
\mathbf{x}\right) ,\mathbf{j}\left( t,\mathbf{x}%
\right) \right\rangle $ натуральное число $s$, клиффордово множество $K$ 
ранга $s$ и комплексный $s$-вектор $\overleftarrow{\varsigma } \left( t,\mathbf{x}%
\right) $ существуют, для которых: $\gamma _n\in K$ и 

\begin{equation}  \label{1}
\overleftarrow{\varsigma } \left( t,\mathbf{x}\right) ^{\dagger } \overleftarrow{\varsigma } \left( t,%
\mathbf{x}\right) =\rho \left( t,\mathbf{x}\right) ,
\end{equation}

\begin{equation}  \label{1'}
\overleftarrow{\varsigma } \left( t,\mathbf{x}\right) ^{\dagger } \gamma _n \overleftarrow{\varsigma }
\left( t,\mathbf{x}\right) =j_n\left( t,\mathbf{x}\right) .
\end{equation}

В этом случае $\overleftarrow{\varsigma } \left( t,\mathbf{x}\right) $ называется 
$s$-{\it спинором} для $\left\langle \rho \left( t,\mathbf{x}\right) ,%
\mathbf{j}\left( t,\mathbf{x}\right) \right\rangle $.

Пусть $\rho _c(t,\mathbf{x}|t_0,\mathbf{x_0})$ - плотность 
условной вероятности события $A\left( t,\mathbf{x}\right)$ по событию 
$B\left( t_0,\mathbf{x}_0\right)$ в $\mu +1$ пространстве-времени. И если

\[
\rho _c(t,\mathbf{x}|t_0,\mathbf{x_0})=g(t,\mathbf{x}%
|t_0,\mathbf{x_0}) \rho (t,\mathbf{x})\mbox{,} 
\]

то функция $g(t,\mathbf{x}|t_0,\mathbf{x_0})$ есть функция
взаимодействия для $A$ и $B$ в $\mu +1$-пространстве-времени.

Пусть $\overleftarrow{\varsigma } _c$ и $\overleftarrow{\varsigma } $ - $s$-спиноры, для которых: $\rho =\overleftarrow{\varsigma }
^{\dagger } \overleftarrow{\varsigma } $ и $\rho _c=\overleftarrow{\varsigma } _c^{\dagger } \overleftarrow{\varsigma } _c$.

Если на этих спинорах определено произведение $\circ $ так, что для каждого 
$\overleftarrow{\varsigma } _c$ и $\overleftarrow{\varsigma } $ существует элемент $\overleftarrow{\varsigma }_g $ этой алгебры, 
для которого: $\overleftarrow{\varsigma } _c=\overleftarrow{\varsigma }_g %
\circ \overleftarrow{\varsigma } $ и $\overleftarrow{\varsigma }_g ^{\dagger}%
 \overleftarrow{\varsigma }_g =g$,
то множество этих спиноров образует нормированную алгебру с делением.

Размерность такой алгебры по теореме Гурвица \cite{O1} (Каждая нормированная 
алгебра с единицей изоморфна одной из следующих: алгебра вещественных чисел $R$, 
алгебра комплексных чисел $C$, алгебра кватернионов $K$, или алгебра октав
$\acute O$) и по обобщенной теореме Фробениуса \cite{O2} (Алгебра с делением 
имеет размерность только 1,2,4 или 8) не больше, чем 8. 

В этом случае размер матриц клиффордова множества не более, чем $4\times 4$ 
(матрицы этого множества комплексные). Такое клиффордово множество содержит не 
более, чем 5 элементов. Диагональные элементы этой пентады определяют 
пространство событий. Размерность этого пространства не больше 3. 
Следовательно, в этом случае мы имеем 3+1 пространство-время.

Если $\mu >3$, то для каждой такой алгебры существует функция взаимодействия, 
не принадлежащая этой алгебре. Я называю такие взаимодействия {\it 
сверхестест-венными} для этой алгебры взаимодействиями.

\subsection{Интерпретация квантовой теории событиями}

Здесь я продолжаю развивать идею интерпретации квантовой теории 
событиями \cite{Brg}:

Как мы видели, понятия и утверждения квантовой теории представляют 
понятия и утверждения о вероятностях точечных событий и их ансамблей.

Поведение физической элементарной частицы в вакууме подобно поведению этих 
вероятностей. В двух-щелевом эксперименте \cite{Mr} если в вакууме между 
источником физической частицы и детектирующим экраном помещается перего-родка с 
двумя щелями, то наблюдается интерференция вероятностей. Но если эту систему 
поместить в камеру Вильсона, то частица будет иметь ясную траекторию, 
обозначенную каплями конденсата, и всякая интерференция исчезнет. Это подобно 
тому, что физическая частица существует только в тот момент когда с нею 
случается какое-нибудь событие. В остальные времена частицы нет, а есть только 
вероятность того, что с нею что-нибудь случится.

Следовательно, если между событием рождения и событием детектирования с 
частицей не случается никаких событий, то поведение частицы - это поведение 
вероятности между точкой рождения и точкой детектирования; при этом наблю-дается 
интерференция. Но в камере Вильсона, где акты ионизации образуют почти 
непрерывную линию, частица имеет ясную траекторию и никакой интерференции. 
Эта частица движется потому, что такая линия не абсолютно непрерывна. Каждая 
точка ионизации имеет соседнюю ионизационную точку, и между этими точками 
никаких событий с частицей не происходит. Следовательно, физическая частица 
движется потому, что соответствующая вероятность распространяется в 
пространстве между этими точками.

Следовательно, частица есть ансамбль событий, связанных вероятностями. А 
заряды, массы, импульсы, среднее число частиц в конденсате и т.п. представляют 
статистические параметры волн этих вероятностей, распространяющихся в 
прост-ранстве-времени. Это объясняет все парадоксы квантовой физики.

\section{Заключение}

1. Основные свойства времени - одномерность и необратимость -, 
метрические свойства пространства и принципы теории относительности выводятся 
из логиче-ских свойств множества рекордеров. Таким образом, если у вас есть  
какое-нибудь множество объектов, имеющих дело с информацией, то "время" и 
"пространство" неизбежны на этом множестве. И не имеет значения, включено ли 
это множество в наш мир или в какие-либо другие миры, не имеющие изначально  
пространственно-временной структуры. Следовательно, пространственно-
временная структура есть следствие логических свойств информации.

2. Вероятность представляет обобщение классической пропозициональной логики. 
Вероятость есть логика еще не произошедших событий.

3. Мы сами выбираем выражение вероятности в форме (\ref{j}). Мы 
сами вводим операторы рождения и операторы уничтожения вероятности события. 
Мы сами добавляем две квазипространственные координаты к нашим трем, etc. То 
есть мы конструируем логическую структуру, в которой частицы, античастицы и 
калибровочные бозоны неизбежны. Каков вопрос - таков ответ.

Таким образом мы сами устанавливаем правила обработки вероятностной информации 
в форме законов квантовой теории. А значения параметров зависят от структуры 
наших приборов.

Похоже на то, что Природа дает нам только вероятности событий. А законы физики, 
оперирующие этими вероятностями, являются результатом свойств конструкции наших 
приборов и логического поведения нашего языка. Если где-нибудь действуют другие 
методы получения и обработки информации, то там физические законы будут иметь 
другую форму. Поэтому квантовая теория представляет один из возможных способов 
обработки вероятностной информации.

-----------

Таким образом, основными сущностями Универсума являются не частицы и поля, а 
логические события и логические вероятности. Универсум - т. е. время, 
пространство и все их содержимое - представляет производное из этой логики.

Информационная система людей с их приборами, в конечном счете образован-ная 
спинорами и клиффордовыми матрицами, оказалась настолько сильной, что 
позволяет изучать себя своими же средствами. Существует ли другая 
информаци-онная структура, сравнимая по теоретической силе с этой?


\end{document}